\documentclass[12pt,preprint]{aastex}
\usepackage{epsfig}
\usepackage{rotating}
\usepackage{natbib}
\def\etal{{et al.\thinspace}}

\def\spose#1{\hbox to 0pt{#1\hss}}

\def\multleft#1{\hbox to size{\vbox {\halign {\lft{##}\cr #1}}\hfill}\par}
\def\multright#1{\hbox to size{\vbox {\halign {\rt{##}\cr #1}}\hfill}\par}

\def\degmark{^\circ}
\def\boxit#1{\vbox{\hrule\hbox{\vrule\kern3pt\vbox{\kern3pt
          #1 \kern3pt}\kern3pt\vrule}\hrule}}

\def\cm{{\rm\thinspace cm}}

\def\erg{{\rm\thinspace erg}}
\def\eV{{\rm\thinspace eV}}

\def\keV{{\rm\thinspace keV}}

\def\ph{{\rm\thinspace ph}}
\def\s{{\rm\thinspace s}}
\def\ks{{\rm\thinspace ks}}

\def\ergcmps{\hbox{$\erg\cm\s^{-1}\,$}}
\def\ergpcmps{\hbox{$\erg\cm^{-1}\s^{-1}\,$}}
\def\ergpcmsqps{\hbox{$\erg\cm^{-2}\s^{-1}\,$}}

\def\pcmsq{\hbox{$\cm^{-2}\,$}}

\def\phpcmsqps{\hbox{$\ph\cm^{-2}\s^{-1}\,$}}

\makeatletter

\let\@internalcite\cite
\def\cite{\@ifstar{\citey}{\citefull}}
\def\citefull{\def\astroncite##1##2{##1\ ##2}\@internalcite}
\def\citey{\def\astroncite##1##2{##1\ (##2)}\@internalcite}
\def\citeyear{\def\astroncite##1##2{##2}\@internalcite}
\def\citename{\def\astroncite##1##2{##1}\@internalcite}
\def\@citex[#1]#2{\if@filesw\immediate\write\@auxout{\string\citation{#2}}\fi
  \def\@citea{}\@cite{\@for\@citeb:=#2\do
    {\@citea\def\@citea{; }\@ifundefined
       {b@\@citeb}{{\bf ??}\@warning
       {Citation `\@citeb' on page \thepage \space undefined}}%
{\csname b@\@citeb\endcsname}}}{#1}}
 
\def\@cite#1#2{#1\if@tempswa #2\fi} 
\def\@biblabel#1{}
 
\def\astroncite#1#2{#1\ #2}
\makeatother

\begin{document}

\title{Relativistic Broadening of Iron Emission Lines in a Sample of AGN}

\author{Laura~W.~Brenneman\altaffilmark{1}, 
Christopher~S.~Reynolds\altaffilmark{2}}

\altaffiltext{1}{NASA's GSFC, mail code 662, Greenbelt MD~20771}
\altaffiltext{2}{Dept. of Astronomy, University of Maryland, College
Park, College Park MD~20742}

\begin{abstract}
We present a uniform X-ray spectral analysis of eight type-1 active
galactic nuclei (AGN) that have been previously observed with
relativistically broadened iron emission lines.  Utilizing data from
the {\it XMM-Newton} European Photon Imaging Camera (EPIC-pn) we
carefully model the spectral continuum, taking complex intrinsic
absorption and emission into account. 
We then proceed to model the broad Fe K$\alpha$ feature in each source
with two different accretion disk emission line codes, as well as a
self-consistent, ionized accretion disk spectrum convolved with
relativistic smearing from the inner disk.  Comparing the results, 
%Our principal assumption, supported by recent simulations of
%geometrically-thin accretion disks, is that no iron line emission (or
%any associated X-ray reflection features) can originate from the disk
%within the innermost stable circular orbit.  Under this assumption,
%which tends to lead to constraints in the form of lower limits on the
%spin parameter, 
we show
that relativistic blurring of the disk emission is required to explain
the spectrum in most sources, even when
one models the full reflection spectrum from the photoionized disk.
\end{abstract}


\section{Introduction}
\label{sec:intro}

Black holes (BHs) are among the simplest objects in nature, able to be
defined completely by mass, spin and charge alone.  But whereas the
mass of a BH is relatively straightforward to measure, provided
material orbiting the BH can be observed (i.e., we are not blinded by
the central engine), spin has proven much more challenging to
constrain.  With the advent of sensitive X-ray spectroscopy, allowing
us to characterize processes very close to the BH in detail,
we are now in a position to search for signatures of BH
spin.  This paper and its forthcoming counterpart (part II) represents
the first survey to quantify the angular
momenta of BHs in a sample of AGN using relativistically broadened
iron lines as our spin diagnostic.  Here, in Part I, we examine the
statistical need for
including relativistic effects when modeling the emission from the
inner accretion disks around BHs.

As all current observables from a BH system are in one way or
another related to the accretion process, any diagnostic of BH
spin must be based on spin-induced changes to the accretion disk
structure.  BH spin manifests
itself by setting the location of the innermost stable circular orbit
(ISCO) which, in turn, sets the effective inner edge of a
geometrically-thin accretion disk.  For the purposes of this study,
the ``inner edge'' of the disk is defined as the location inside of
which no X-ray reflection signatures (e.g., iron line emission and
Compton reflection hump) are produced either due to the disk becoming
optically-thin or too highly ionized.  As the magnitude of the
(prograde) spin increases, frame-dragging stabilizes otherwise unstable
orbits, causing the ISCO to move closer to the BH.  The
location of the inner
edge of the disk can therefore be used to measure BH spin
\citep[e.g.,][]{RN2003,Brenneman2006,RF08}.

The position of the ISCO can be constrained through detailed modeling
of the X-ray spectrum of the disk.  Two popular modeling schemes
include fitting the thermal continuum of the disk \citep[e.g.,][]{NMS08} and
fitting the shape of the fluorescent Fe K line emitted by the disk in
response to irradiation by the X-ray power-law \citep[e.g.,][]{Brenneman2006}.
The former method relies on {\it a priori} knowledge of the BH mass
and distance as well as the disk inclination angle, however, and hence
is most readily applied to Galactic BH systems (GBHs) in thermally
dominant states as opposed to supermassive BHs found in the centers of
AGN.  By contrast, the shape of the iron line
profile depends upon the gravitational redshift and fluid velocity
field, both of which depend only upon the mass and radius of the black
hole ($M/R$).  Because the radius of the event horizon is directly
dependent of the mass of the BH, the iron line profile is therefore mass-independent,
making it a useful technique for deriving spin in any BH system
\citep{RN2003}.  This is an 
especially useful technique in AGN in which the BH mass is often poorly constrained.

The shape of the iron line profile is determined by several physical
parameters: the BH angular momentum, the radial extent of the disk,
the emissivity profile of the disk and the inclination angle of the
disk to the observer's line of sight.  A proper model for the iron
line must also take into account the effects of the Doppler shift,
Special Relativity (e.g., beaming) and General Relativity (e.g.,
gravitational redshift), each self-consistently calculated for the given
value of BH spin.  The two widely available relativistic line
models within the {\sc xspec} spectral analysis software were computed
only for single values of spin; the {\tt diskline} model describes
only non-spinning Schwarzschild BHs \citep{Fabian1989}, and the {\tt
laor} model describes near-maximally ($a=0.998$) Kerr BHs
\citep{Laor1991}.  Thus, neither of these models allow for the
self-consistent computation of a line profile from an arbitrary spin
BH, even if one allows the inner edge of the line emitting
region to vary.  If one wishes to constrain BH spin using
self-consistent models, therefore, it is necessary to create a new,
fully-relativistic model code which enables the BH spin to be fit as a
free parameter.

%
% Shorten next three parags to de-emph variable spin stuff in this paper?
%
We have developed such a model as both a line code ({\tt kerrdisk})
and a convolution kernel ({\tt kerrconv}) that can be applied to a
full reflection spectrum expected from the disk.\footnote{These models
  are now available for
public use in {\sc xspec12}.  Future updates will be made available
initially at
http://www.astro.umd.edu/$\sim$chris/kerrdisk/kerrdisk\_model.html.}  
To simplify the
models, we have assumed that the BHs have no net charge (a
very good approximation for all likely astrophysical BHs) and
that there is a negligible amount of mass in the surrounding matter
compared to the mass of the BH.  Thus, the spacetime is
described by the usual Kerr metric.  These models were introduced and
detailed in \citet{Brenneman2006}, hereafter BR06.  BR06 also contains
detailed comparisons of the {\tt kerrdisk} and {\tt kerrconv} models
to similar codes developed by other groups \citep[e.g.,][]{Dovciak2004}.

Our models assume that, in the co-moving frame of the disk surface,
the emitted line/reflection spectrum is limb-darkened using the same
prescription as \citet{Laor1991}.  In fact, limb-darkening effects are
usually completely unimportant to studies such as that conducted here;
they only become significant for very high inclination angles, and
furthermore it is only light-bending induced {\it differences} in the
angle at which various parts of the disk are viewed that change the
profile of the iron line.

BR06 also presents a series of model fits to the $330\ks$ {\it
XMM-Newton}/EPIC-pn spectrum of the canonical broad iron line Sy 1 AGN
MCG--6-30-15, noting that the best fit is achieved with a {\tt
kerrconv} model acting on an ionized disk reflection spectrum, where
the inner edge of the disk was $r_{\rm min} \leq 1.62\,r_{\rm g}$.
These data have allowed us to place formal constraints on the BH spin
in MCG--6-30-15 of $a \geq 0.987$ ($90\%$ confidence level).  Because this
fit was performed under the assumption of a strict truncation of the
iron line region at the innermost stable circular orbit (ISCO) in the
disk, this constraint is somewhat weakened ($a \geq 0.92$) if the iron line
emission region does in fact bleed inside of the ISCO \citep{RF08}.
Given that MCG--6-30-15 harbors the broadest iron line in an AGN
observed to date, this model fit reinforces our expectation that this
AGN contains a BH of near-maximal spin.  Many other, similar Sy 1
galaxies are also observed to have broad iron lines, however
\citep{Guainazzi2006,Nandra2007,Miller2007}, which begs two important
questions: (1) can spectral observations of broad iron lines
be used to robustly constrain BH spin, and (2) how is spin distributed
between and among different types of BH systems?

To begin addressing these questions, we have collected a sample of
eight other Sy 1 AGN from the {\it XMM-Newton} archive that have
previously been observed to have broad iron lines.  Using the same
technique we used to model the spectrum of MCG--6-30-15 by first
addressing the continuum and absorption and then considering the broad
line region, we have assessed each AGN independently, with the initial goal of
establishing whether we can definitively show that relativistic
effects from the inner accretion disk are evident in these spectra
(Part I, this paper).  Following this conclusion, we will examine the
archived {\it Suzaku} spectra of the AGN in our sample in order to take
advantage of the increased spectral resolution and extended
energy range that this observatory offers.  The goal of this second study is to 
obtain statistically robust BH spin constraints whenever possible
(Part II, forthcoming).

Section~2 of this work discusses the observations of the sources in our sample and
the method used to reduce the EPIC-pn data and extract spectra.  A
summary of our modeling technique is given in \S3.  Section~4 contains the
detailed spectral fitting results for each source.  Results and tables
comparing the relevant best-fit parameters for each source can be
found in \S5.  Implications of these results, conclusions we have
drawn and a prognosis for future work in this field are discussed in
\S6.  Part II will discuss the {\it Suzaku} results and
provide the first formal constraints on BH spin in our sample of AGN.

\section{Observations and Data Reduction}
\label{sec:reduction}

Our source selection was based on work by \citet{Nandra2007}
and the review by \citet{Miller2007}, both of whom consider the
robustness of broad iron line detections across different source
populations.  The Miller review, in particular, collects all the
recent X-ray data published on Seyfert sources and categorizes these
objects by the strength and robustness of their broad iron line
features.  We have used these two studies as a guide for selecting the
candidates for our sample, which we list in Table~\ref{tab:tab1.tex}.

For this work, our data have been collected from the {\it XMM-Newton} Science
Archive, using only publicly available observations.  {\it XMM-Newton}
data were preferentially used as opposed to {\it Chandra} data because
of the superior collecting area of {\it XMM-Newton} in the $2-10 \keV$
band where the broad Fe K$\alpha$ feature is the most prominent.
%Although there is a growing body of broad iron line results from {\it
%Suzaku}, 
There were insufficient numbers of publicly available {\it Suzaku}
datasets at the time Part I of this study began (early 2007) to allow
this work to be undertaken using {\it Suzaku}.  The {\it Suzaku} study
will be the focus of Part II in a forthcoming paper.

In this paper, we exclusively focus on the EPIC-pn data to avoid
cross-calibration issues between the pn, MOS and RGS instruments.
Though the RGS spectra in particular can provide detailed insight into
the $0.3-2.0 \keV$ spectrum and help illuminate the nature of the soft
absorption and excess seen in so many Sy 1 AGN, the mismatch in
resolution and signal-to-noise as well as the possibility of
cross-calibration errors severely reduce the utility of including RGS
data in our formal spectral fitting.
The basic parameters characterizing the observations of each of our
sources are shown in Table~\ref{tab:tab1.tex}.

\begin{sidewaystable}
\begin{center}
%\begin{scriptsize}
\begin{tabular} {|l|l|l||l|l|l|l|l|l|l|} 
\hline \hline
{\bf AGN} & {\bf Date Obs.} & {\bf Exposure (s)} & {\bf Counts} & {\bf pn Mode} & {\bf Filter} & 
{\bf RA (h)} & {\bf Dec. ($\degmark$)} & {\bf z} & {\bf Source Type}\\  
\hline \hline
{\bf MCG--5-23-16} & 12/8/2005 & $121578$ & $1.2 \times 10^6$ & LW & Medium & $9.7945$ & $-30.7489$ 
& $0.0085$ & Sy 1.9 \\
\hline
{\bf Mrk~766} & 5/20/2001 & $111789$ & $2.4 \times 10^6$ & PSW & Medium & $12.3074$ & $29.8128$ 
& $0.0129$ & Sy 1.5, NLS1 \\
\hline
{\bf NGC~3783} & 12/19/2001 & $131025$ & $1.6 \times 10^6$ & PSW & Medium & $11.6505$ & $-37.7385$ 
& $0.0097$ & Sy 1 \\
\hline
{\bf NGC~4051} & 5/16/2001 & $121889$ & $2.5 \times 10^6$ & PSW & Medium & $12.0527$ & $44.5313$ 
& $0.0023$ & Sy 1.5, NLS1 \\
\hline
{\bf Fairall~9} & 7/5/2000 & $29209$ & $2.0 \times 10^5$ & PFW & Medium & $1.3961$ & $-58.8058$ 
& $0.0470$ & Sy 1 \\
\hline
{\bf Ark~120} & 8/24/2003 & $112130$ & $2.3 \times 10^6$ & PSW & Thin1 & $5.2699$ & $-0.1502$ 
& $0.0327$ & Sy 1 \\
\hline
{\bf NGC~2992} & 5/19/2003 & $28917$ & $3.4 \times 10^5$ & PFW & Medium & $9.7617$ & $-14.3264$ 
& $0.0077$ & Sy 1.9 \\
\hline
{\bf 3C~273} & 6/30/2004 & $19861$ & $3.2 \times 10^5$ & PSW & Thin1 & $12.4852$ & $2.0524$ 
& $0.1583$ & QSO, Sy 1 \\ 
\hline \hline
\end{tabular}
%\end{scriptsize}
\end{center}
\caption[Sy 1 AGN Sample Observations]{Observation parameters for the eight AGN in our sample 
of broad iron line sources.  Exposure times are after filtering, and
counts are measured from $2-10 \keV$ after filtering.  \label{tab:tab1.tex}}
\end{sidewaystable}

For each source in question we performed reprocessing (when necessary)
and data reduction with the SAS version 7.0.0 software, including the
latest CCF calibration files.  We have used the SAS {\tt epatplot}
task to compute the fraction of single, double, triple and quadruple
events as a function of energy and compared these fractions to their
nominal values as measured from weak source observations. For sources
that are affected by pile-up, these fractions deviate from the nominal
values due to the higher probability of wrong pattern
classification. No significant deviation from the nominal single and
double distributions was found for any of our sources, indicating that
our EPIC-pn observations are not affected by pile-up.  Source and
background spectra and light curves were extracted using the {\tt
xmmselect} task from the SAS GUI.  Response matrices and ancillary
response files were generated using the {\tt rmfgen} and {\tt arfgen}
tasks.  These were then grouped with the spectral files using the {\sc
ftools} package {\tt grppha} with a minimum of 25 counts/bin in order
to use $\chi^2$ as a meaningful goodness-of-fit statistic.  Spectral
analysis was carried out as before for MCG--6-30-15 (BR06) with {\sc
xspec} version 12.4.0. 
%with {\tt kerrdisk} model
%installed as ``local models''\footnote{These models are available for
%download and use in {\sc xspec} from
%http://www.astro.umd.edu/$\sim$chris/kerrdisk/kerrdisk\_model.html.}.

\section{Spectral Analysis Methodology}
\label{sec:analysis}

In the following Section we outline the method used for the data
reduction and spectral analysis of the AGN in our sample.  We then
discuss the spectral modeling of each source, culminating in a
relativistically smeared ionized disk reflection spectrum model from
which we infer the importance of taking GR into account when fitting
each source.  This is a critical first step toward measuring BH spin.

\subsection{Continuum Fitting}
\label{sec:continuum}

We have endeavored to make the spectral analysis method as uniform as
possible for all the AGN examined.  There are, however, certain
intrinsic physical differences between the sources that must be taken
into account in order to properly model the continuum and isolate the
iron line(s) in each case.  Though all of our sources are Sy 1 AGN (or
otherwise observed in states where they manifest broad disk lines),
each system is unique in its physical properties.  Some have higher
signal-to-noise, either because of higher flux, longer observation
time or both.  Some exhibit evidence for complex, multi-zone warm
absorbers (WAs) intrinsic to the central engine, whereas some show
only cold absorption from neutral hydrogen along our line of sight.  A
soft excess is seen in some sources but not others.  And of course,
the strength and breadth of the Fe K$\alpha$ line varies from source
to source as well, though all have been chosen because they have
previously shown evidence for broad iron lines.

In order to examine the iron line region in detail the continuum must
first be accurately modeled.  After excluding the energy ranges
relevant to the iron line ($4.0-8.0 \keV$) and the mirror edges ($\sim
1.5-2.5 \keV$), we examine the rest of the $0.6-10.0 \keV$ spectrum,
fitting it with a power-law continuum typical of an AGN which is then
modified by cold photoabsorption from neutral hydrogen.  We set the
minimum value of $N_{\rm H}$ equal to the Galactic column density
along our line of sight to the source\footnote{This parameter is
determined from H\,{\sc i} $21 \cm$ surveys.  See
http://heasarc.gsfc.nasa.gov/cgi-bin/Tools/w3nh/w3nh.pl for the column
density calculator as well as \citet{Dickey1990} and
\citet{Kalberla2005}.}.  If further absorption is necessary, we employ
a second neutral absorber (either {\tt zphabs} or {\tt zpcfabs}) and
allow the column (and covering fraction, if applicable) to vary as necessary to
accommodate the cold absorption intrinsic to the system.  Following
the prior published analyses of these AGN, some sources require a
second power-law component as well to properly model the
basic continuum shape.  We disregard energies below $\sim 0.6 \keV$
due to calibration uncertainties in this range for the EPIC-pn
instrument.

If significant residual features remain after fitting this continuum
model which indicate the presence of a soft excess and/or warm
absorption, we include these components one by one as long as they
make a significant difference in the global goodness-of-fit according
to the {\it F}-test: $\Delta\chi^2 \leq -4$ for each new parameter
introduced into the fit \citep{Bevington}.  The soft excess is
typically parameterized by a blackbody component representing thermal
emission in the central engine, but may also be modeled by
bremsstrahlung emission or Comptonized emission from a thermal disk if
either of these more complex forms gives a significantly better
reduction in $\chi^2$.  The warm absorption is modeled using the same
{\sc xstar}-generated multiplicative table model described in
\citet{Brenneman2007} and BR06 in the analyses of NGC~4593 and
MCG--6-30-15, respectively.  Solar abundances are assumed for all elements and the
redshift of the warm absorber(s) is set to that of the AGN host
galaxy.  Some AGN show no evidence for a warm absorber, while others
statistically require up to two physically separate WA components,
each exhibiting a distinct column density ($N_{\rm H}$) and ionization
parameter ($\xi$).  We note that the EPIC-pn does not have sufficient
spectral resolution to discern velocity differences between the warm
absorber and the host galaxy; thus we do not need to include the warm
absorber velocity as a free parameter.

Once the continuum has been properly modeled, we initially freeze the parameter
values for all components except the power-law spectral index
($\Gamma$) and normalization (flux in $\phpcmsqps$) and restrict our
energy range from $2.5-10.0 \keV$ in order to focus on the hard X-ray
spectrum and the iron line region.  Including the energies from
$4.0-8.0 \keV$ again, we check for residual emission (or absorption)
lines from $6.4-6.97 \keV$, indicating the presence of unmodeled
neutral and/or partly ionized Fe K$\alpha$ in the spectrum.  Because
these sources have been pre-determined to possess significant,
broadened iron emission, it is not surprising that such residual
emission features are seen in each case.  We begin by attempting to
fit the $6.4 \keV$ line of neutral iron (and any other ionized iron lines)
with a simple Gaussian feature, with the line width frozen at
$\sigma=0 \keV$ (i.e., intrinsically narrow) and the redshift again
set to the cosmological value for the host galaxy.  Such narrow lines
are expected to originate from the outer accretion disk, or perhaps
from reprocessing in a Compton-thick toroidal region surrounding the
nucleus.  As expected, in each case we note
significant residual ``wings'' remaining around the Fe K$\alpha$
feature after this narrow core was fit, indicating the presence of a
broad component to the line originating from closer to the BH.

\subsection{Fitting the Iron Line Profile}
\label{sec:Fe_line}

To assess the robustness of the broad iron component and the
importance of relativistic smearing in the spectrum, we follow our continuum
fitting procedure (including modeling the narrow core of the iron
line) by running each source through two additional steps of more
complex modeling in the broad iron line region.

We begin by adding to the spectral model a broad iron line
    described by: (1) a broad Gaussian line ($\sigma=$free), replaced
    by (2) the {\tt diskline} model (spin hardwired at
    $a=0.0$), then replaced by (3) the {\tt laor} model (spin hardwired at $a=0.998$).
    The broad Gaussian is used primarily to gauge the energy and
    velocity width of the line in order to infer its radius of origin,
    whereas the relativistic line models are thought to be
    more physically accurate representations of the line morphology.
    The energy of the line is initially set to a rest-frame energy of $6.4
    \keV$, corresponding to
    the energy of the cold Fe K$\alpha$ line.  We use the
    change in $\chi^2/\nu$ with the addition of the broad line
    component to verify whether a
    broad line is statistically warranted in our global model.  According to
    this criterion, all of our sources required a broad line, and relativistic
    broadening was robustly preferred over a simple Gaussian in half
    of the cases.  This is unsurprising since our objects
    were preselected to possess reported broad iron lines.
    Using the simple {\tt diskline} and {\tt laor} models, we also note the inner radius
    of disk emission: a broad line with the potential to diagnose BH
    spin should have an inner disk radius of roughly $r_{\rm min} \leq
    6\,r_{\rm g}$.  We relax this restriction here to allow for
    differences in fit that may be achieved through different models,
    requiring that $r_{\rm min} \lesssim 20\,r_{\rm g}$ to infer that
    relativistic broadening is important.

We next perform a study to assess the degree to which the X-ray
reflection continuum may be contaminating (or even mimicking) the
broad iron line.  To begin with, we replace the broad iron line with a
{\it static} (unsmeared) ionized reflection spectrum \citep{Ross2005}
and refit.  We then note the change in the goodness-of-fit criterion
when the reflection spectrum is subjected to smearing characterizing a
relativistic disk by convolving the spectrum with a line profile
smearing kernel.  The first such kernel we use is for a non-spinning
BH ($a=0.0$) and is known as {\tt rdblur}. The second convolution
model assumes a maximally-spinning BH ($a=0.998$) and is known as {\tt
kdblur}.

We force the outer radius of disk emission to be $r_{\rm
max}=400\,r_{\rm g}$ in all the relativistic broad line
fits.  The value itself is somewhat arbitrary; as long as $\alpha
\gtrsim 2$
(where $\alpha$ here denotes the emissivity index of the disk and
the disk radiates as a function of radius: $r^{-\alpha}$), emission
from this outer part of the disk is negligible, and we have already
incorporated a separate narrow line core component whose width
dictates an origin in the outer disk or torus region.  For the ionized
disk component, we begin with the assumptions of solar iron abundance
and relative neutrality (${\rm Fe/solar}=1$ and $\xi=30 \ergpcmps$).
We then relax these assumptions and allow these two parameters to fit
freely, provided that reasonable statistical error bars can still be
obtained for each parameter.  We also note that all the models used
above assume that the disk radiation is truncated at the ISCO; that
is, that no contribution to the continuum or iron line profile
originates from within this region.  
%For {\tt kerrdisk} and {\tt kerrconv},
%we initially assume that the inner and outer portions of the disk emit
%under the same emissivity index: $\alpha_1=\alpha_2$, where the disk
%radiates as $r^{-\alpha}$ at any given radius.  Again, if the
%signal-to-noise if sufficient the two indices are allowed to each fit
%freely, but due to limitations in the number of photons we have for
%each data set, all of our sources in this sample have their indices
%linked at the same value.  The only source we have analyzed so far
%with enough counts to support a
%scenario where $\alpha_1 \neq \alpha_2$ is MCG--6-30-15
%\citep{Brenneman2006}.

\section{Source-by-source Model Fitting}
\label{sec:sources}

\subsection{Target selection}

As mentioned in the previous Section, we have selected the AGN for our
sample primarily based on the research performed by other groups on
the robustness of broad iron line observations in various AGN and the
likelihood of being able to identify relativistic blurring effects in these sources
using the iron line method \citep{Nandra2007,Miller2007}.  In
particular, \citet{Miller2007} discusses recent results from $30$ Seyfert AGN
in which relativistic disk lines have been detected and published, and
divides those sources up into three ``tiers'' based on the nature and
robustness of the detections.  Many of the sources overlap with those
presented in Nandra et al.  Though we have examined observed spectra
from all of the $9$ AGN listed in Tier~1 (the most robust cases), NGC~3516
and NGC~4151 were not included in our study due to their extremely
complex soft spectra \citep{Turner2005,Schurch2004}.  We did analyze
the other $7$ sources in full: 3C~120, MCG--6-30-15, MCG--5-23-16,
NGC~2992, NGC~4051, NGC~3783 and Mrk~766.  3C~120 is not included in
this work, however, as our spectral fitting results for this object
were inconclusive in several ways due in part to a paucity of
photons.  We have also examined $3$ other sources from Tier~3 that are
mentioned prominently in Nandra \etal: Fairall~9, Ark~120 and 3C~273.

\subsection{Source-by-source results}

We present the best-fitting results of our spectral analysis for each source in
its own subsection below.  In many cases the {\tt kdblur} and {\tt rdblur}
results were very similar, implying that little can be
inferred about the spin of the BH in these AGN without more data.  A more detailed
comparison of all of our models is taken up in
\S\ref{sec:compare_all}.  Note that the reflection fractions quoted
are not fitted values, but rather are estimates calculated 
%using the method described in \S\ref{sec:compare_all} 
for the purpose of
comparison to other studies which may model disk reflection
differently (e.g., with {\tt pexrav} rather than {\tt reflion}).
We expand on this point in greater detail in \S\ref{sec:compare_all}.
Detailed spectral fitting results for MCG--6-30-15
can be found in BR06 and are therefore not included herein.  

A table of best-fit parameters for each source is included, as are
figures representing the residual features left behind after initially
fitting the spectrum with an absorbed power-law, the final best
fit to the spectrum, and the best-fit model components for each
object.  
%The line residuals represent the data-to-model ratio for the
%best-fit model after removal of the component(s)
%used to fit the broad Fe K$\alpha$ line; the narrow line core is left in the
%model.  

\subsubsection{MCG--5-23-16}
\label{sec:MCG5}

MCG--5-23-16 is a moderately absorbed Seyfert galaxy of intermediate
type (Sy 1.9).  It is relatively nearby at a redshift of $z=0.0085$,
and with a typical $2-10 \keV$ flux of $F_{2-10} \sim 8 \times 10^{-11} \ergpcmsqps$
it is one of the brightest known Seyfert galaxies \citep{Reeves2006}.
The source has
been observed previously to have a soft excess below $\sim 1 \keV$ and
an absorbing column of $N_{\rm H} \sim 10^{22} \pcmsq$
\citep{Dewangan2003,Balestra2004}, and {\it ASCA} observations have
indicated the presence of a 
broad Fe K$\alpha$ line of $EW \sim 200 \eV$
\citep{Weaver1997,Weaver1998}.  The line was successfully modeled
with a narrow core at the rest-frame energy of $6.4 \keV$ and a broad
component superposed on it.  This feature was modeled with similar
success in the {\it XMM-Newton} observations of Dewangen \etal and
Reeves \etal, as well as the {\it Suzaku} observation also published
by these authors \citep{Reeves2007}.

In December 2005, MCG--5-23-16 was simultaneously observed with {\it
  XMM-Newton}, {\it Chandra}, {\it Suzaku} and {\it RXTE}.  The
{\it XMM-Newton} results are reported by Reeves \etal as well as
Braito \etal, and our data
reduction followed those works \citep{Reeves2006,Braito2006}.  The EPIC-pn
instrument had a net exposure of $96 \ks$ and returned $\sim 1.2 \times
10^6$ photons after filtering.

As suggested by Reeves \etal, a simple photoabsorbed power-law fit
does not model the continuum well, especially below $\sim 1 \keV$.
Adding in a source of soft thermal emission also leaves prominent
residuals and does not adequately account for the shape of the soft
excess, so following the lead of Reeves \etal we have employed a two-power-law 
model \citep{Reeves2006}: one of the power-law components is
affected only by Galactic photoabsorption ($N_{\rm H}=8 \times 10^{20}
\pcmsq$ as per Reeves \etal), and hence leaves the AGN system
effectively unscattered, while the other component is also
subject to absorption intrinsic to the AGN, in this
case with a fitted absorbing column density of $N_{\rm H}=1.19 \pm
0.01 \times 10^{22}
\pcmsq$.  This component also experiences scattering within the
system.  We set both power-law photon indices to be equal, indicating that
the two power-law components originate from the same physical reservoir of
photons (with $\Gamma=1.66 \pm 0.01$).  Calculating the ratio of the flux of the
scattered power-law
component to the unscattered power-law component (from $0.6-10.0
\keV$, fitting only the continuum)
yielded an estimate of
the optical depth of the scattering plasma.  In this case $\tau=7.37
\times 10^{-4} \phpcmsqps / 1.98 \times 10^{-2} \phpcmsqps
\approx 0.04$, so the scattered fraction is low, implying that the electron
plasma is optically thin.  No evidence for spectral curvature due to
the presence of a WA was found in this source.

The spectrum does show evidence for two prominent, narrow emission
lines of iron K$\alpha$ and K$\beta$ (at $E_{K\alpha}=6.40 \pm 0.01
\keV$ and $E_{K\beta}=7.06^{+0.02}_{-0.01} \keV$,
with $EW_{K\alpha}=167^{+15}_{-13} \eV$ and
$EW_{K\beta}=104^{+32}_{-30} \eV$, respectively), as well as three narrow
absorption lines of intermediately ionized iron or nickel at $E_1=7.24^{+0.03}_{-0.04} \keV$,
$E_2=7.48 \pm 0.04 \keV$ and $E_3=7.85^{+0.02}_{-0.05} \keV$
($EW_1=-79^{+30}_{-29} \eV$, $EW_2=-121^{+49}_{-50} \eV$, and
$EW_3=-143 \pm 43 \eV$, respectively).   

After successfully modeling the continuum and narrow line parameters
for MCG--5-23-16, we restricted our attention to the hard spectrum
($2.5-10.0 \keV$) and analyzed the broad line component of this source
using the method outlined above in \S\ref{sec:Fe_line}.  The initial
broad Gaussian fit implied that the broad line originates in neutral
iron with $E=6.34 \pm 0.02 \keV$ and $\sigma=0.28 \pm 0.03 \keV$.  The
width of the line suggests an origin within $d \sim 100\,r_{\rm g}$ of
the BH.
A {\tt laor} line component fit the data marginally better than a {\tt
  diskline}, with an improvement of $\Delta\chi^2/\Delta\nu=-145/-4$ in global
goodness-of-fit over a model without a broad iron line component.
Here, $\nu$ is the number of degrees of freedom in the fit, defined by
the total number of spectral channels in the data minus the number of
free model parameters, so a decrease in $\nu$ implies an increase in
the number of free model parameters.

\begin{figure}
\begin{center}
\includegraphics[width=0.33\textwidth,angle=270]{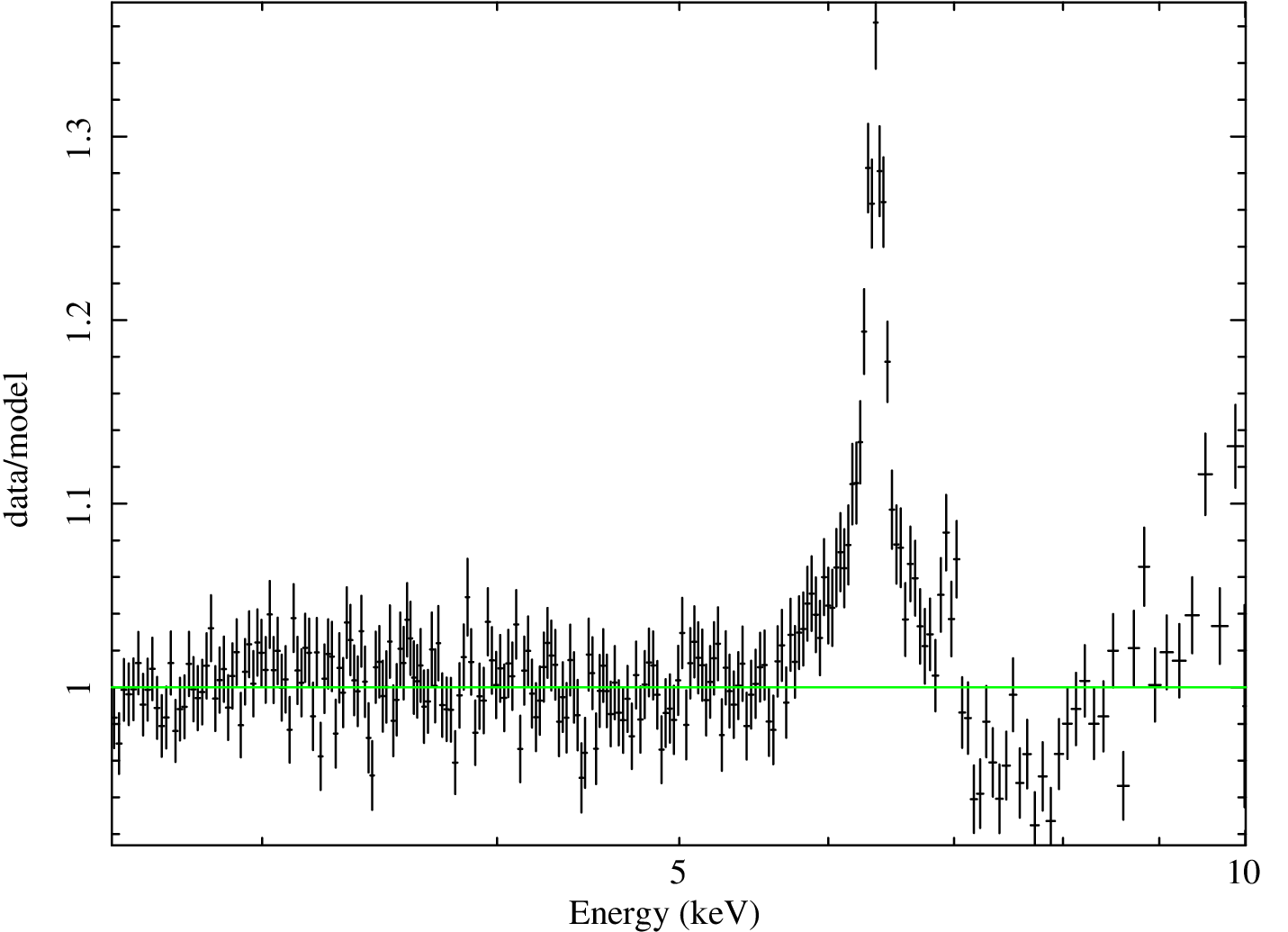}
\includegraphics[width=0.33\textwidth,angle=270]{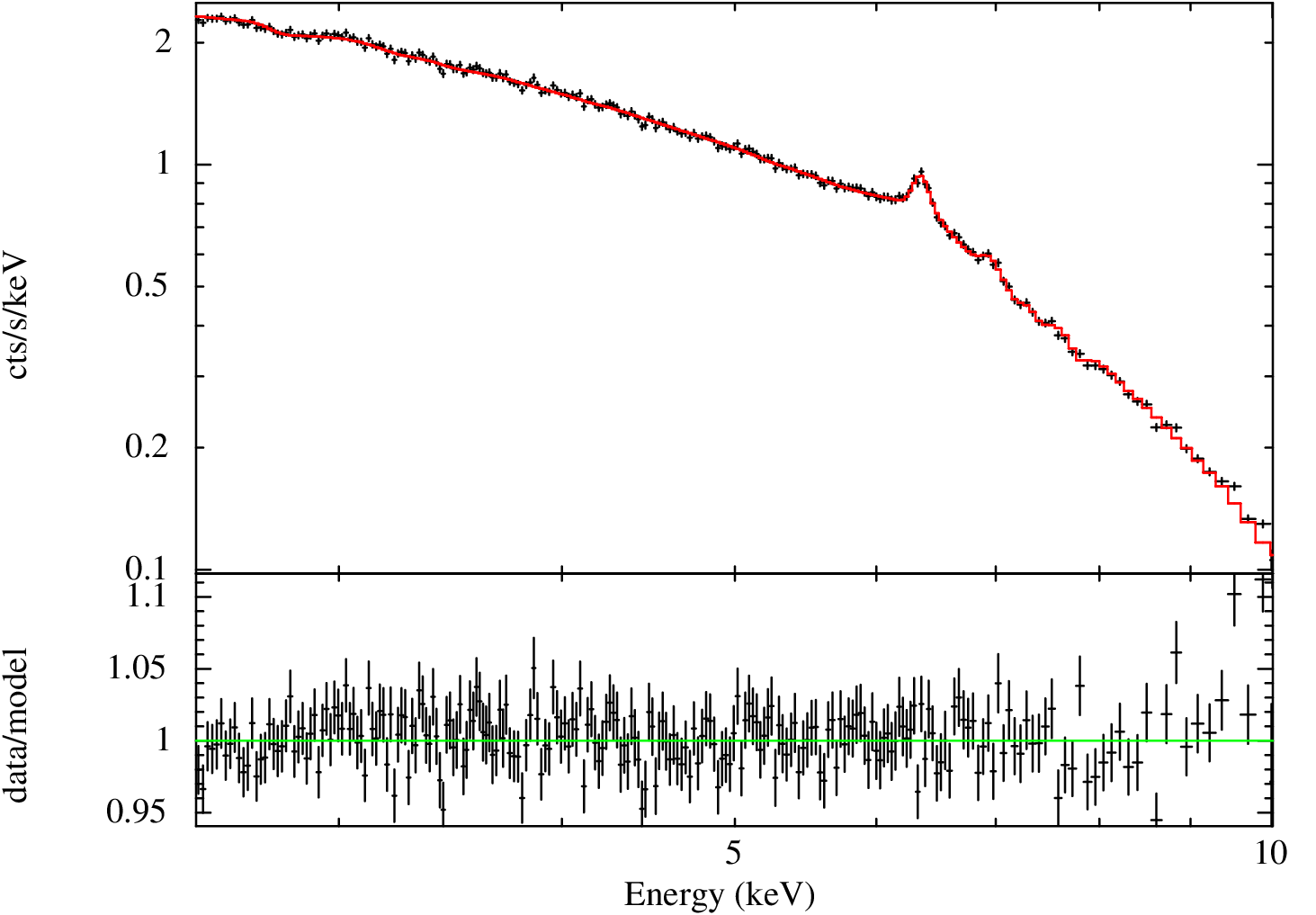}
\end{center}
\begin{center}
\includegraphics[width=0.33\textwidth,angle=270]{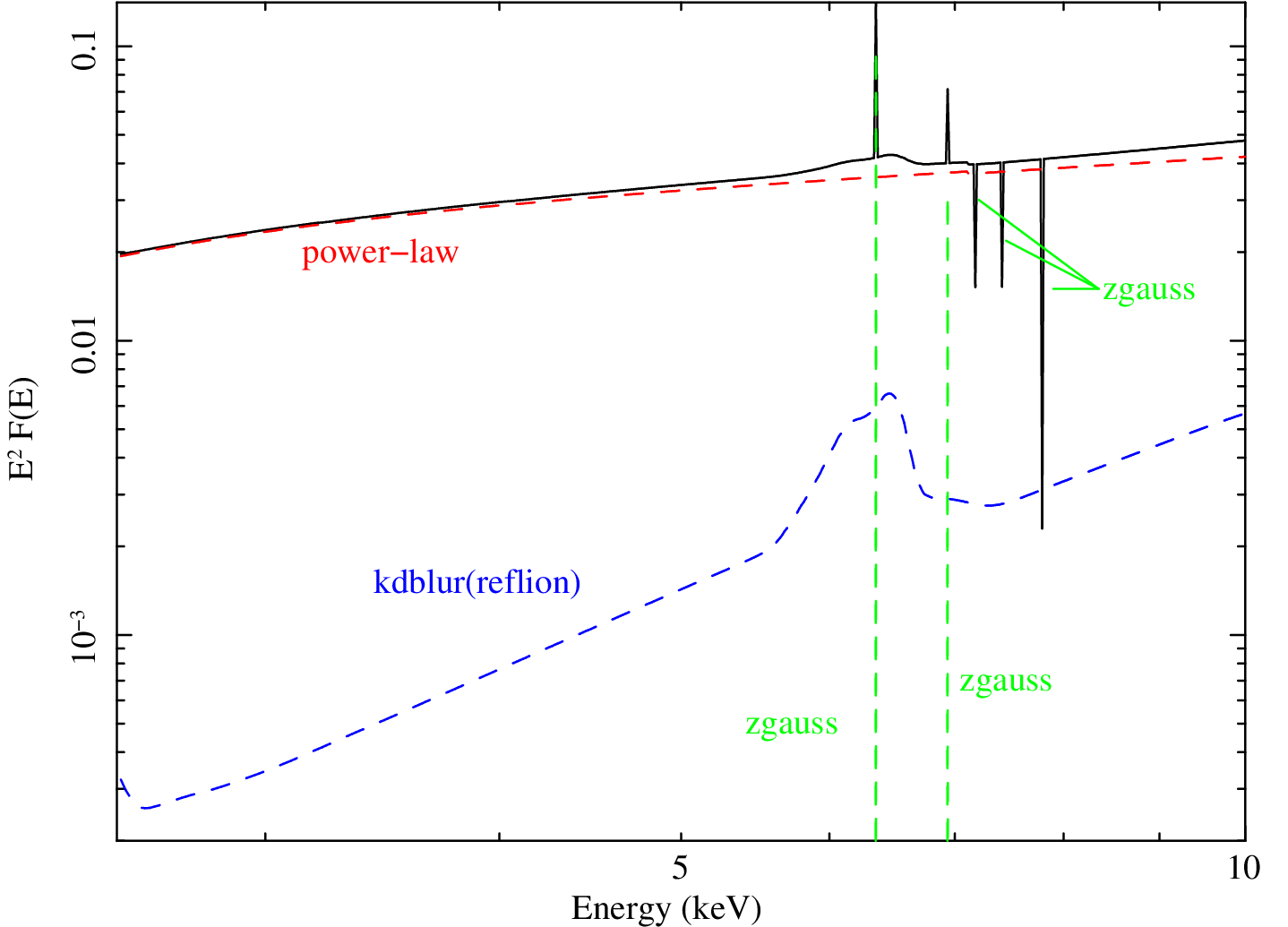}
\end{center}
\caption[Spectral fit for MCG--5-23-16.]{{\it Top left:} The $2.5-10.0
  \keV$ spectrum of MCG--5-23-16 fit with a two-power-law model
  modified by Galactic photoabsorption; data-to-model ratio.  Note the
  prominent Fe K line and the large residual emission component above
  $7 \keV$, which likely indicates the presence of ionized disk
  reflection.  $\chi^2/\nu=2675/1499\,(1.78)$.  {\it Top right:}
  Best-fit model for MCG--5-23-16, including our continuum
  model, two narrow emission lines, and three narrow absorption lines
  as well as an ionized disk reflection spectrum convolved with our
  {\tt kdblur} relativistic smearing kernel.
  $\chi^2/\nu=1485/1488\,(1.00)$.  {\it Bottom:} An ${\rm
  E}^2\,{\rm F(E)}$ plot depicting the relative flux
  in each of the best-fit model components in the spectrum.}
\label{fig:mcg5_figs}
\end{figure}

We determined
that an ionized disk reflection spectrum convolved with {\tt kdblur}
provides the best model fit, though the inner disk radius could only
be weakly constrained ($r_{\rm min} \leq 63\,r_{\rm g}$).  This
prevents us from being able to place any meaningful constraints on the
spin of the BH in this source with these data, in spite of the presence of
relativistic broadening in the iron line profile.  Note that
eliminating soft energies from the fit removed the statistical need
for a second, soft power-law in the fit.  
%This power-law component
%therefore appears to have zero slope in Fig.~\ref{fig:mcg5_figs}c.  
For a full
listing of best-fit model parameters and error bars for MCG--5-23-16,
see Table~\ref{tab:tab2.tex}.
The residual iron line feature, best-fit model and best-fit model
components for the hard spectrum are 
shown in Fig.~\ref{fig:mcg5_figs}.  The {\tt diskline} and {\tt laor}
fits are shown in Table~\ref{tab:tab10.tex}, while comparisons of the
ionized disk reflection spectrum with and without relativistic
smearing can be found in Table~\ref{tab:tab11.tex}.
%The best-fitting model for MCG--5-23-16 is the ionized disk reflection
%spectrum \citep{Ross2005} convolved with {\tt kdblur}.  
%Although we
%were not able to constrain the BH spin in this source, we did achieve
%a fit constraint on the inner radius of the disk emission: $r_{\rm min}
%\sim 16\,r_{\rm g}$.  This corresponds to a range of $\sim
%36-96\,r_{\rm g}$ for a BH spin between $a=0.0-0.998$.  Given that the effective inner edge of the disk
%is not within the radius of marginal stability for a Schwarzschild BH,
%it is not surprising that we were not able to constrain the value of
%the BH spin in this source.

\begin{table}
\begin{center}
%\begin{scriptsize}
\begin{tabular} {|l|l|l|} 
\hline \hline
{\bf Model Component} & {\bf Parameter} & {\bf Value} \\
\hline \hline
{\tt phabs}    & $N_{\rm H1}\,(\pcmsq)$ & $8 \times 10^{20}$ \\
\hline
{\tt zphabs}   & $N_{\rm H2}\,(\pcmsq)$ & $1.19^{+0.01}_{-0.01} \times 10^{20}$ \\
\hline
{\tt zpo}      & $\Gamma_{\rm po}$ & $1.66^{+0.01}_{-0.01}$ \\
               & ${\rm flux}\,(\phpcmsqps)$ & $1.98^{+0.04}_{-0.13} \times 10^{-2}$ \\
\hline
{\tt zpo}      & $\Gamma_{\rm po}$ & $1.66^{+0.01}_{-0.01}$ \\
               & ${\rm flux}\,(\phpcmsqps)$ & $7.37^{+0.09}_{-0.21} \times 10^{-4}$ \\
\hline
{\tt zgauss}   & ${\rm E}\,(\keV)$ & $6.40^{+0.01}_{-0.01}$ \\
               & ${\rm flux}\,(\phpcmsqps)$ & $4.00^{+0.35}_{-0.32} \times 10^{-5}$ \\
               & ${\rm EW}\,(\eV)$ & $167^{+15}_{-13}$ \\
\hline
{\tt zgauss}   & ${\rm E}\,(\keV)$ & $7.06^{+0.02}_{-0.01}$ \\
               & ${\rm flux}\,(\phpcmsqps)$ & $9.80^{+2.98}_{-2.87} \times 10^{-6}$ \\
               & ${\rm EW}\,(\eV)$ & $104^{+32}_{-30}$ \\
\hline
{\tt zgauss}   & ${\rm E}\,(\keV)$ & $7.24^{+0.03}_{-0.04}$ \\
               & ${\rm flux}\,(\phpcmsqps)$ & $-7.39^{+2.84}_{-2.73} \times 10^{-6}$ \\
               & ${\rm EW}\,(\eV)$ & $-79^{+30}_{-29}$ \\
\hline
{\tt zgauss}   & ${\rm E}\,(\keV)$ & $7.48^{+0.04}_{-0.04}$ \\
               & ${\rm flux}\,(\phpcmsqps)$ & $-6.93^{+2.83}_{-2.84} \times 10^{-6}$ \\
               & ${\rm EW}\,(\eV)$ & $-121^{+49}_{-50}$ \\
\hline
{\tt zgauss}   & ${\rm E}\,(\keV)$ & $7.85^{+0.02}_{-0.05}$ \\
               & ${\rm flux}\,(\phpcmsqps)$ & $-9.62^{+2.90}_{-2.86} \times 10^{-6}$ \\
               & ${\rm EW}\,(\eV)$ & $-143^{+43}_{-43}$ \\
\hline
{\tt kdblur} & $\alpha$ & $2.38^{+2.02}_{-1.33}$ \\
               & $i\,(\degmark)$ & $28^{+9}_{-6}$ \\
               & $r_{\rm min}\,(r_{\rm g})$ & $33^{+31}_{-33}$ \\
               & $r_{\rm max}\,(r_{\rm g})$ & $400$ \\
\hline
{\tt reflion}  & {\rm Fe/solar} & $0.61^{+0.08}_{-0.05}$ \\
               & $\xi_{\rm refl}\,(\ergpcmps)$ & $30^{+1}_{-0}$ \\
               & $\Gamma_{\rm refl}$ & $1.66^{+0.01}_{-0.01}$ \\
               & $R_{\rm refl}$ & $0.41^{+0.02}_{-0.02}$ \\
               & ${\rm flux}\,(\phpcmsqps)$ & $1.64^{+0.10}_{-0.10} \times 10^{-5}$ \\
\hline
{\bf $\chi^2/\nu$} & & $1485/1488\,(1.00)$ \\
\hline \hline 
\end{tabular}
%\end{scriptsize}
\end{center}
\caption[Best fit parameters for MCG--5-23-16]{Best-fitting model
  parameters for the $2.5-10.0 \keV$ spectrum of MCG--5-23-16,
  including components and parameter values from $0.6-1.5 \keV$ for
  completeness.  The energies from $1.5-2.5 \keV$ were not included
  due to the presence of absorption edges from the {\it XMM-Newton}
  mirrors.  Error bars are quoted at $90\%$ confidence.  For the {\tt
    zgauss} lines, we required each to be intrinsically narrow, i.e.,
  $\sigma=0 \keV$.  Redshifts were frozen at the cosmological value for
  the source, in this case $z=0.0085$.  Note the absence of warm
  absorption and a soft excess in this object.  When no error bars are
quoted, the parameter in question is frozen at the given value.\label{tab:tab2.tex}}
\end{table}

\subsubsection{NGC~3783}
\label{sec:n3783}

NGC~3783 is a bright, nearby Sy 1 galaxy at a redshift of $z=0.0097$.
It was first detected in X-rays with the {\it Ariel-V} all-sky survey
\citep{McHardy1981} and subsequently in the high Galactic latitude
survey conducted by {\it HEAO-1} \citep{Piccinotti1982}.  Since these
early detections, there have been many observations of NGC~3783 with
higher resolution instruments:  A {\it ROSAT} observation of the
source showed evidence of an ionized absorber in the soft band
\citep{Turner1993}, which was confirmed during {\it ASCA} observations
\citep{George1995,George1998_2}.  High-resolution grating observations of
NGC~3783 with {\it Chandra} and {\it XMM-Newton} followed, unveiling
the soft spectrum of this source in detail 
\citep{Kaspi2000,Kaspi2001,Kaspi2002,Blustin2002,Behar2003}.  The higher
signal-to-noise of {\it XMM-Newton}, in particular, has also enabled the
iron line to be studied extensively in this source
\citep{Reeves2004}.  Using two observations from December 2001, Reeves
\etal have obtained $\sim 240 \ks$ of data on NGC~3783.
Their global fits to the merged EPIC-pn spectrum are currently second
in length only to the $\sim 330 \ks$ observation of
MCG--6-30-15 \citep{Fabian2002}.  In these observations, the source has
an average $2-10 \keV$ flux of $F_{2-10}=6.8 \times 10^{-11}
\ergpcmsqps$.  The spectrum is
noted to have iron line peaks at $6.39 \pm 0.01 \keV$ and $7.00 \pm 0.02 \keV$,
representing neutral Fe K$\alpha$ ($EW=123 \pm 6 \eV$) and a blend of
ionized Fe K$\alpha$ and Fe K$\beta$ ($EW=34 \pm 5 \eV$), respectively.
A strong absorption line of intermediately ionized iron is also observed at
$6.67 \pm 0.04 \keV$ ($EW=17 \pm 5 \eV$) that exhibits
variability in direct correlation with that
of the continuum flux.  A weak red wing of the $6.39 \keV$ line is
noted by the authors, but once warm absorption is taken into account
in the system the requirements for a broad Fe K$\alpha$ component are
significantly reduced \citep{Reeves2004}.

We analyzed the first of the two December 2001 {\it XMM-Newton}
observations.  We chose not to merge the event files of both data sets
to avoid the uncertainties inherent in so doing, instead focusing on
only one data set with an effective exposure time of $\sim 120 \ks$
and a total of $\sim 1.6 \times 10^6$ photons after filtering.  This
number of counts
is still large enough to obtain valid statistical fits to the model
parameters used.  We follow the data reduction steps of Reeves \etal,
excluding the last part of the observation due to contamination from a
background flare.  Our analysis follows in the same spirit as these
authors, but in order to remain consistent with the method described
above in \S\ref{sec:analysis} we do not make any assumptions about the
soft spectrum and continuum emission based on prior RGS results
\citep{Blustin2002}.  

We began our analysis of the time-averaged spectrum with a simple
photoabsorbed power-law fit using the Galactic absorbing column of
$N_{\rm H}=8.5 \times 10^{20} \pcmsq$, and neglecting the energy
ranges associated with the mirror edges and the iron lines, as in
\S\ref{sec:analysis}.  This fit leaves obvious residuals in all parts of
the spectrum, however, and is especially poor for the soft energies
below $\sim 1.5 \keV$.  Continuum curvature associated with a soft
excess and warm absorption are both clearly evident.  As discussed by
Reeves \etal, the fit was much improved with the addition of a thermal
blackbody component ($kT \sim 0.07 \pm 0.01 \keV$, as compared with their
value: $kT \sim 0.09 \pm 0.01 \keV$).  We approached the question of
the warm absorption in the same manner as Reeves \etal, using our {\sc
  xstar} table model to parameterize the column density and ionization
level of the absorbing medium.  As with these authors, we found
statistical evidence for a multi-zone warm absorption structure, though
our fitted values for the column densities and ionization parameters of these
zones vary from those of Reeves et al.  These authors find $N_{\rm
  H1}=1.1 \times 10^{21} \pcmsq$ and $\log \xi_1=-0.1$ (for this zone,
${\rm Fe/solar}=10$, in contrast to the other zones with solar iron
abundance), $N_{\rm
  H2}=1.2 \times 10^{22} \pcmsq$ and $\log \xi_2=2.1$, with $N_{\rm
  H3}=4.4 \times 10^{22}\pcmsq$ and $\log \xi_3=3.0$ (no final error
bars were given).  By contrast, we find $N_{\rm
  H1}=1.80^{+0.23}_{-0.29} \times 10^{20} \pcmsq$ and $\log \xi_1=0$, 
$N_{\rm H2}=1.58^{+0.65}_{-0.62} \times 10^{23} \pcmsq$ and $\log \xi_2=2.82
\pm 0.94$, with $N_{\rm H3}=8.43^{+0.07}_{-0.25} \times 10^{23} \pcmsq$
and $\log \xi_3=3.79^{+0.02}_{-0.14}$.  As with the results of Reeves
\etal the third, neutral absorber of smaller column density was required
to obtain a quality fit, but a very high iron abundance was necessary:
${\rm Fe/solar}=10$, where the abundance was approximated iteratively
and held constant in the final fit because it was at the endpoint of
its parameter space.  We used the same technique to constrain the
ionization parameter of this absorbing zone for the same reason.
The differences between our results and those of Reeves \etal
are not surprising, however, because Reeves \etal base their fits on
the RGS results for this source \citep{Blustin2002}.  We do not use
any {\it a priori} information to augment or guide the EPIC-pn
spectral fits due to calibration uncertainties.  Finally,
we allowed for a second neutral hydrogen absorbing column intrinsic to
the source (using the {\tt zpcfabs} model component) in an effort to
improve the goodness-of-fit and found that the
spectrum preferred a higher value than the Galactic column: $N_{\rm
  H}=2.77^{+0.11}_{-0.13} \times 10^{21} \pcmsq$ with a covering
fraction of $f=87 \pm 2 \%$.  This was also in contrast to the
Reeves \etal result, where the cold hydrogen column was held at the
Galactic value and no intrinsic neutral hydrogen absorption was
allowed.  The differences in
absorbing structures between our model and that of Reeves \etal also
explain the difference in photon indices of the power-law continuum in
the two models: our best-fit value was marginally softer, at
$\Gamma=1.83^{+0.03}_{-0.04}$, whereas Reeves \etal found a harder
photon index of $\Gamma=1.73 \pm 0.01$. 

Reeves \etal found two emission peaks in the spectrum at $6.39 \pm 0.01 \keV$
($EW=123 \pm 4 \eV$) and $7.00 \pm 0.02\keV$ ($EW=34 \pm 5 \eV$), representing
cold Fe K$\alpha$ and likely a blend of ionized Fe K$\alpha$ and
Fe K$\beta$, respectively.  The latter line, in particular, showed no
appreciable variance over the observation, leading the authors to
postulate a relatively distant origin for the lines away from the
central parts of the accretion disk.  A $6.67 \pm 0.04 \keV$ absorption feature
($EW=17 \pm 5 \eV$) of highly ionized iron was also seen which did appear to vary with
time and was strongest when the continuum flux was highest, suggesting
an origin in the region of the warm absorber within $0.1$ parsec of
the nucleus.  Though inclusion of the WA structure (tailored to
include this absorption feature) did lessen
the statistical case for a broad iron line in NGC~3783, Reeves \etal
nonetheless identified a residual broad feature which they fit using a
{\tt diskline} model with $\alpha=3.3 \pm 0.5$, $i=19 \pm 9 \degmark$,
$r_{\rm min}=6.0\,r_{\rm g}$ (held constant, with $r_{\rm
  max}=100\,r_{\rm g}$), and $EW=58 \pm 12 \eV$.  

\begin{figure}
\begin{center}
\includegraphics[width=0.33\textwidth,angle=270]{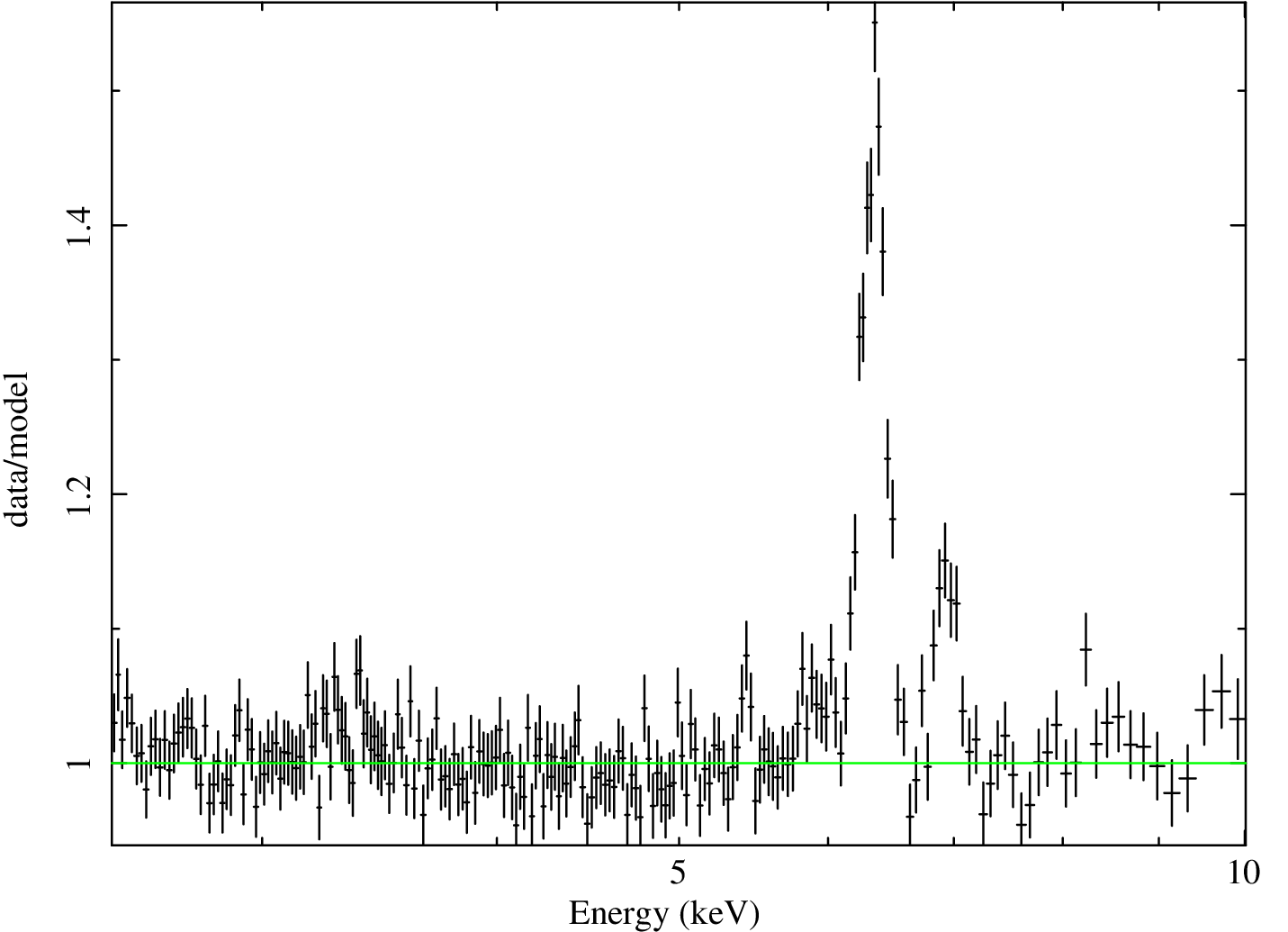}
\includegraphics[width=0.33\textwidth,angle=270]{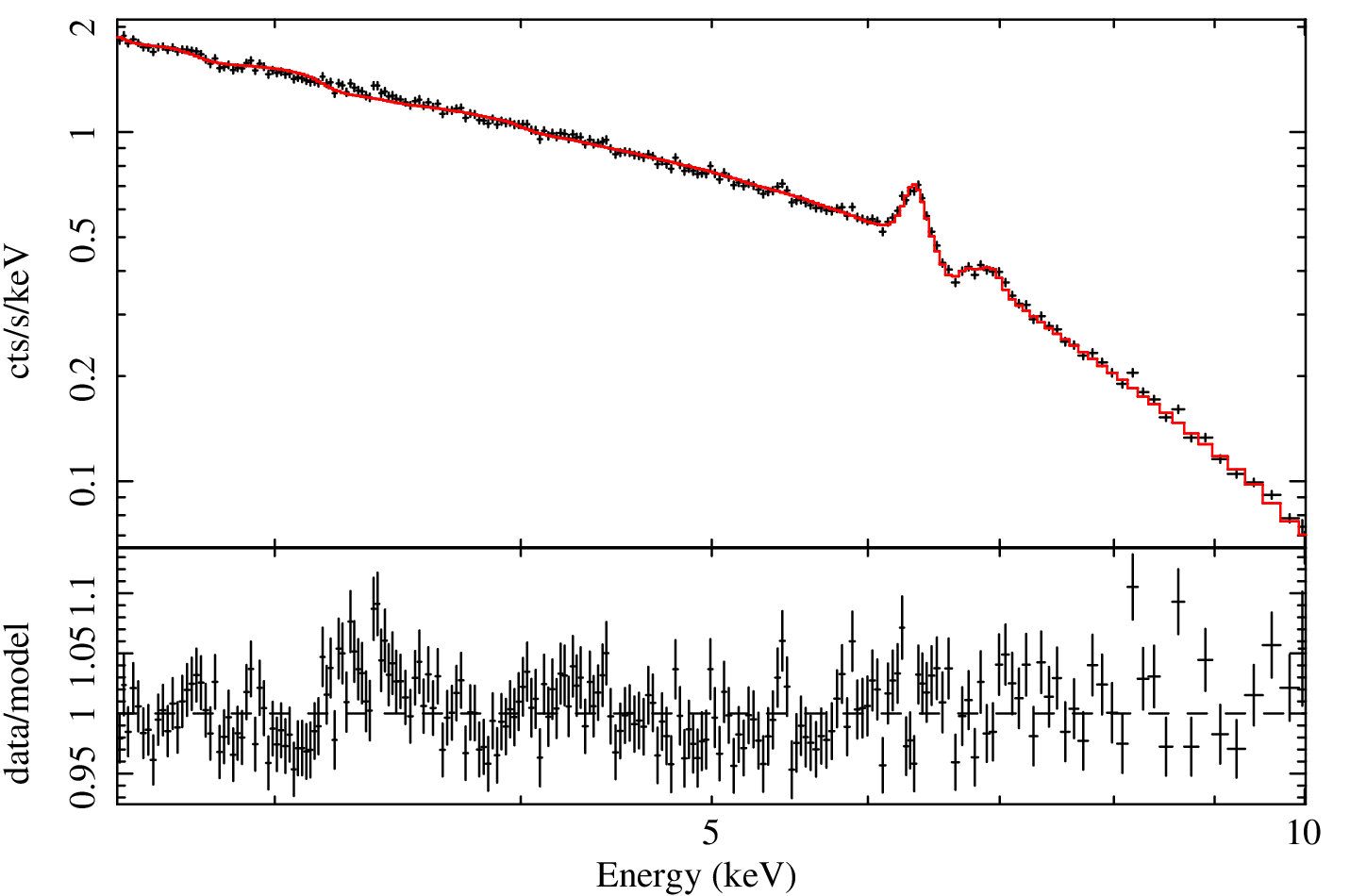}
\end{center}
\begin{center}
\includegraphics[width=0.33\textwidth,angle=270]{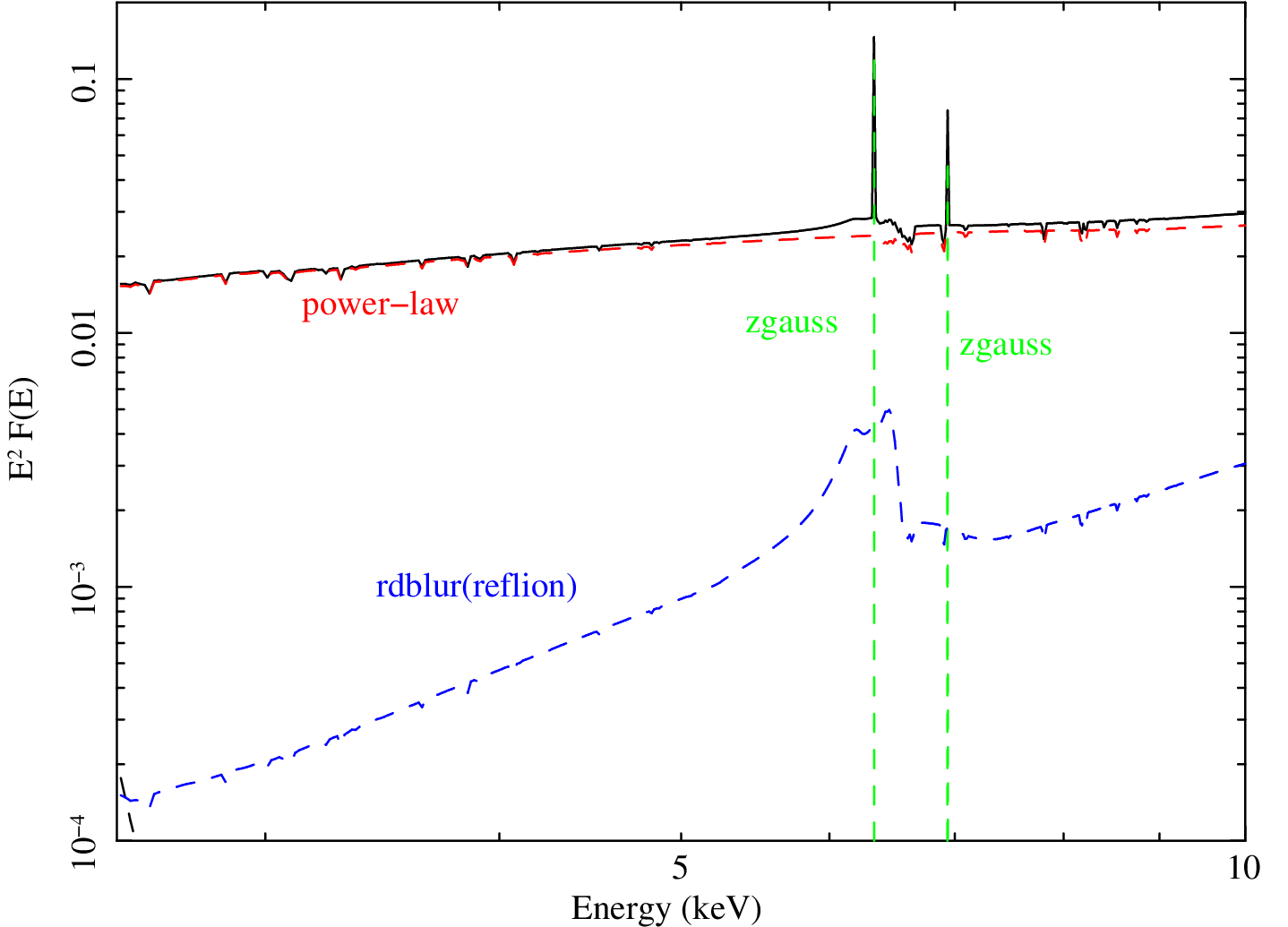}
\end{center}
\caption[Spectral fit for NGC~3783.]{{\it Top left:} The $2.5-10.0
  \keV$ spectrum of NGC~3783 fit with a power-law model
  modified by Galactic photoabsorption, along with a three-zone warm
  absorber and a blackbody soft excess.
  $\chi^2/\nu=2751/1487\,(1.85)$.  {\it Top right:} The $2.5-10.0
  \keV$ best-fit model for NGC~3783, including our continuum
  model and two narrow emission lines as well as an ionized disk
  reflection spectrum convolved with the
  {\tt rdblur} relativistic smearing kernel.
  $\chi^2/\nu=1480/1487\,(1.00)$.  {\it Bottom:} An ${\rm
  E}^2\,{\rm F(E)}$ plot depicting the relative flux
  in each of the best-fit model components in the spectrum.}
\label{fig:n3783_figs}
\end{figure}

We also detected the emission and absorption features discussed by
Reeves \etal, though we found that the equivalent widths for the
Fe K$\alpha$ and Fe K$\alpha$/K$\beta$ blend differed from the fits performed by these authors:
$EW_{K\alpha}=65 \pm 12 \eV$, $EW_{K\alpha/K\beta}=28 \pm 3 \eV$.  As mentioned
above, the inclusion of the three-zone WA structure eliminated the
need to include a separate Gaussian absorption line at $6.67 \keV$.
It should also be noted that, contrary to the Reeves \etal
analysis, we required the core of the $6.4 \keV$ line to be
intrinsically narrow, so it is not unusual that we obtain a smaller
equivalent width than Reeves et al.  When we included a broad {\tt
  diskline} component
of the neutral K$\alpha$ line, the goodness-of-fit improved
dramatically: $\Delta\chi^2/\Delta\nu=-170/-4$.  This fit proved
superior to both a {\tt laor} line model and a broad Gaussian
component, the latter of which particularly did not model the shape of
the line well.  Our best
fit to the hard spectrum of NGC~3783 convolved an ionized disk
reflection spectrum with {\tt rdblur}, as can be seen in
Table~\ref{tab:tab3.tex} below.  The inclusion of the ionized disk
reflection spectrum component ({\tt reflion}) accounted for the soft
excess self-consistently, without the need to include the
phenomenological blackbody component.
Adding in the effects of relativistic smearing on this reflection
spectrum, our constraints on the radial extent of the disk
showed that $r_{\rm min}\leq 46\,r_{\rm g}$, making it difficult to
distinguish a rapidly-spinning BH from a slowly-spinning one, even
though relativistic disk effects are clearly important in broadening
the iron line profile.
The residual iron line feature, best-fit model and best-fit model
components for the hard spectrum are 
shown in Fig.~\ref{fig:n3783_figs}.  The {\tt diskline} and {\tt laor}
fits are shown in Table~\ref{tab:tab10.tex}, while comparisons of the
ionized disk reflection spectrum with and without relativistic
smearing can be found in Table~\ref{tab:tab11.tex}.   

\begin{table}
\begin{center}
%\begin{scriptsize}
\begin{tabular} {|l|l|l|} 
\hline \hline
{\bf Model Component} & {\bf Parameter} & {\bf Value} \\
\hline \hline
{\tt phabs}    & $N_{\rm H}\,(\pcmsq)$ & $8.50 \times 10^{20}$ \\
\hline
{\tt zpcfabs}  & $N_{\rm H2}\,(\pcmsq)$ & $2.77^{+0.11}_{-0.13} \times
               10^{21}$ \\ 
               & $f_{\rm cov}\,(\%)$ & $88^{+2}_{-2}$ \\ 
\hline
{\tt WA 1}     & $N_{\rm WA1}\,(\pcmsq)$ & $1.80^{+0.23}_{-0.29} \times 10^{20}$ \\
               & $\log \xi_{\rm WA1}$ & $0$ \\
\hline
{\tt WA 2}     & $N_{\rm WA2}\,(\pcmsq)$ & $1.58^{+0.65}_{-0.62} \times 10^{23}$ \\
               & $\log \xi_{\rm WA2}$ & $2.82^{+0.94}_{-0.94}$ \\
\hline
{\tt WA 3}     & $N_{\rm WA3}\,(\pcmsq)$ & $8.43^{+0.07}_{-0.25} \times 10^{23}$ \\
               & $\log \xi_{\rm WA3}$ & $3.79^{+0.02}_{-0.14}$ \\
\hline
{\tt zpo}       & $\Gamma_{\rm po}$ & $1.83^{+0.03}_{-0.04}$ \\
               & ${\rm flux}\,(\phpcmsqps)$ & $1.33^{+0.05}_{-0.05}
               \times 10^{-4}$ \\
\hline
%{\tt bbody}    & $kT\,(\keV)$ & $7.00^{+0.11}_{-0.17} \times 10^{-2}$ \\
%               & ${\rm flux}\,(\phpcmsqps)$ & $5.49^{+1.10}_{-0.25} \times 10^{-5}$ \\
%\hline
{\tt zgauss}   & ${\rm E}\,(\keV)$ & $6.40^{+0.02}_{-0.01}$ \\
               & ${\rm flux}\,(\phpcmsqps)$ & $3.25^{+0.37}_{-0.37}
               \times 10^{-7}$ \\
               & ${\rm EW}\,(\eV)$ & $65^{+7}_{-7}$ \\
\hline
{\tt zgauss}   & ${\rm E}\,(\keV)$ & $7.01^{+0.02}_{-0.03}$ \\
               & ${\rm flux}\,(\phpcmsqps)$ & $1.11^{+0.20}_{-0.20}
               \times 10^{-7}$ \\
               & ${\rm EW}\,(\eV)$ & $28^{+5}_{-5}$ \\
\hline
{\tt rdblur} & $\alpha$ & $1.90^{+1.40}_{-0.87}$ \\
               & $i\,(\degmark)$ & $23^{+3}_{-3}$ \\
               & $r_{\rm min}\,(r_{\rm g})$ & $17^{+29}_{-17}$ \\
               & $r_{\rm max}\,(r_{\rm g})$ & $400$ \\
\hline
{\tt reflion}  & {\rm Fe/solar} & $0.83^{+0.12}_{-0.88}$ \\
               & $\xi_{\rm refl}\,(\ergpcmps)$ & $30^{+24}_{-0}$ \\
               & $\Gamma_{\rm refl}$ & $1.83^{+0.03}_{-0.04}$ \\
               & $R_{\rm refl}$ & $0.49^{+0.22}_{-0.37}$ \\
               & ${\rm flux}\,(\phpcmsqps)$ & $7.46^{+2.88}_{-4.07}
               \times 10^{-8}$ \\
\hline
{\bf $\chi^2/\nu$} & & $1480/1487\,(1.00)$ \\
\hline \hline 
\end{tabular}
%\end{scriptsize}
\end{center}
\caption[Best fit parameters for NGC~3783.]{Best-fitting model
  parameters for the $2.5-10.0 \keV$ spectrum of NGC~3783,
  including components and parameter values from $0.6-1.5 \keV$ for
  completeness.  The energies from $1.5-2.5 \keV$ were not included
  due to the presence of absorption edges from the {\it XMM-Newton}
  mirrors.  Error bars are quoted at $90\%$ confidence.  For the {\tt
    zgauss} lines, we required each to be intrinsically narrow, i.e.,
  $\sigma=0 \keV$.  Redshifts were frozen at the cosmological value for
  the source, in this case $z=0.0097$.  Note the multi-zone warm
  absorber present in this object, similar
  to many other Sy 1 sources.\label{tab:tab3.tex}}
\end{table}

\subsubsection{Mrk~766}
\label{sec:Mrk766}

Mrk~766 is designated as a classic, bright NLS1 galaxy (Sy 1.5) with a
redshift of $z=0.0129$ and a typical $2-10 \keV$ flux of $F_{2-10}
\sim 2.5 \times
10^{-11} \ergpcmsqps$ \citep{Pounds2003a}.  Previous X-ray
observations of this source have provided contradictory evidence on
the detection of a broad Fe K$\alpha$ feature.  Although not one of
the most convincing cases, Mrk~766 was included in an {\it ASCA}
spectral survey of bright Seyfert galaxies showing evidence of 
relativistic iron lines \citep{Nandra1997}.  A separate analysis of
simultaneous {\it ROSAT} and {\it ASCA} observations
\citep{Leighly1996}, however, showed the X-ray spectrum to be described by a
power-law of index increasing strongly with flux from $\Gamma \sim
1.6$, but with only a
narrow Fe K$\alpha$ emission line ($EW \sim 100 \eV$ at $6.4 \keV$).
A later observation with {\it BeppoSAX} found a steeper power-law
($\Gamma \sim 2.2$), and evidence for an absorption edge at $\sim 7.4
\keV$ \citep{Matt2000}, implying strong reflection from
intermediately ionized material.  Interestingly, based on {\it
  XMM-Newton} observations of the source, Turner \etal found that
energy-time maps of Mrk~766
reveal a periodic energy shift in an ionized component of Fe K$\alpha$
emission, with a period of $\sim 165 \ks$.  This can be interpreted as
evidence for emission from orbiting gas within $\sim 100\,r_{\rm g}$
of the central BH.  A likely explanation is that this gas represents a
hot spot on the disk illuminated by magnetic reconnection
\citep{Turner2006}.  

We examine the May 2001 {\it XMM-Newton} observation of Mrk~766 taken
by Mason \etal, from which the RGS results were published in 2003
\citep{Mason2003} and the EPIC spectra were explored more thoroughly
by other authors \citep{Pounds2003a,Turner2006}.  In our re-analysis,
the EPIC-pn
data have an effective exposure time of $128 \ks$ and the filtered
data yield $\sim 2.4 \times 10^6$ photons.

Mrk~766 bears many spectral similarities to MCG--6-30-15, showing
significant complexity beyond a simple photoabsorbed power-law fit.
There is clear evidence for a thermal soft excess below $\sim 1 \keV$,
which we initially parameterize with a phenomenological blackbody of
$kT=0.18^{+0.00}_{-0.01} \keV$.
There is also evidence for two distinct physical zones of low ($N_{\rm H1}=3.29^{+0.18}_{-0.28} \times
10^{21} \pcmsq$, $\log \xi_1=0.69^{+0.07}_{-0.03}$) to moderately
ionized ($N_{\rm H2}=2.73^{+0.51}_{-0.42} \times
10^{21} \pcmsq$, $\log \xi_2=2.17^{+0.21}_{-0.20}$) intrinsic
absorption.  These components exist in addition to cold absorption by the
Milky Way ($N_{\rm H}=1.71 \times 10^{20} \pcmsq$).  We find that the power-law
in this source has a spectral index of $\Gamma=2.13 \pm 0.01$,
which is slightly harder than the value determined for the time-averaged
spectrum by \citet{Pounds2003a} ($\Gamma=2.21 \pm 0.01$).
Unfortunately, these authors
did not probe the spectrum below $3 \keV$, so we cannot compare our
soft excess or warm absorption parameters to theirs.  

\begin{figure}
\begin{center}
\includegraphics[width=0.33\textwidth,angle=270]{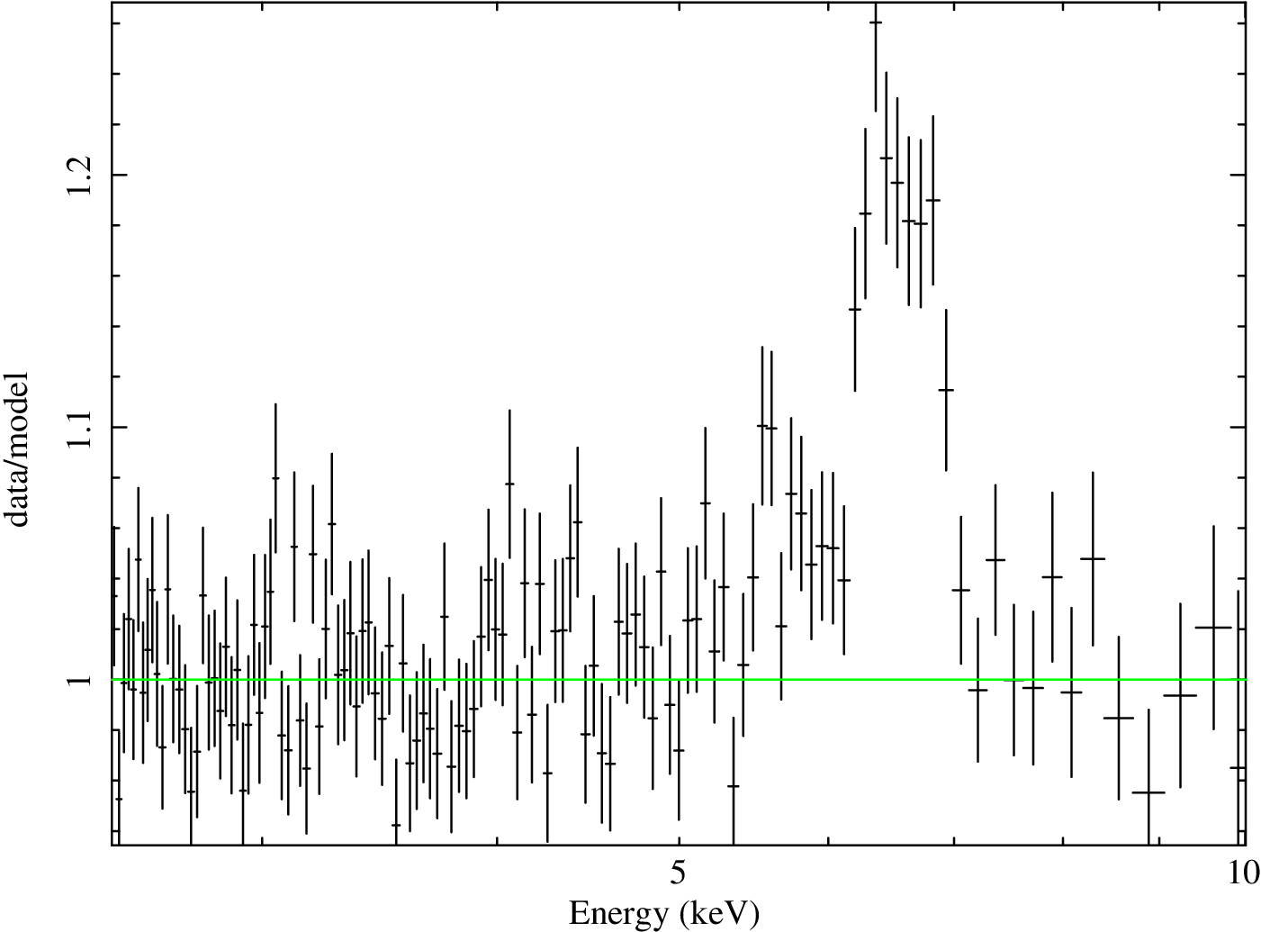}
\includegraphics[width=0.33\textwidth,angle=270]{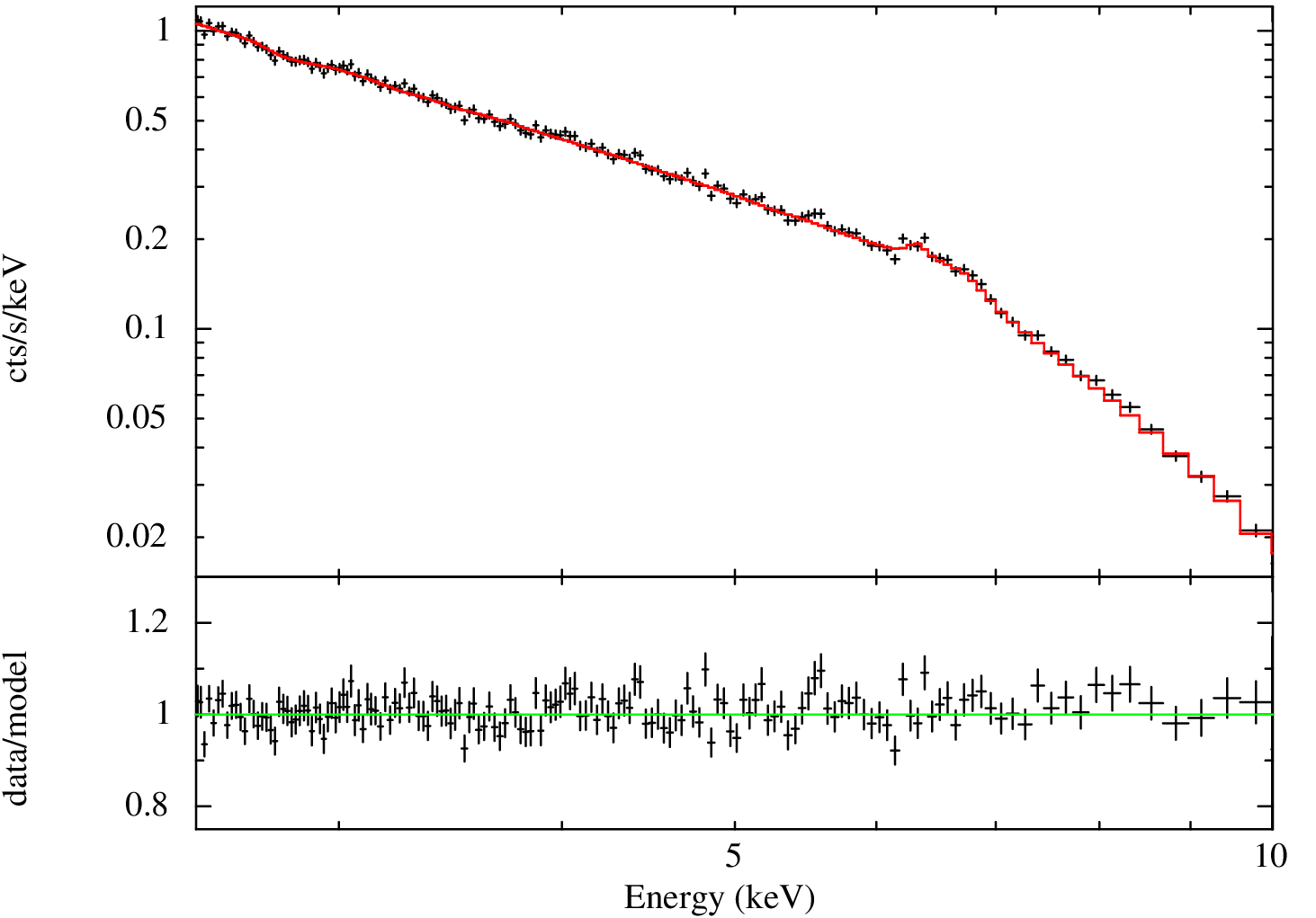}
\end{center}
\begin{center}
\includegraphics[width=0.33\textwidth,angle=270]{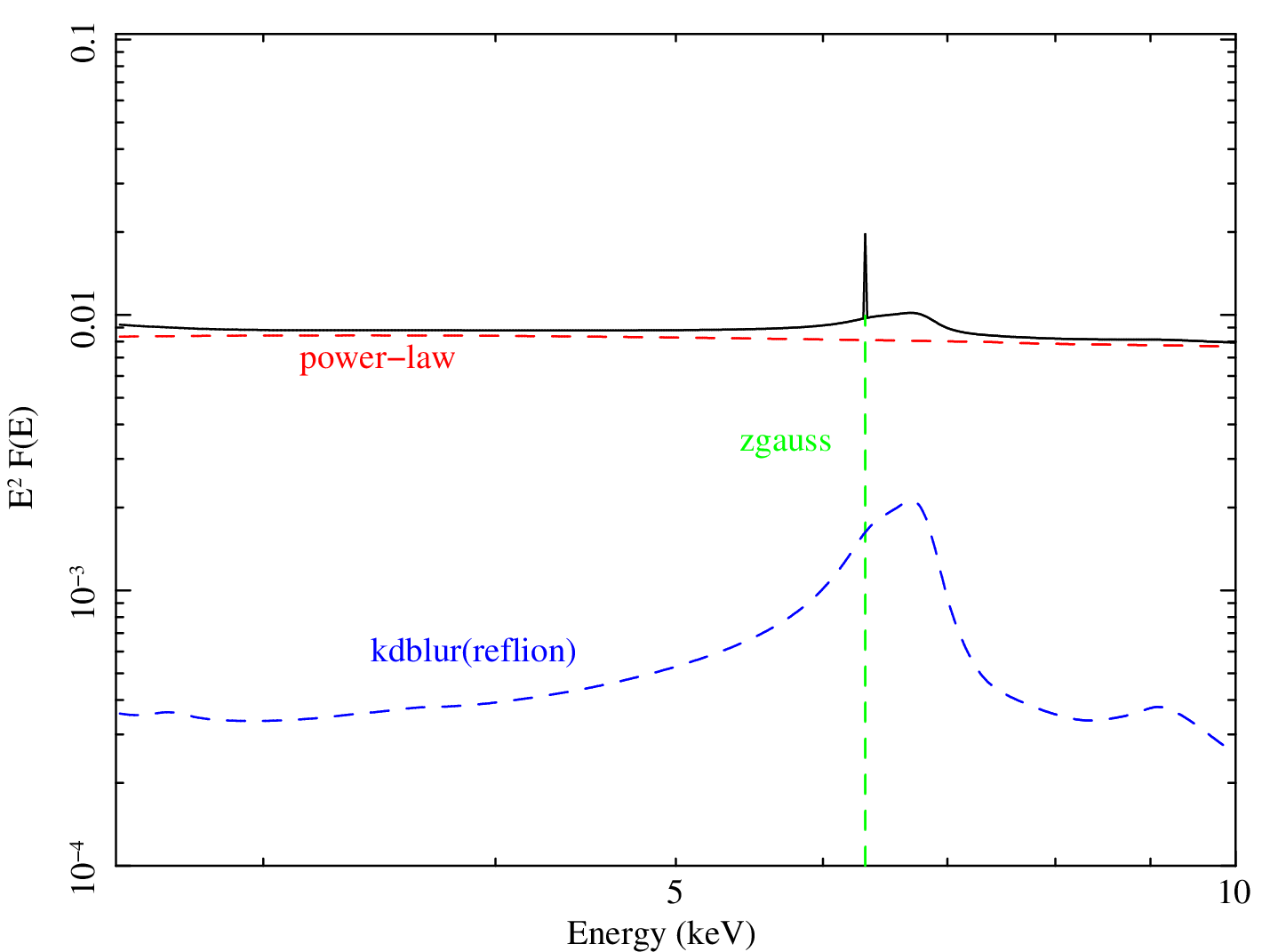}
\end{center}
\caption[Spectral fit for Mrk~766.]{{\it Top left:} The $2.5-10.0
  \keV$ spectrum of Mrk~766 fit with a continuum composed of a power-law model
  modified by Galactic photoabsorption, a two-zone warm absorber model
  and a blackbody soft excess.  $\chi^2/\nu=1681/1315\,(1.28)$.  {\it Top right:} The $2.5-10.0
  \keV$ best-fit model for Mrk~766, including our continuum
  model and an ionized disk reflection spectrum convolved with the
  {\tt kdblur} relativistic smearing kernel.
  $\chi^2/\nu=1426/1308\,(1.09)$.  {\it Bottom:} An ${\rm
  E}^2\,{\rm F(E)}$ plot depicting the relative flux
  in each of the model components in the spectrum.}
\label{fig:mrk766_figs}
\end{figure}

The narrow Fe K$\alpha$ core of Mrk~766 has a rest-frame energy of
$6.40 \pm 0.01 \keV$, consistent with neutral iron, and an $EW=47 \pm
8 \eV$, consistent with the strength of the narrow line of neutral
iron found by Pounds et al.  Fitting this line
with a Gaussian component leaves significant residuals strongly
indicative of the presence of a broad line.  When this residual
feature is fit with a broad Gaussian component its rest-frame energy
is indicative of ionized iron: $E=6.62 \pm 0.07 \keV$.  Its width
($\sigma=0.30 \pm 0.08 \keV$) is indicative of an origin $\sim 100 \,r_{\rm g}$
from the black hole.  This broad component is
most successfully fit here with a {\tt kdblur(reflion)} model
(removing the now-unnecessary blackbody component), yielding
comparable results to the disk reflection model employed to model the
broad iron line in Pounds \etal (reduced $\chi^2=1.09$ in our work,
compared with $1.11$ in Pounds \etal), though with slightly tighter
parameter constraints (e.g., $r_{\rm min} \leq 14\,r_{\rm g}$ in our
work vs. $70 \pm 12\,r_{\rm g}$ in Pounds \etal).  Our model also detects
no absorption features above $8 \keV$, as discussed by Pounds \etal, 
possibly due to our different approach toward modeling the reflection
spectrum, i.e., the use of {\tt reflion} vs. the {\tt xion} model
employed by Pounds et al.  We note, however, that there is no
significant difference between our {\tt kdblur} and {\tt rdblur} fits,
or between out {\tt diskline} and {\tt laor} fits, indicating that we
are not able to distinguish between a spinning and non-spinning black
hole in this source using these data.  
%The BH spin, to $90\%$ confidence, is relatively high at
%$a>0.85$ and the inner edge of the accretion disk is
%constrained to $r_{\rm min}\leq 2.25\,r_{\rm ms}$.  
The final best-fit results
are shown in Table~\ref{tab:tab4.tex}.
Fig.~\ref{fig:mrk766_figs} shows the iron line
residual, best fit to this residual, and relative contributions of the
individual model components, respectively.  The {\tt diskline} and {\tt laor}
fits are shown in Table~\ref{tab:tab10.tex}, while comparisons of the
ionized disk reflection spectrum with and without relativistic
smearing can be found in Table~\ref{tab:tab11.tex}.   

\begin{table}
\begin{center}
%\begin{scriptsize}
\begin{tabular} {|l|l|l|} 
\hline \hline
{\bf Model Component} & {\bf Parameter} & {\bf Value} \\
\hline \hline
{\tt phabs}    & $N_{\rm H}\,(\pcmsq)$ & $1.71 \times 10^{20}$ \\
\hline
{\tt WA 1}     & $N_{\rm WA1}\,(\pcmsq)$ & $3.29^{+0.18}_{-0.28}\times 10^{21}$ \\
               & $\log \xi_{\rm WA1}$ & $0.69^{+0.07}_{-0.03}$ \\
\hline
{\tt WA 2}     & $N_{\rm WA2}\,(\pcmsq)$ & $2.73^{+0.51}_{-0.42}\times 10^{21}$ \\
               & $\log \xi_{\rm WA2}$ & $2.17^{+0.21}_{-0.20}$ \\
\hline
{\tt zpo}       & $\Gamma_{\rm po}$ & $2.13^{+0.01}_{-0.01}$ \\
               & ${\rm flux}\,(\phpcmsqps)$ & $1.05^{+0.02}_{-0.04} \times 10^{-2}$ \\
\hline
%{\tt bbody}    & $kT\,(\keV)$ & $0.18^{+0.00}_{-0.01}$ \\
%               & ${\rm flux}\,(\phpcmsqps)$ & $3.63^{+1.03}_{-1.51} \times 10^{-3}$ \\
%\hline
{\tt zgauss}   & ${\rm E}\,(\keV)$ & $6.40^{+0.01}_{-0.01}$ \\
               & ${\rm flux}\,(\phpcmsqps)$ & $3.80^{+1.94}_{-2.03} \times 10^{-6}$ \\
               & ${\rm EW}\,(\eV)$ & $47^{+8}_{-8}$ \\
\hline
{\tt kdblur} & $\alpha$ & $2.29^{+1.35}_{-0.45}$ \\
               & $i\,(\degmark)$ & $28^{+4}_{-5}$ \\
               & $r_{\rm min}\,(r_{\rm g})$ & $14^{+0}_{-14}$ \\
               & $r_{\rm max}\,(r_{\rm g})$ & $400$ \\
\hline
{\tt reflion}  & {\rm Fe/solar} & $1.0$ \\
               & $\xi_{\rm refl}\,(\ergpcmps)$ & $961^{+329}_{-465}$ \\
               & $\Gamma_{\rm refl}$ & $2.13^{+0.01}_{-0.01}$ \\
               & $R_{\rm refl}$ & $0.01^{+0.17}_{-0.01}$ \\
               & ${\rm flux}\,(\phpcmsqps)$ & $1.38^{+8.25}_{-0.64} \times 10^{-7}$ \\
\hline
{\bf $\chi^2/\nu$} & & $1426/1308\,(1.09)$ \\
\hline \hline 
\end{tabular}
%\end{scriptsize}
\end{center}
\caption[Best fit parameters for Mrk~766]{Best-fitting model
  parameters for the $2.5-10.0 \keV$ spectrum of Mrk~766,
  including components and parameter values from $0.6-1.5 \keV$ for
  completeness.  The energies from $1.5-2.5 \keV$ were not included
  due to the presence of absorption edges from the {\it XMM-Newton}
  mirrors.  Error bars are quoted at $90\%$ confidence.  We required
  the {\tt zgauss} line representing the core of the Fe K$\alpha$
  feature to be intrinsically narrow, i.e.,
  $\sigma=0 \keV$.  Redshifts were frozen at the cosmological value for
  the source, in this case $z=0.0130$.  A two-zone warm absorber is
  present in this object, as in MCG--6-30-15
  and many other Sy 1 sources.  We were not able to constrain the iron
  abundance for the ionized disk model, so we elected to fix it at the
  solar value.\label{tab:tab4.tex}}
\end{table}

\subsubsection{3C~273}
\label{sec:3c273}

3C~273 is the most distant source in our sample at a redshift of
$z=0.1583$.  Classified as a bright variable quasar, this AGN 
displays a strong jet and during epochs of radio-loudness it
possesses the flat X-ray spectrum of a blazar, with highly beamed jet
emission \citep{Turler2006}.  When the jet is
reduced in strength, however, this source has been observed to exhibit
Sy 1-type accretion disk signatures such as a broad iron line.  A
jet-minimum state occurred in March 1986 and allowed Robson \etal (1986)
to identify a new near-infrared spectral component.
An even better opportunity arose in early 2004, when the
sub-millimeter flux of 3C~273 was observed to be almost two times lower
than in 1986.  This new minimum triggered a slew of simultaneous
observations with instruments in all wavebands such as {\it INTEGRAL},
{\it XMM-Newton} and {\it RXTE}, among several other optical, radio
and sub-millimeter telescopes.  The $2-10 \keV$ flux during this
period was $F_{2-10}=6.7 \times 10^{-11} \ergpcmsqps$.  Such a low flux in 3C~273
has only been measured twice in the past, by {\it Ginga} in July 1987
\citep{Turner1990} and by {\it BeppoSAX} on 18 July 1996
\citep{Haardt1998}, the latter of which was coincident with the low
sub-mm flux mentioned above.  This X-ray/sub-mm correlation strongly
supports a synchrotron self-Compton origin for the X-ray jet emission
in 3C~273.   

T{\"u}rler \etal obtained a $20 \ks$ {\it XMM-Newton} observation of
3C~273 during this jet-minimum state in June 2004.  We use their (reprocessed)
thin-filter EPIC-pn observation in the interest of collecting as many photons
as possible, though due to photon pile-up it is necessary to exclude
the centralmost region of the source, as detailed by the authors
\citep{Turler2006}.  In our filtered data set we capture $\sim 3.2
\times 10^5$ photons.

T{\"u}rler \etal note the inadequacy of a simple photoabsorbed
power-law fit to the data.  Their best fit to the continuum is achieved
using two power-law components: $\Gamma_{\rm hard}=1.63 \pm 0.02$ and
$\Gamma_{\rm soft}=2.69 \pm 0.06$, with the hard component flux
$\sim 2.3$ times the soft flux.  Both components were modified by
Galactic photoabsorption with $N_{\rm H}=1.79 \times 10^{20} \pcmsq$.
We based our initial continuum fit on theirs, with two power-law
components: $\Gamma_{\rm hard}=1.72^{+0.11}_{-0.28}$ and
$\Gamma_{\rm soft}=3.01^{+1.29}_{-0.71}$, both consistent with the
T{\"u}rler \etal results.  The ratio we calculate between the fluxes
of the hard and soft power-law components is $\sim 4.18$,
however. This is significantly higher than that calculated by T{\"u}rler
\etal, and the error bars on the spectral indices, in particular, led us to consider
eliminating the soft component.  We did so, but still found it
necessary to parameterize the remaining soft excess with a thermal
{\tt bbody} model at first ($kT=0.14^{+0.03}_{-0.02} \keV$).  Such a
component is not entirely phenomenological
in nature, however: in many radio-loud sources a soft thermal
excess has been noted, and is often interpreted as arising from the
interaction of the jet with the surrounding ISM \citep[e.g.,][]{Brenneman2009}.  Our best-fitting
continuum model, therefore, has a blackbody and only one power-law component at
$\Gamma=1.84^{+0.46}_{-0.10}$ and exhibits the same statistical goodness-of-fit as the
two power-law model.

Evidence for excess emission from $2.5-7.0 \keV$ was also found by
T{\"u}rler \etal, supporting the presence of a broad Fe K$\alpha$ line
when 3C~273 is in a jet-minimum state.  The authors quote this excess
as significant at the $6\sigma$ level with an integrated flux of $2.6
\pm 0.4 \times 10^{-4} \phpcmsqps$, corresponding to an $EW=166 \pm 26
\eV$.  These values are consistent with those reported from previous
observations of the source in a jet-minimum state
\citep{Kataoka2002,Page2004,Yaqoob2000}.  The breadth of the excess is
not satisfactorily fitted by a Gaussian line or a {\tt diskline}
component from a non-rotating BH.  If the line extent is real, the
only remaining explanation, according to the authors, is that it is
emitted from around a near-extreme
Kerr BH.  The sharp edge of the iron line at $7 \keV$ suggests that
the angle of inclination of the accretion disk is $35-40 \degmark$
\citep{Turler2006}.  This is contrary to radio observations of the
source which have measured superluminal velocity
in the jet consistent with it having an inclination angle of $\leq 14
\degmark$ to the line of sight \citep{Ballantyne2004}.  Such a marked
difference between the disk and jet inclination angles may imply a
warped disk, or perhaps it may mean that the jet inclination is
different on smaller, unresolvable scales.  Deeper observations are
needed in order to resolve this discrepancy. 

\begin{figure}
\begin{center}
\includegraphics[width=0.33\textwidth,angle=270]{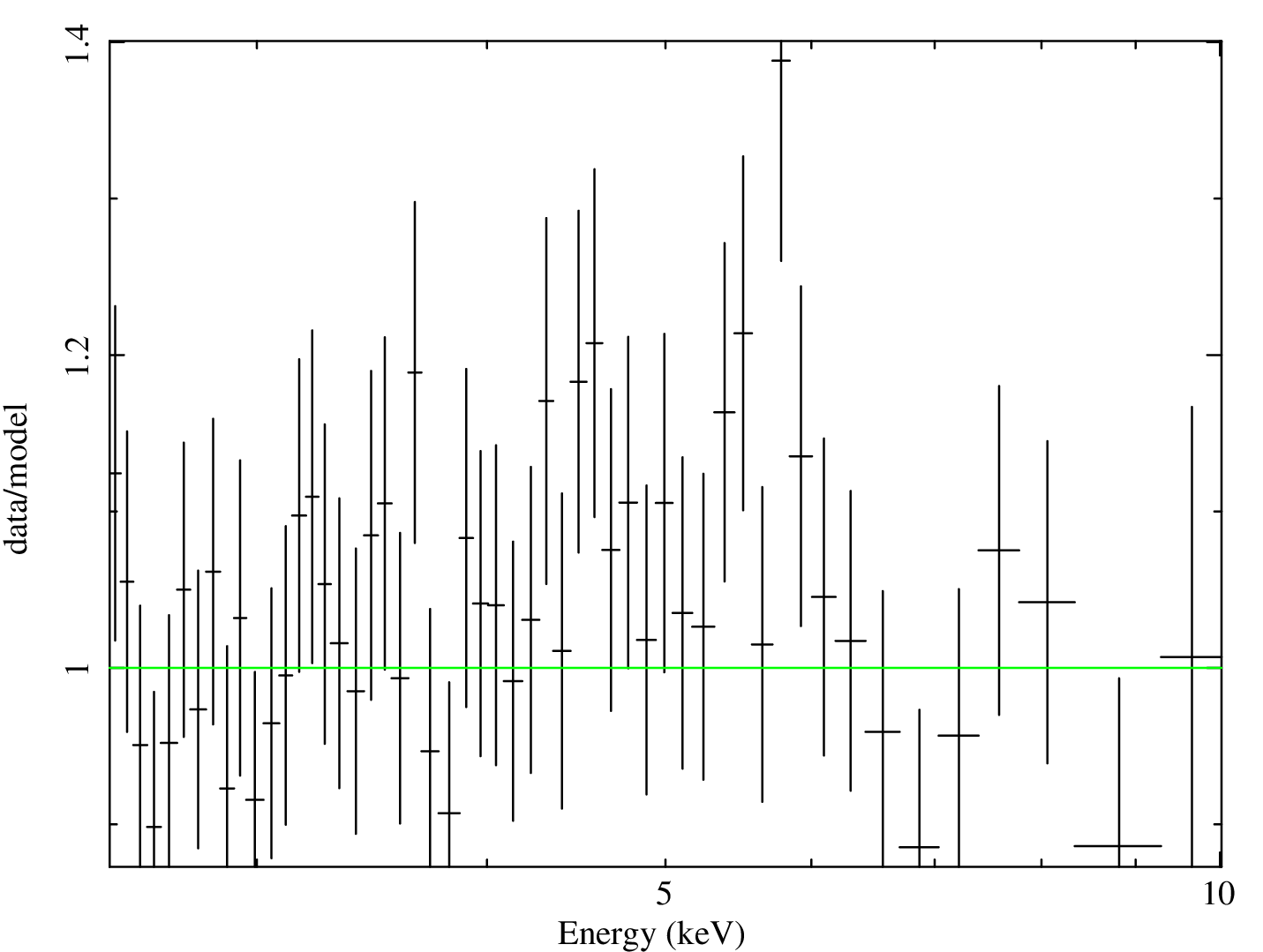}
\includegraphics[width=0.33\textwidth,angle=270]{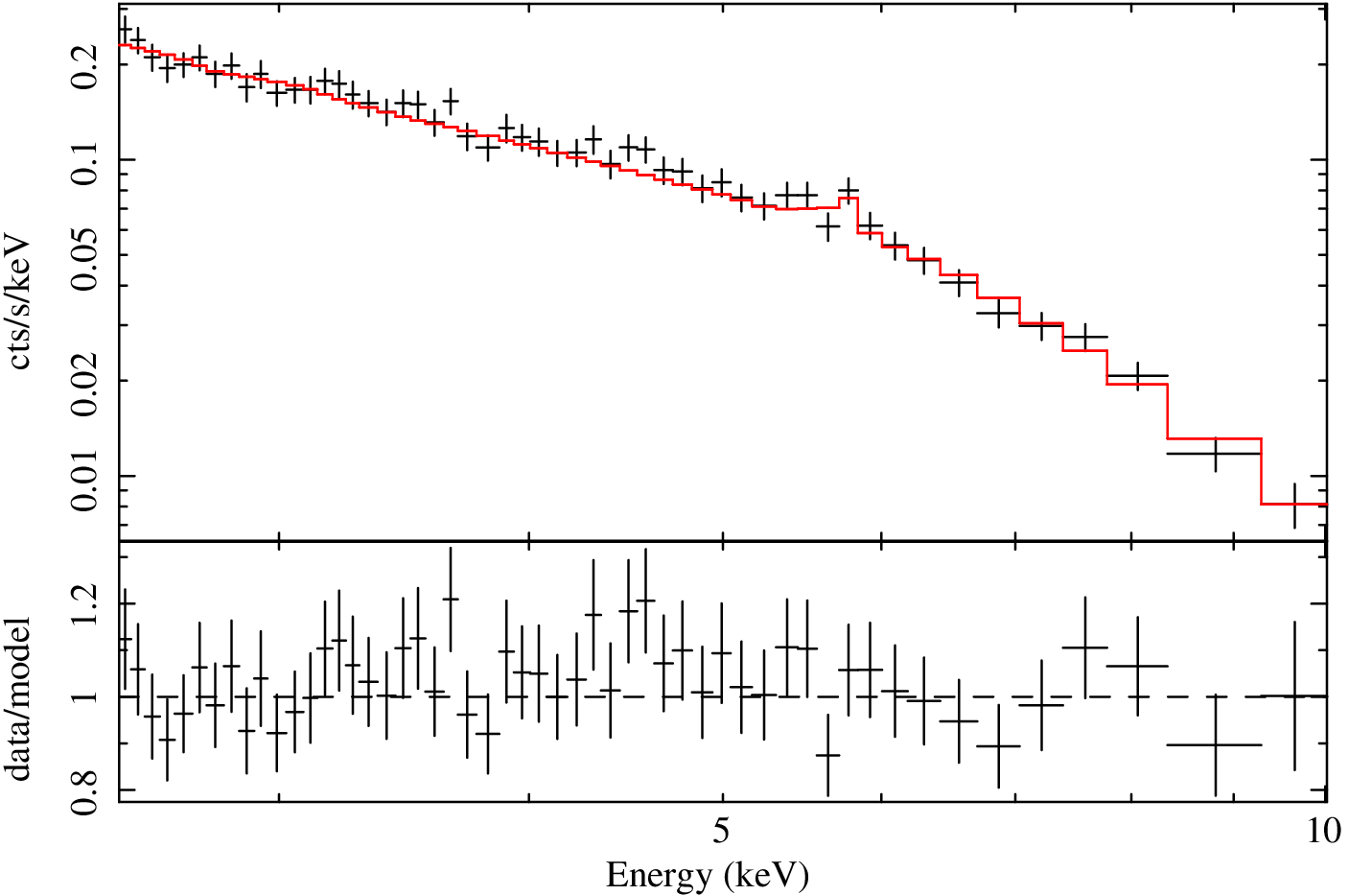}
\end{center}
\begin{center}
\includegraphics[width=0.33\textwidth,angle=270]{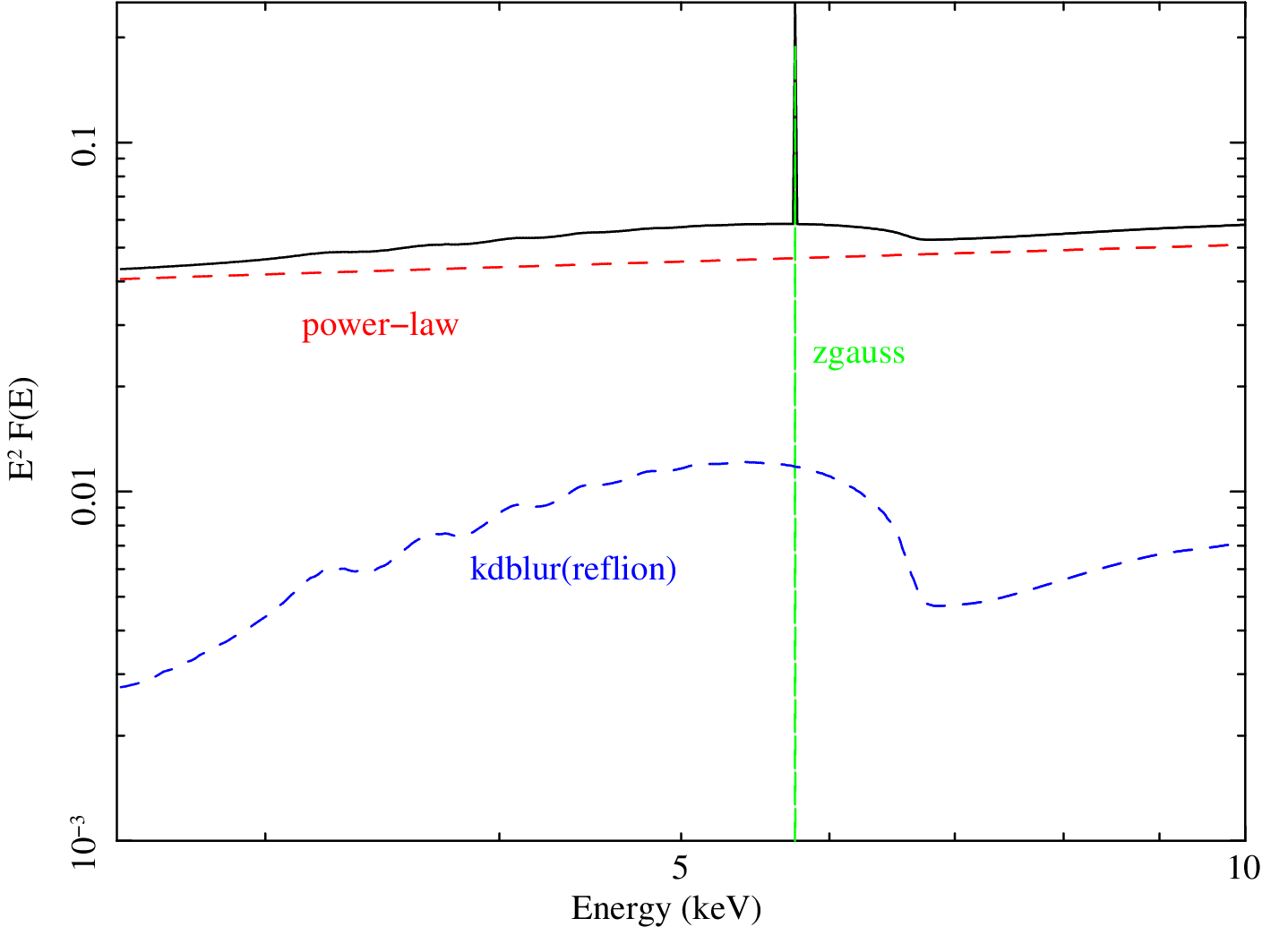}
\end{center}
\caption[Spectral fit for 3C~273.]{{\it Top left:} The $2.5-10.0
  \keV$ spectrum of 3C~273 fit with a power-law + blackbody model
  modified by Galactic photoabsorption.
  $\chi^2/\nu=248/270\,(0.92)$.  {\it Top right:} The $2.5-10.0
  \keV$ best-fit model for 3C~273, including our continuum
  model and an ionized disk reflection spectrum convolved with the
  {\tt kdblur} relativistic smearing kernel.
  $\chi^2/\nu=229/264\,(0.86)$.  {\it Bottom:} An ${\rm
  E}^2\,{\rm F(E)}$ plot depicting the relative flux
  in each of the model components in the spectrum.}
\label{fig:3c273_figs}
\end{figure}

We find evidence for a similar excess around the
Fe K region in our analysis.  Interestingly, we do {\it
 not} see
a narrow emission line at $6.4 \keV$, but rather the narrow line core
seems to coincide with the He-like line of Fe K$\alpha$ at $6.68 \pm 0.07 \keV$
with an $EW=49^{+37}_{-35} \eV$.  This may mean that the gas in this system is
significantly ionized, which would not be surprising in such an active
source.  Indeed, this hypothesis is lent credence by fitting a full {\tt
  reflion} disk spectrum to the data (which also renders the soft blackbody
component obsolete): the disk ionization parameter is
only constrained to $\xi<5689 \ergcmps$.
Visually, the residuals left over after fitting this narrow line
suggest the presence of a broad component.  Modeling this component
with our analysis method provides, at best, only a marginal improvement in
the global goodness-of-fit ($\Delta\chi^2/\Delta\nu=-9/-6$), perhaps due to the
distance of the source and the paucity of photons in the data as
compared with the other sources in our sample.  With a $20 \ks$
observation we have collected $\sim 3 \times 10^5$ photons from $2.5-10
\keV$.  Given more observing time and a larger number of (pile-up
free) photons, our
line analysis would be significantly improved.  Taking into account
these caveats about the robustness of the broad line, the {\tt
  kdblur(reflion)} model still provides the best fit, as shown in
Table~\ref{tab:tab5.tex}, though this fit is statistically no better
than that of a broad Gaussian or {\tt laor} line.  The disk emission is highly
centrally concentrated with $\alpha=6.04^{+3.94}_{-2.06}$, consistent
with the small inner radius of emission fit by the model: $r_{\rm
  min} \leq 3.73\,r_{\rm g}$.  The X-ray constraints on the inclination
angle of the disk are consistent with those obtained by T{\"u}rler
\etal, within errors: $i=58^{+15}_{-22}$ degrees.  The amount of
reflection in this source is difficult to constrain without better
photon statistics: $R_{\rm refl}=0.89^{+6311}_{-0.89}$.  
Fig.~\ref{fig:3c273_figs} shows the iron line
residual, best fit to this residual, and relative contributions of the
individual model components in the best fit, respectively.
%To $90\%$ confidence, $a>0.72$ and $r_{\rm min}\leq 2.30\,r_{\rm ms}$, but
%longer, more sensitive observations are necessary in order to
%further substantiate these results.  
The {\tt diskline} and {\tt laor}
fits are shown in Table~\ref{tab:tab10.tex}, while comparisons of the
ionized disk reflection spectrum with and without relativistic
smearing can be found in Table~\ref{tab:tab11.tex}.

\begin{table}
\begin{center}
%\begin{scriptsize}
\begin{tabular} {|l|l|l|} 
\hline \hline
{\bf Model Component} & {\bf Parameter} & {\bf Value} \\
\hline \hline
{\tt phabs}    & $N_{\rm H}\,(\pcmsq)$ & $1.79 \times 10^{20}$ \\
%\hline
%{\tt bbody}    & $kT\,(\keV)$ & $0.14^{+0.03}_{-0.02}$ \\
%               & ${\rm flux}\,(\phpcmsqps)$ & $2.10^{+0.49}_{-0.50} \times 10^{-4}$ \\
\hline
{\tt zpo}       & $\Gamma_{\rm po}$ & $1.84^{+0.46}_{-0.10}$ \\
               & ${\rm flux}\,(\phpcmsqps)$ & $4.61^{+0.35}_{-0.59} \times 10^{-2}$ \\
\hline
{\tt zgauss}   & ${\rm E}\,(\keV)$ & $6.68^{+0.07}_{-0.07}$ \\
               & ${\rm flux}\,(\phpcmsqps)$ & $1.00^{+0.75}_{-0.72} \times 10^{-4}$ \\
               & ${\rm EW}\,(\eV)$ & $49^{+37}_{-35}$ \\
\hline
{\tt kdblur} & $\alpha$ & $6.04^{+3.94}_{-2.06}$ \\
               & $i\,(\degmark)$ & $58^{+15}_{-22}$ \\
               & $r_{\rm min}\,(r_{\rm g})$ & $3.73^{+0.00}_{-3.73}$ \\
               & $r_{\rm max}\,(r_{\rm g})$ & $400$ \\
\hline
{\tt refl}     & {\rm Fe/solar} & $1.0$ \\
               & $\xi_{\rm refl}\,(\ergpcmps)$ & $82^{+5607}_{-82}$ \\
               & $\Gamma_{\rm refl}$ & $1.84^{+0.46}_{-0.10}$ \\
               & $R_{\rm refl}$ & $0.89^{+6311}_{-0.89}$ \\
               & ${\rm flux}\,(\phpcmsqps)$ & $4.78^{+10755}_{-3.65} \times 10^{-5}$ \\
\hline
{\bf $\chi^2/\nu$} & & $229/264\,(0.86)$ \\
\hline \hline 
\end{tabular}
%\end{scriptsize}
\end{center}
\caption[Best fit parameters for 3C~273]{Best-fitting model
  parameters for the $2.5-10.0 \keV$ spectrum of 3C~273,
  including components and parameter values from $0.6-1.5 \keV$ for
  completeness.  As before, the energies from $1.5-2.5 \keV$ were not included
  due to the presence of absorption edges from the {\it XMM-Newton}
  mirrors.  Error bars are quoted at $90\%$ confidence.  We required
  the {\tt zgauss} line representing the core of the Fe K$\alpha$
  feature to be intrinsically narrow, i.e.,
  $\sigma=0 \keV$.  Interestingly, this line was found at a moderate
  ionization state of iron ($6.68 \keV$) rather than at the neutral
  rest-frame energy
  of $6.4 \keV$.  Redshifts were frozen at the cosmological value for
  the source, in this case $z=0.1583$.  No warm absorption is seen in
  this object.  Note the large error bars on the value of $R_{\rm
  refl}$ that we calculate.  These are 
  likely an artifact of the relatively low photon count for this
  observation.\label{tab:tab5.tex}}
\end{table}

\subsubsection{NGC~2992}
\label{sec:n2992}

NGC~2992 is a Sy 1.9 galaxy ($z=0.0077$) that appears to show
a broad iron line even though it is highly obscured.  This source has
been the subject of intense study due to the variability of its X-ray
emission \citep{Gilli2000}.  In 1997 and 1998,
{\it BeppoSAX} caught NGC~2992 transitioning from a Compton-thick to a
Compton-thin state, resulting in an order of magnitude increase in its
X-ray luminosity as well as a qualitative difference in its spectral
appearance, with more disk features (e.g., broad lines) being seen in
several wavelengths.  In the
two {\it BeppoSAX} pointings, NGC~2992 displays 
$2-10 \keV$ X-ray fluxes of $0.63$ and $7.4 \times 10^{-11}
\ergpcmsqps$, respectively \citep{Gilli2000}.

We use the May 2003 {\it XMM-Newton} observation of NGC~2992, totaling
$29 \ks$, which translates to $\sim 3.4 \times 10^5$ photons in the
filtered data set.
This observation appears to be unpublished, so all the fit
values referenced herein are our own, compiled using the reduction
and analysis methods detailed in \S\ref{sec:analysis}.

Upon first inspection, it is immediately clear that NGC~2992 is
heavily absorbed below $\sim 2 \keV$.  Though statistical evidence
exists for a soft excess, none of the telltale signatures of intrinsic warm
absorption are present, and the flux decreases precipitously at soft
energies.  When a simple photoabsorbed power-law model 
is applied to the continuum, the remaining residuals on the
soft end indicate that some absorption remains unaccounted for. 
Including a {\tt zpcfabs} component to model this intrinsic absorption
with a partial-coverer in addition to the 
Galactic hydrogen column of $N_{\rm H}=5.26 \times 10^{20} \pcmsq$ neatly
corrects the discrepancy.  The final continuum parameters are
$N_{\rm H2}=5.39^{+0.13}_{-0.08} \times 10^{21} \pcmsq$ with a
covering fraction of $f_{\rm cov}=95^{+1}_{-0} \%$ absorbing a power-law of
photon index of $\Gamma=1.71^{+0.06}_{-0.04}$.

\begin{figure}
\begin{center}
\includegraphics[width=0.33\textwidth,angle=270]{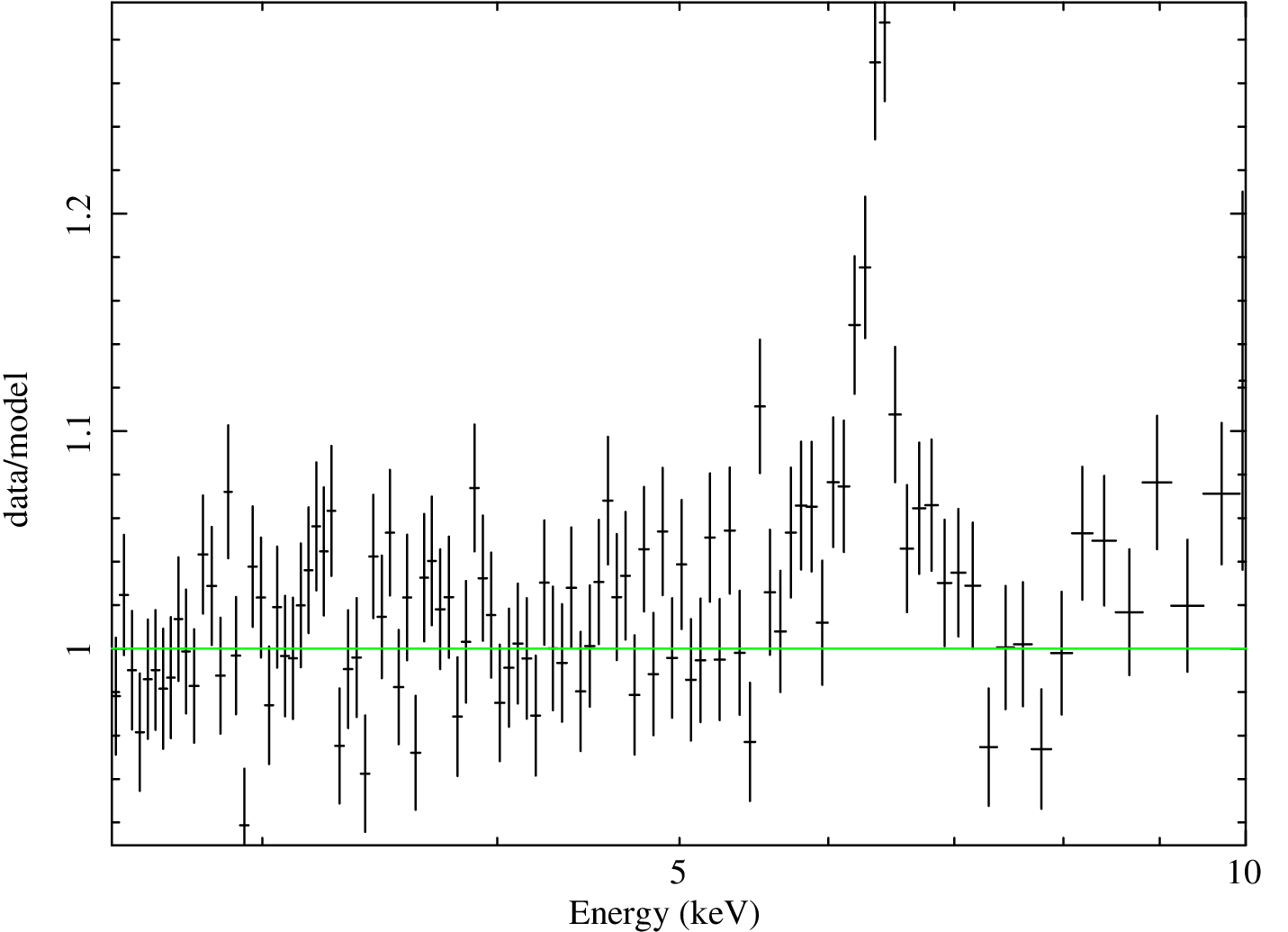}
\includegraphics[width=0.33\textwidth,angle=270]{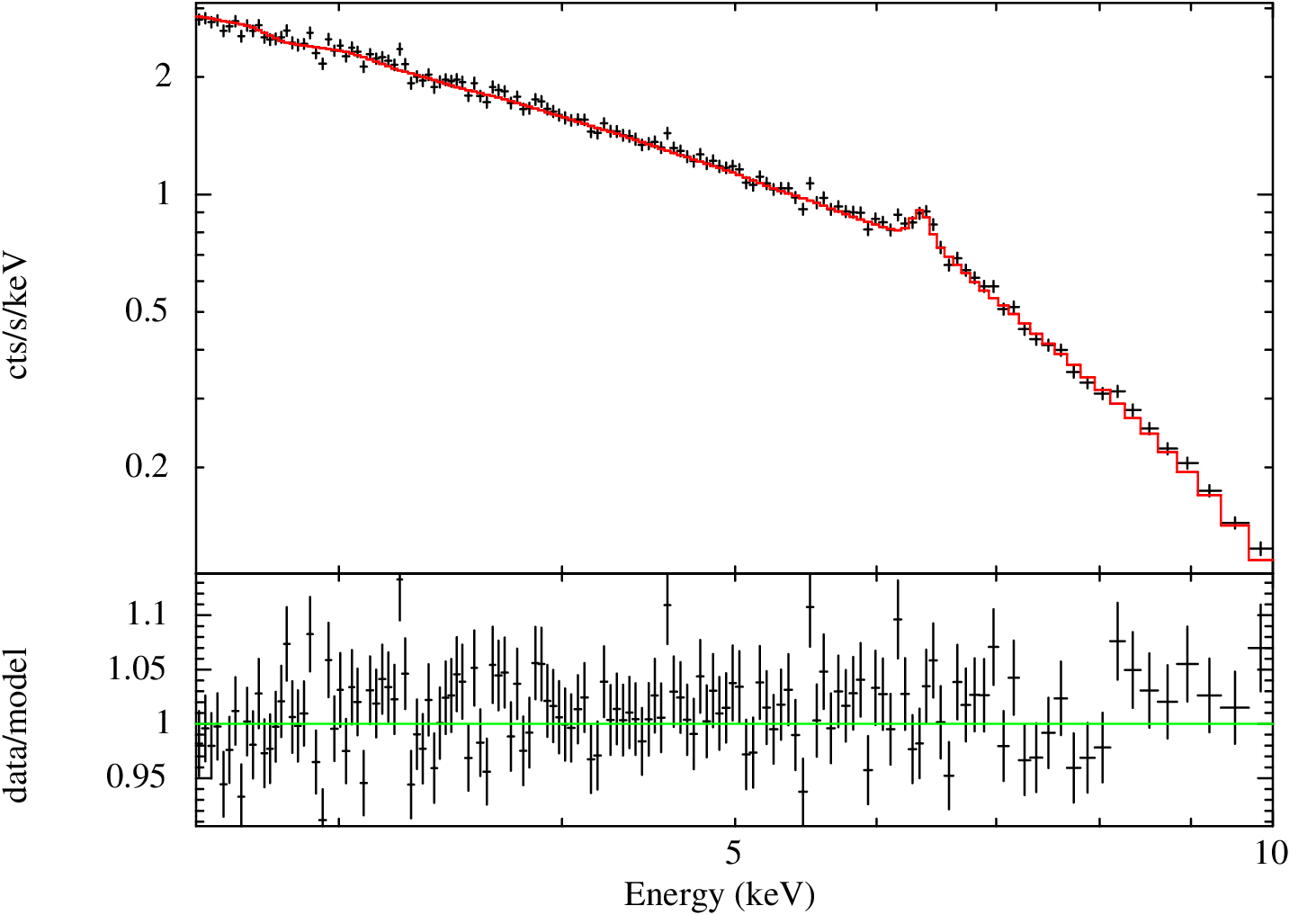}
\end{center}
\begin{center}
\includegraphics[width=0.33\textwidth,angle=270]{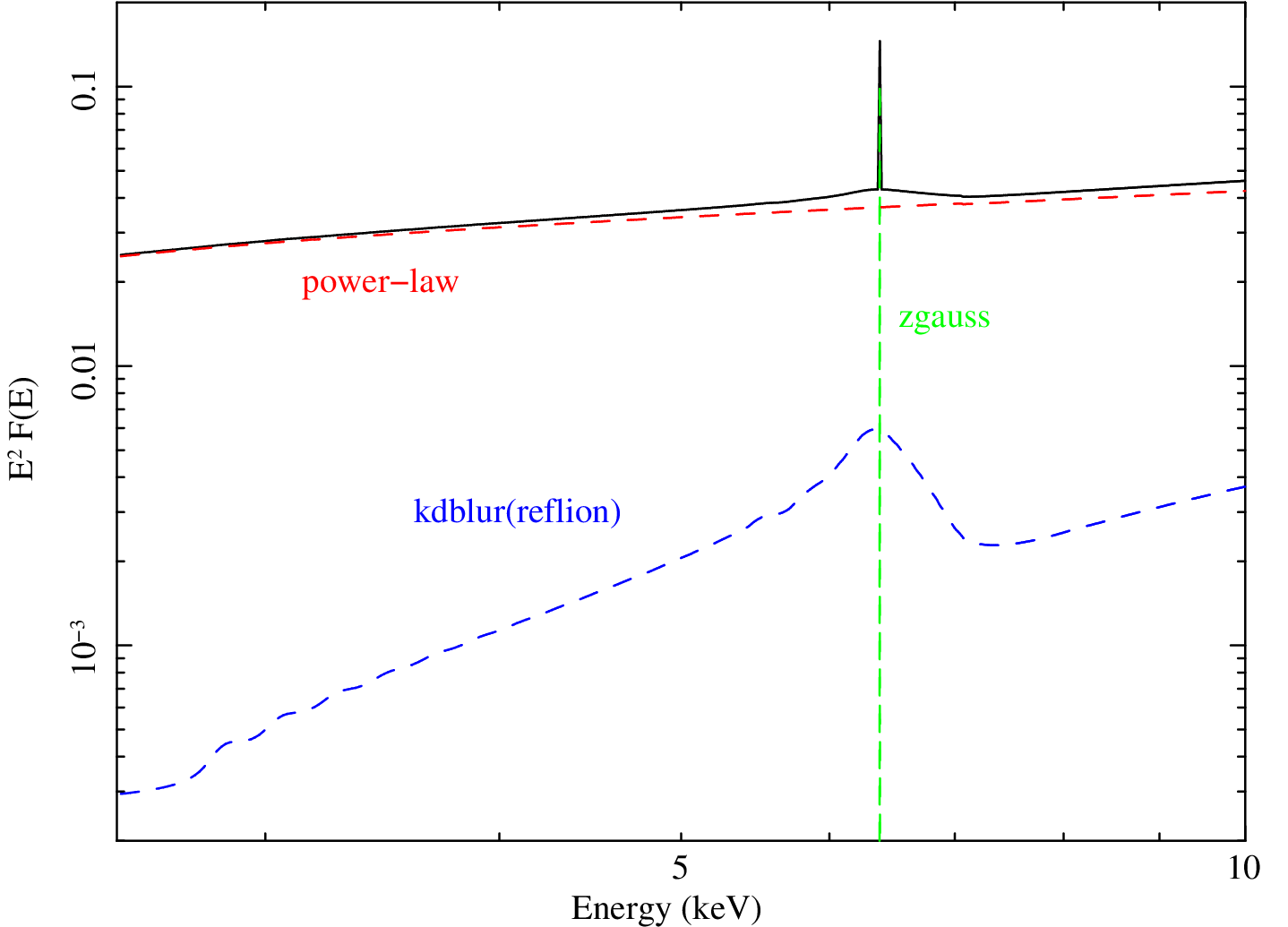}
\end{center}
\caption[Spectral fit for NGC~2992.]{{\it Top left:} The $2.5-10.0
  \keV$ spectrum of NGC~2992 fit with a continuum composed of a power-law model
  modified by Galactic photoabsorption as well as intrinsic absorption
  by neutral hydrogen.  $\chi^2/\nu=1581/1317\,(1.20)$.  {\it Top right}: The $2.5-10.0
  \keV$ best-fit model for NGC~2992, including our continuum
  model and an ionized disk reflection spectrum convolved with the
  {\tt kdblur} relativistic smearing kernel.
  $\chi^2/\nu=1396/1310\,(1.07)$.  {\it Bottom:} An ${\rm
  E}^2\,{\rm F(E)}$ plot depicting the relative flux
  in each of the model components in the spectrum.}
\label{fig:n2992_figs}
\end{figure}

The Fe K$\alpha$ line is present with a narrow core at
$6.43^{+0.01}_{-0.02} \keV$ and
$EW=36^{+12}_{-9} \eV$.  Significant residuals remain surrounding the core,
however, suggestive of the presence of a broad component to the Fe
K$\alpha$ line.  Adding a broad Gaussian component at $6.40 \keV$
results in a significant improvement in the global goodness-of-fit of
$\Delta\chi^2/\Delta\nu=-25/-3$, and the fitted width of $\sigma=0.20-0.51
\keV$ implies an origin of $\sim 30-190\,r_{\rm g}$ from the black
hole assuming a Keplerian accretion disk.  The fit is only marginally
improved by modeling this broad component with a {\tt diskline}, {\tt
  laor} or relativistically smeared ionized disk reflection model,
though the best statistical fit is technically achieved with a {\tt
  kdblur(reflion)} model.  Nonetheless, due to a lack of significant
difference between the {\tt rdblur} and {\tt kdblur} fits, we cannot
distinguish between a spinning and non-spinning black hole in this
source, nor can we definitively affirm the presence of relativistic
effects on the Fe K$\alpha$ line profile.   
See Table~\ref{tab:tab6.tex} for model parameters and error bars.
Fig.~\ref{fig:n2992_figs} shows the iron line
residual, best fit to this residual, and relative contributions of the
individual model components for the best fit.
%Note that the BH spin cannot be constrained.  Constraints are also
%difficult to achieve on the disk emissivity index $\alpha$,
%especially, though we have constrained the effective inner disk radius
%to $r_{\rm min}\leq 11.78\,r_{\rm ms}$.  
The {\tt diskline} and {\tt laor}
fits are shown in Table~\ref{tab:tab10.tex}, while comparisons of the
ionized disk reflection spectrum with and without relativistic
smearing can be found in Table~\ref{tab:tab11.tex}.

\begin{table}
\begin{center}
%\begin{scriptsize}
\begin{tabular} {|l|l|l|} 
\hline \hline
{\bf Model Component} & {\bf Parameter} & {\bf Value} \\
\hline \hline
{\tt phabs}    & $N_{\rm H}\,(\pcmsq)$ & $5.26 \times 10^{20}$ \\
\hline
{\tt zpcfabs}  & $N_{\rm H2}\,(\pcmsq)$ & $5.39^{+0.13}_{-0.08}\times 10^{21}$ \\
               & $F_{\rm cov}\,(\%)$ & $ 95^{+1}_{-0}$ \\
\hline 
{\tt zpo}       & $\Gamma_{\rm po}$ & $1.71^{+0.06}_{-0.04}$ \\
               & ${\rm flux}\,(\phpcmsqps)$ & $2.20^{+0.13}_{-0.06} \times 10^{-2}$ \\
\hline
{\tt zgauss}   & ${\rm E}\,(\keV)$ & $6.43^{+0.01}_{-0.02}$ \\
               & ${\rm flux}\,(\phpcmsqps)$ & $3.84^{+1.24}_{-0.98} \times 10^{-5}$ \\
               & ${\rm EW}\,(\eV)$ & $36^{+12}_{-9}$ \\
\hline
{\tt kdblur} & $\alpha$ & $3.00^{+7.00}_{-3.00}$ \\
               & $i\,(\degmark)$ & $43^{+4}_{-10}$ \\
               & $r_{\rm min}\,(r_{\rm g})$ & $3.64^{+1.29}_{-0.73}$ \\
               & $r_{\rm max}\,(r_{\rm g})$ & $400$ \\
\hline
{\tt reflion}  & {\rm Fe/solar} & $1.0$ \\
               & $\xi_{\rm refl}\,(\ergpcmps)$ & $30^{+37}_{-0}$ \\
               & $\Gamma_{\rm refl}$ & $1.71^{+0.06}_{-0.04}$ \\
               & $R_{\rm refl}$ & $0.49^{+0.10}_{-0.37}$ \\
               & ${\rm flux}\,(\phpcmsqps)$ & $1.82^{+0.34}_{-0.73} \times 10^{-5}$ \\
\hline
{\bf $\chi^2/\nu$} & & $1396/1310\,(1.07)$ \\
\hline \hline 
\end{tabular}
%\end{scriptsize}
\end{center}
\caption[Best fit parameters for NGC~2992]{Best-fitting model
  parameters for the $2.5-10.0 \keV$ spectrum of NGC~2992,
  including components and parameter values from $0.6-1.5 \keV$ for
  completeness.  The energies from $1.5-2.5 \keV$ were not included
  due to the presence of absorption edges from the {\it XMM-Newton}
  mirrors.  Error bars are quoted at $90\%$ confidence.  
%We required
%  the $6.4 \keV$ {\tt zgauss} line core to be intrinsically narrow, i.e.,
%  $\sigma=0 \keV$.  
Redshifts were frozen at the cosmological value for
  the source, in this case $z=0.0077$.  No evidence for warm
  absorption is detected.  The iron abundance of the ionized disk
  could not be constrained, so it was held fixed at the solar value.\label{tab:tab6.tex}}
\end{table}

\subsubsection{NGC~4051}
\label{sec:n4051}

NGC~4051, like NGC~2992, is a heavily absorbed Seyfert AGN (Sy 1.5;
NLS1) seen to vary significantly in flux over the course of several
observations \citep{Guainazzi1996,Lamer2003}.  The source is at a
redshift of $z=0.0023$ and has a
typical flux on the order of a few times $10^{-11} \ergpcmsqps$,
though as stated above, this flux can vary significantly on a variety
of time scales, along with the spectral characteristics of the
source.  Unusually low flux states in this object can last for weeks
to months, during which time the X-ray spectrum shows a hard continuum
power-law of spectral slope $\Gamma \sim 1$, but is dominated by a
softer component
at lower energies with $\Gamma \sim 3$ \citep{Uttley2004}.  A highly
broadened and redshifted iron line has also been noted in the low flux
state of NGC~4051 with {\it RXTE}, suggesting that reflection features
from the accretion disk close to the BH may remain constant in this
source in spite of the large variations in continuum properties
\citep{Uttley2003}.  

NGC~4051 was observed with {\it XMM-Newton} in May 2001 for a duration
of $117 \ks$, yielding $\sim 2.5 \times 10^6$ photons in the filtered
data set.  The
EPIC-pn results were first reported by Mason \etal,
and suggested a continuum described by a power-law ``pivoting'' around
$100 \keV$, according to a simultaneous observation with {\it RXTE}.
Ultraviolet emission from the Optical Monitor on {\it XMM-Newton} was
found to lag the X-ray emission by $\sim 0.2$ days, indicating that it
is likely reprocessed X-ray emission.  The X-ray
emission itself showed variability on time scales as small as $1-2$
hours \citep{Mason2002}.  These results were expanded upon by Pounds
\etal, who noted the intermediate flux of the source at this time and
validated the presence of both an iron line and a thermal soft excess
during the May 2001 observation \citep{Pounds2004}.  Ponti \etal
recently re-analyzed this observation as well as a lower-flux pointing
from November 2002 and reinforced the veracity of this model, while
also considering the comparable efficacy of a model dominated by ionized
reflection from radii quite close to the BH \citep{Ponti2006}.

We have also focused on the May 2001 observation of NGC~4051,
following the reduction and analysis of Mason \etal and Ponti \etal
but using modern calibration files.
Because we do not employ high energy data above $10 \keV$ in our
analysis, we find that the underlying $0.6-10 \keV$ continuum is well
fit by a single power-law component with photon index
$\Gamma=2.36^{+0.36}_{-0.20}$.  This power-law is modified by
Galactic photoabsorption from neutral hydrogen with $N_{\rm H}= 1.32 \times
10^{20}\pcmsq$ as well as ionized intrinsic absorption with $N_{\rm
  H}=2.77^{+3.08}_{-0.51} \times 10^{22} \pcmsq$ and $\xi=81^{+1}_{-2}
\ergcmps$.  A soft excess is seen below $\sim 1 \keV$, as
in so many other Seyfert galaxies.  The best fit to this feature is
initially achieved with a thermal blackbody component, with $kT=0.10
\pm 0.01$.

\begin{figure}
\begin{center}
\includegraphics[width=0.33\textwidth,angle=270]{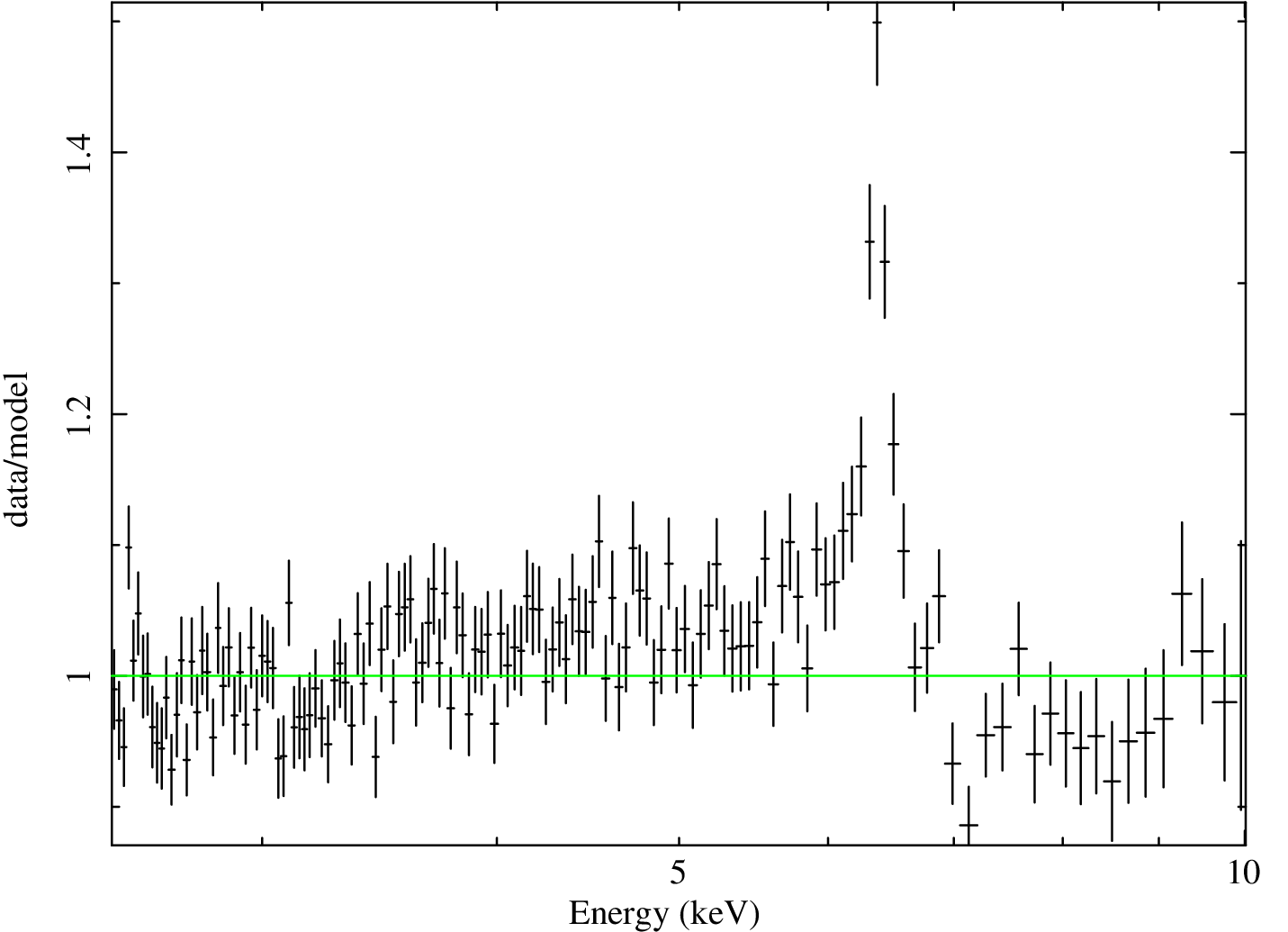}
\includegraphics[width=0.33\textwidth,angle=270]{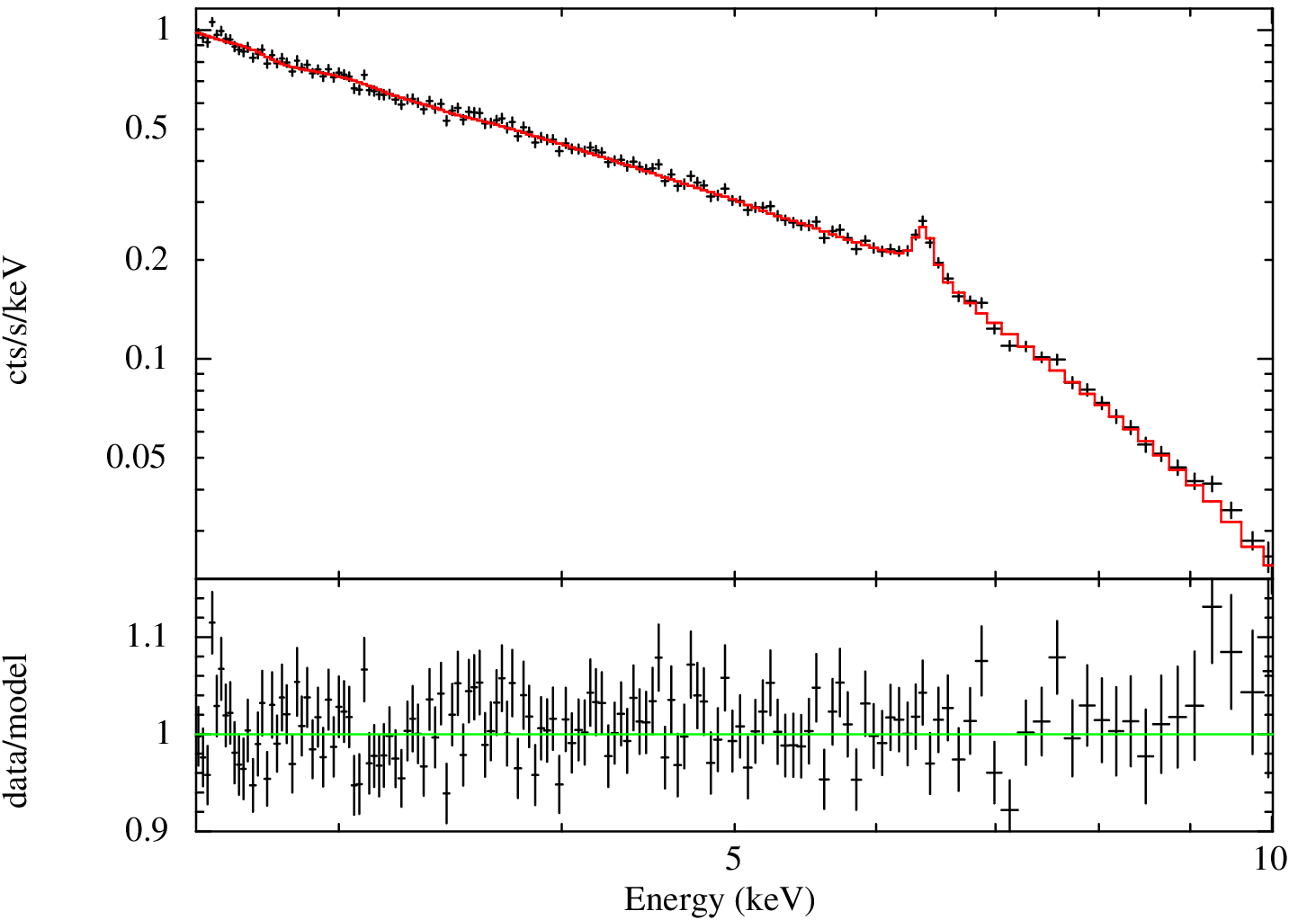}
\end{center}
\begin{center}
\includegraphics[width=0.33\textwidth,angle=270]{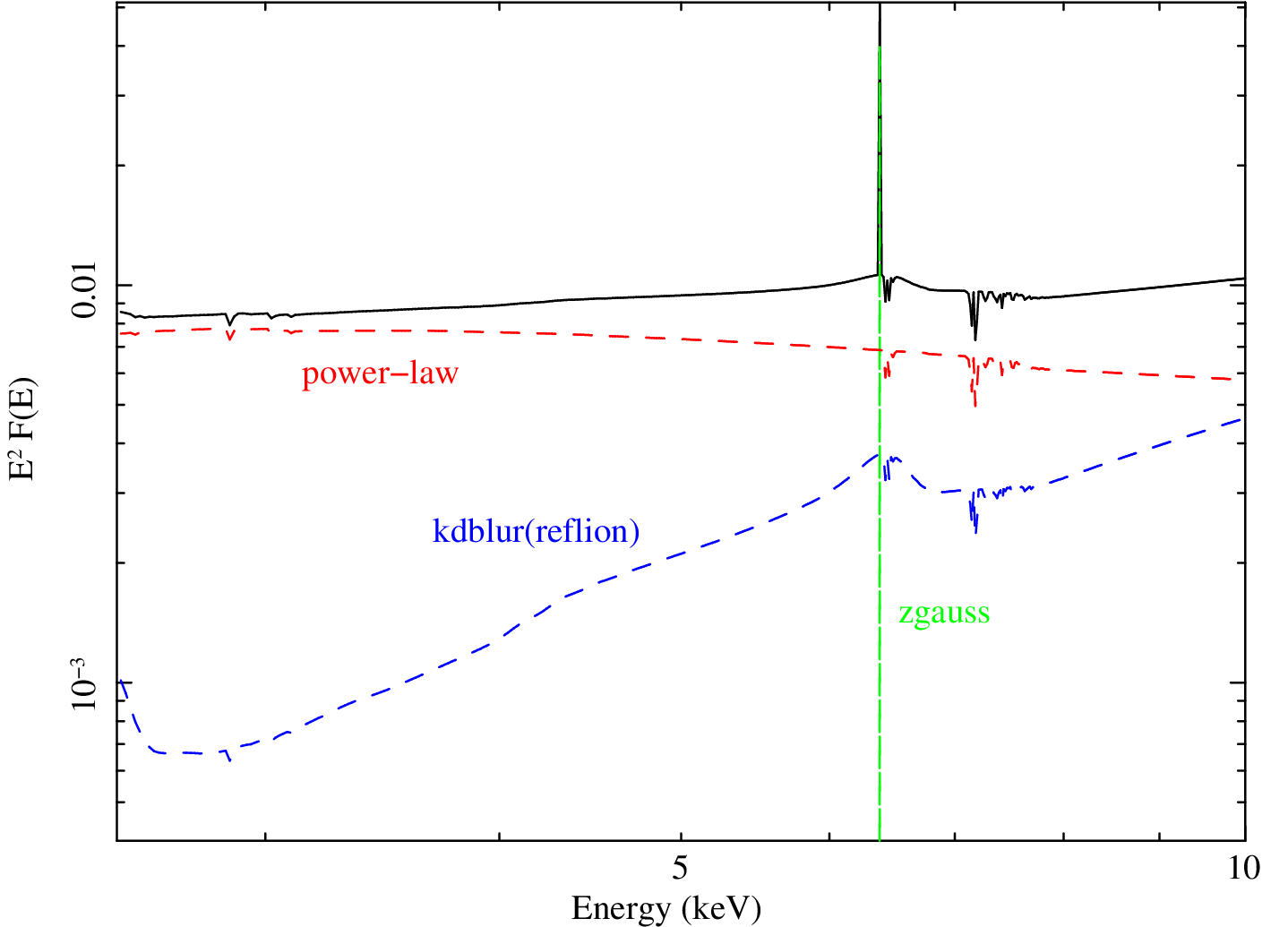}
\end{center}
\caption[Spectral fit for NGC~4051.]{{\it Top left:} The $2.5-10.0
  \keV$ spectrum of NGC~4051 fit with a continuum composed of a power-law model
  modified by Galactic photoabsorption, a warm absorber model
  and a blackbody soft excess.  $\chi^2/\nu=1837/1402\,(1.31)$.  {\it Top right:} The $2.5-10.0
  \keV$ best-fit model for NGC~4051, including our continuum
  model, a narrow Fe K$\alpha$ line core and an ionized disk
  reflection spectrum convolved with the
  {\tt kdblur} relativistic smearing kernel.
  $\chi^2/\nu=1468/1392\,(1.05)$.  {\it Bottom:} An ${\rm
  E}^2\,{\rm F(E)}$ plot depicting the relative flux
  in each of the model components in the spectrum.}
\label{fig:n4051_figs}
\end{figure}

A narrow $6.40^{0.01}_{-0.03} \keV$ Gaussian was successfully fit to the core of the
Fe K$\alpha$ line in this source, though we note significant residuals marking
the red and blue wings of a broadened line.  The narrow core has an
equivalent width of 
$EW= 59^{+14}_{-11} \eV$.  The remaining broad line is best fit here
with either a {\tt diskline} or a {\tt laor} model with $EW \sim 35
\eV$, but in each case fitting this single relativistic disk line
leaves several parameters poorly constrained.  A better fit to the
entire spectrum is achieved by convolving a static ionized disk
reflection spectrum model ({\tt reflion}) with a relativistic smearing
kernel.  This more physical model also has the advantage of
self-consistently accounting for the observed soft excess without the
need to include a phenomenological blackbody component.  The spectrum
of NGC~4051 is equally well fit by a 
{\tt kdblur(reflion)} or an {\tt rdblur(reflion)} model, though the {\tt
  kdblur} model is cited as the best fit due to slightly tighter
parameter constraints on the inner radius of disk emission ($r_{\rm
  min}=6.94^{+2.41}_{-2.95}\,r_{\rm g}$).  
%having a BH spin of $a>0.67$ and an $r_{\rm
%  min}=2.25-4.62\,r_{\rm ms}$.  
Full parameter values and $90\%$
confidence error bars are listed in Table~\ref{tab:tab7.tex}.
Fig.~\ref{fig:n4051_figs} shows the iron line
residual, best fit to this residual, and relative contributions of the
individual model components, respectively.  The {\tt diskline} and {\tt laor}
fits are shown in Table~\ref{tab:tab10.tex}, while comparisons of the
ionized disk reflection spectrum with and without relativistic
smearing can be found in Table~\ref{tab:tab11.tex}.

\begin{table}
\begin{center}
%\begin{scriptsize}
\begin{tabular} {|l|l|l|} 
\hline \hline
{\bf Model Component} & {\bf Parameter} & {\bf Value} \\
\hline \hline
{\tt phabs}    & $N_{\rm H}\,(\pcmsq)$ & $1.32 \times 10^{20}$ \\
\hline
{\tt WA}       & $N_{\rm WA}\,(\pcmsq)$ & $2.77^{+3.08}_{-0.51} \times 10^{22}$ \\
               & $\log \xi_{\rm WA}$ & $1.91^{+0.17}_{-0.27}$ \\
\hline
{\tt zpo}       & $\Gamma_{\rm po}$ & $2.36^{+0.36}_{-0.20}$ \\
               & ${\rm flux}\,(\phpcmsqps)$ & $1.39^{+0.09}_{-0.28} \times 10^{-2}$ \\
\hline
{\tt zgauss}   & ${\rm E}\,(\keV)$ & $6.40^{+0.01}_{-0.03}$ \\
               & ${\rm flux}\,(\phpcmsqps)$ & $1.53^{+0.37}_{-0.29} \times 10^{-5}$ \\
               & ${\rm EW}\,(\eV)$ & $59^{+14}_{-11}$ \\
\hline
{\tt kdblur} & $\alpha$ & $8.29^{+1.71}_{-8.29}$ \\
               & $i\,(\degmark)$ & $33^{+6}_{-9}$ \\
               & $r_{\rm min}\,(r_{\rm g})$ & $6.94^{+2.41}_{-2.95}$ \\
               & $r_{\rm max}\,(r_{\rm g})$ & $400$ \\
\hline
{\tt reflion}  & {\rm Fe/solar} & $0.42^{+0.20}_{-0.08}$ \\
               & $\xi_{\rm refl}\,(\ergpcmps)$ & $99^{+26}_{-61}$ \\
               & $\Gamma_{\rm refl}$ & $2.36^{+0.36}_{-0.20}$ \\
               & $R_{\rm refl}$ & $0.88^{+12.94}_{-0.43}$ \\
               & ${\rm flux}\,(\phpcmsqps)$ & $2.77^{+10.66}_{-0.88} \times 10^{-5}$ \\
\hline
{\bf $\chi^2/\nu$} & & $1468/1392\,(1.05)$ \\
\hline \hline 
\end{tabular}
%\end{scriptsize}
\end{center}
\caption[Best fit parameters for NGC~4051]{Best-fitting model
  parameters for the $2.5-10.0 \keV$ spectrum of NGC~4051,
  including components and parameter values from $0.6-1.5 \keV$ for
  completeness.  The energies from $1.5-2.5 \keV$ were not included
  due to the presence of absorption edges from the {\it XMM-Newton}
  mirrors.  Error bars are quoted at $90\%$ confidence.  We required
  the $6.4 \keV$ {\tt zgauss} line to be intrinsically narrow, i.e.,
  $\sigma=0 \keV$.  Redshifts were frozen at the cosmological value for
  the source, in this case $z=0.0023$.  A warm absorber 
  is present in this object, as in many
  other Sy 1 sources.\label{tab:tab7.tex}}
\end{table}

\subsubsection{Ark~120}
\label{sec:ark120}

Ark~120 is a bright Sy 1 AGN 
%with an estimated BH mass of $\sim 2
%\times 10^8 \Msun$ \citep{Wandel1999} and a bolometric luminosity of $L_{\rm bol}
%\geq 10^{45} \ergps$ \citep{Edelson1986}.  
at a redshift of $z=0.0327$,
and has a relatively constant $2-10 \keV$ X-ray flux of
$F_{2-10} \sim 2.50 \times 10^{-11} \ergpcmsqps$
\citep{Vaughan2004_2}.  The source is radio-quiet, and due to a lack
of observed evidence for intrinsic absorption, Ark~120 has been
labeled a ``bare'' Sy 1 nucleus \citep{Ward1987,Vaughan2004_2}.  
%Its host galaxy is
%an early-type spiral of Hubble type S0/a with an inclination angle of
%$i \approx 26 \degmark$ \citep{Nordgren1995}.  
Ark~120 has been
observed by most of the major X-ray observatories.  An {\it EXOSAT}
observation showed the source to have a steep soft X-ray spectrum
\citep{Turner1989}, as did a subsequent {\it ROSAT} observation
\citep{Brandt1993}.  Furthermore, as mentioned above, these X-ray
observations showed no indication of any warm absorption features.
Similar findings were seen in ultraviolet observations
\citep{Crenshaw1999,Crenshaw2001}.  

The source was observed by {\it XMM-Newton} in August 2003 for an
effective duration of $100 \ks$ ($\sim 2.3 \times 10^6$ photons by
\citet{Vaughan2004_2}.  We follow the data
reduction and analysis of these authors, using the latest calibration
files, though we restrict our
attention to the EPIC-pn camera data.  Vaughan \etal find that the
continuum emission is well described by a simple photoabsorbed
power-law model with no evidence for complex absorption intrinsic to
the system.  This finding is substantiated by a comparison to the
spectrum of 3C~273 in its radio-loud state, when the source is
dominated by jet emission and displays a very flat, featureless X-ray
spectrum lacking intrinsic emission and absorption features.  Vaughan
\etal calculated the ratio between the spectra of Ark~120 and 3C~273,
then normalized it by a spectral model for 3C~273.  The resulting spectrum
clearly shows that Ark~120, too, possesses a flat, featureless
continuum above $\sim 1 \keV$ and that its only significant emission line is that of
Fe K$\alpha$ at $6.4 \keV$.  The authors use
a multiple blackbody component (among many other models, with varying
degrees of success) for the soft excess to reduce
residuals left over from the power-law fit below $\sim 1 \keV$, and a slight
curvature of the continuum is noted at higher energies that is thought to indicate the
presence of disk reflection in this source \citep{Vaughan2004_2}.  

Our continuum fit is consistent with that of Vaughan \etal, though we
find a statistical need for only two blackbody components to parameterize the
soft excess emission, rather than three: $kT_1=0.19 \pm 0.01 \keV$ and
$kT_2= 0.08^{+0.01}_{-0.00}\keV$.  These two temperatures are within errors
of the first two blackbodies of Vaughan \etal, but we do not detect
the third component at $kT \sim 0.67 \keV$ robustly.  The hard X-ray
continuum is well represented by a power-law of
$\Gamma=2.11^{+0.03}_{-0.07}$, again consistent with the Vaughan \etal
value within errors.  The value of the Galactic absorbing column is
taken as $N_{\rm H}=1.26 \times 10^{21}\pcmsq$.  

Vaughan \etal detected the presence of broad and narrow components to
the $6.4 \keV$ Fe K$\alpha$ line ($EW_{\rm broad} \sim 100 \eV$,
$EW_{\rm narrow} \sim 40 \eV$).  The narrow
component could easily be fit with a $6.4 \keV$ Gaussian profile, as was our
Fe K$\alpha$ narrow core, but Vaughan \etal found that a {\tt
  diskline} model worked best for their broad component ($r_{\rm min}
\approx 144\,r_{\rm g}$).  Interestingly, this broad component
fit to a rest-frame energy of $6.56 \keV$ rather than $6.4 \keV$, with
the latter yielding a substantially worse global fit
($\Delta\chi^2=+21.5$) \citep{Vaughan2004_2}.  This energy
corresponds to the line emission
being dominated by intermediately ionized iron (Fe\,{\sc xx-xxii}), meaning
that resonant trapping and the Auger effect should destroy the line
\citep{Ross1996}.  

\begin{figure}
\begin{center}
\includegraphics[width=0.33\textwidth,angle=270]{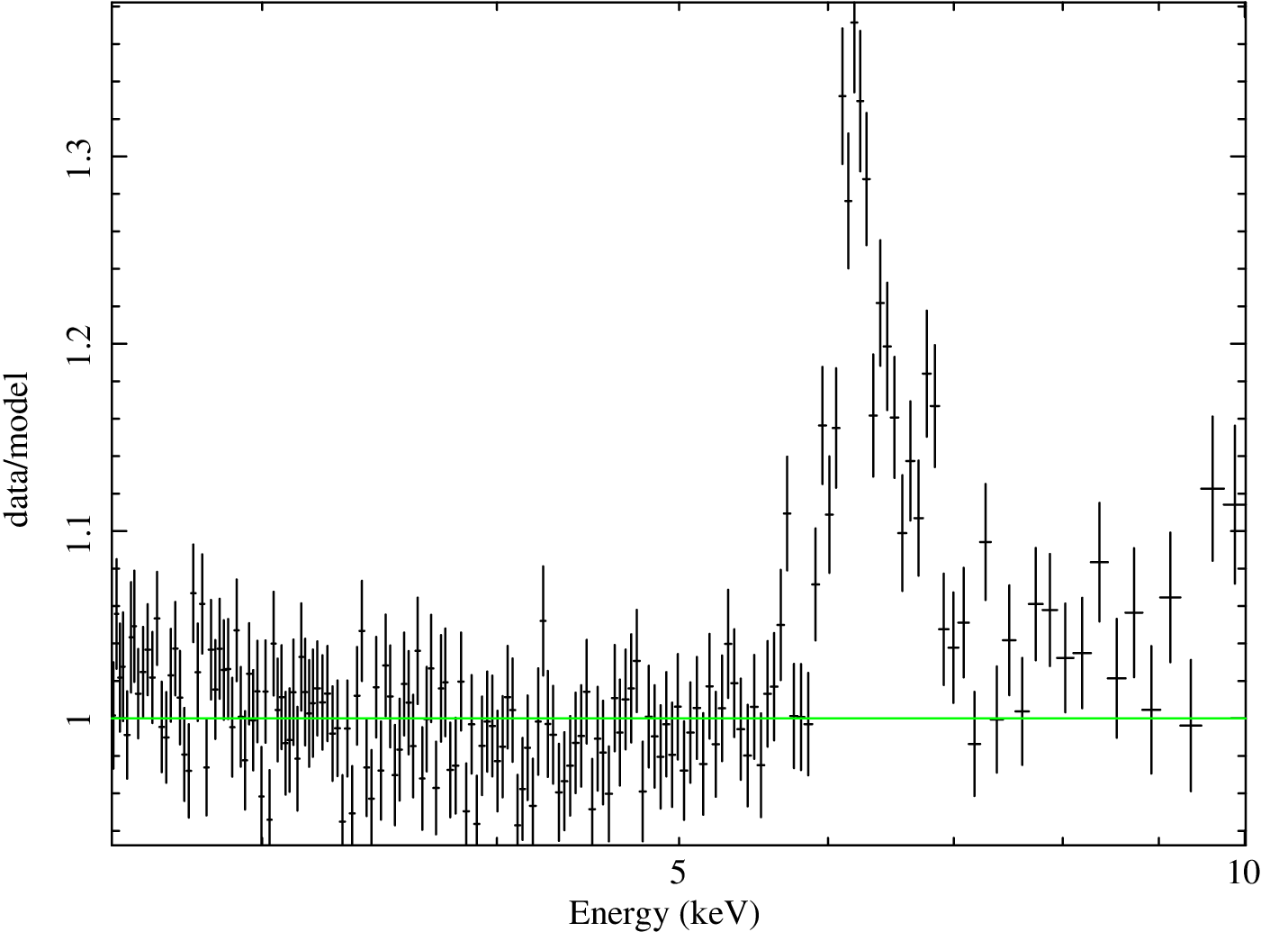}
\includegraphics[width=0.33\textwidth,angle=270]{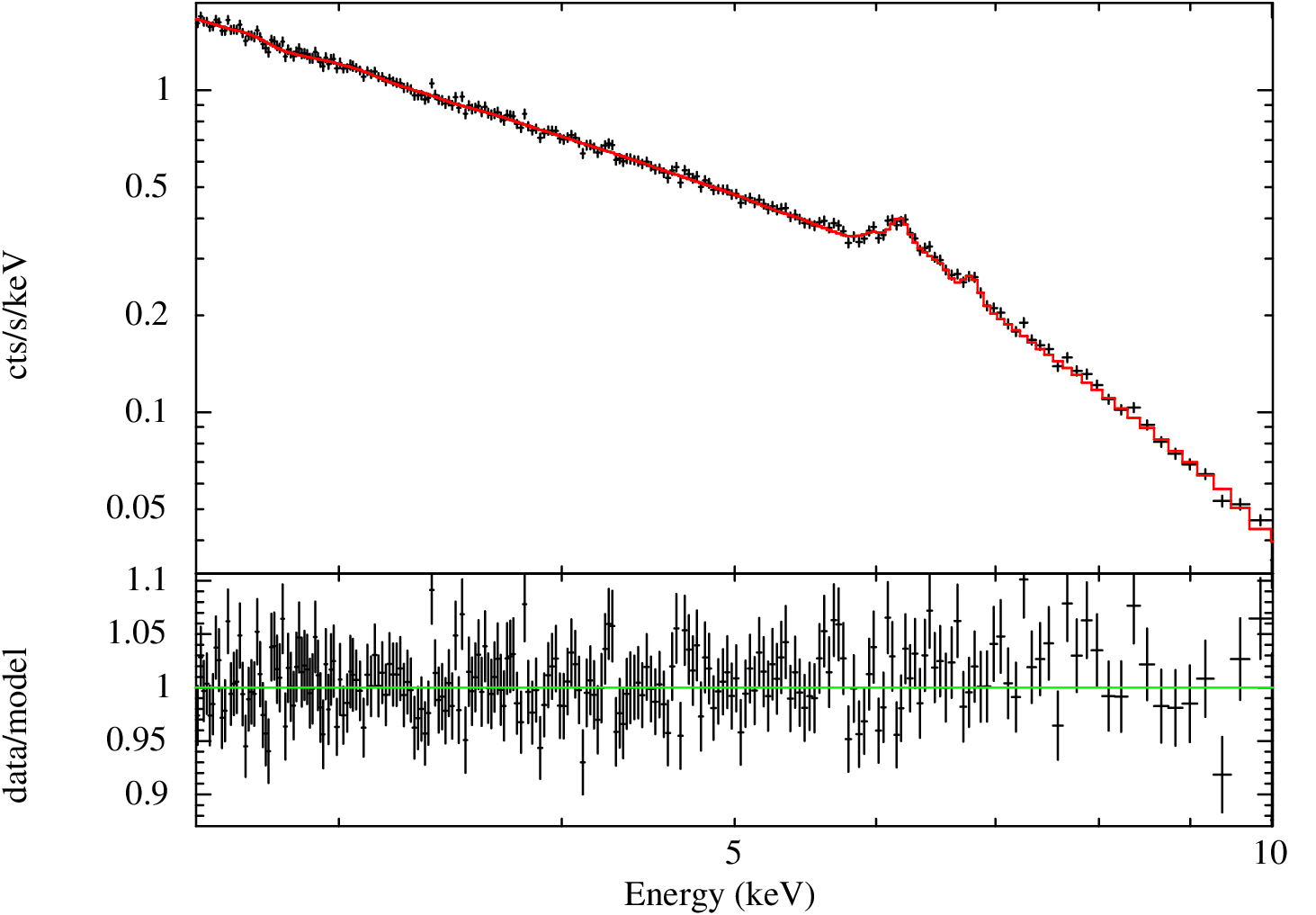}
\end{center}
\begin{center}
\includegraphics[width=0.33\textwidth,angle=270]{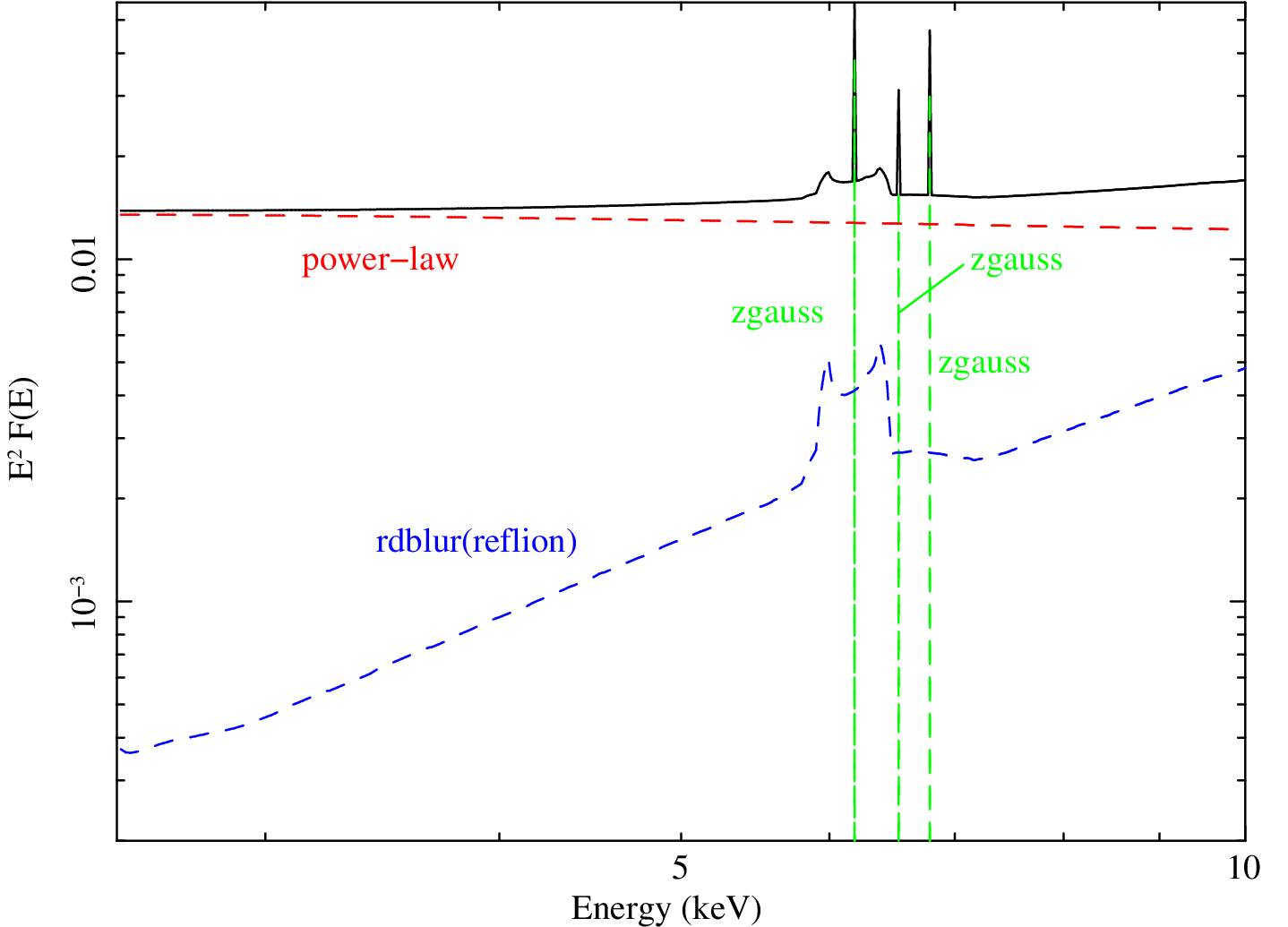}
\end{center}
\caption[Spectral fit for Ark~120.]{{\it Top left:} The $2.5-10.0
  \keV$ spectrum of Ark~120 fit with a continuum composed of a power-law model
  modified by Galactic photoabsorption
  and a two-zone blackbody soft excess.
  $\chi^2/\nu=2153/1428\,(1.51)$.  {\it Top right:} The $2.5-10.0
  \keV$ best-fit model for Ark~120, including our continuum
  model and an ionized disk reflection spectrum convolved with the
  {\tt rdblur} relativistic smearing kernel.
  $\chi^2/\nu=1511/1418\,(1.07)$.  {\it Bottom:} An ${\rm
  E}^2\,{\rm F(E)}$ plot depicting the relative flux
  in each of the model components in the spectrum.}
\label{fig:ark120_figs}
\end{figure}

In an effort to solve this puzzle we have fit our iron line profile
with three narrow Gaussian components, the first two at $6.40 \pm 0.01 \keV$ and 
$7.00^{+0.02}_{-0.03} \keV$, representing neutral iron and a likely
blend of highly ionized H-like Fe K$\alpha$ and Fe K$\beta$, respectively.  The lines
have $EW_{\rm cold}=36^{+17}_{-8} \eV$ and $EW_{\rm ion}=28^{+7}_{-8}
\eV$.  After an initial attempt to fit the broad residual remaining,
there is evidence for a third narrow line as well at
$6.75^{+0.04}_{-0.07} \keV$ ($EW=14^{+8}_{-4} \eV$), indicating the presence of
moderately ionized iron.  The addition of this component 
improves the fit by $\Delta\chi^2=-19$ for two additional degrees of
freedom.  
This fit leaves a residual broad feature nicely centered at $6.4 \keV$,
which renders this scenario physically consistent with other systems
and removes the need to explain the presence of the broad line of iron
in an intermediate state of ionization.  We then achieve a best fit to
the broad component using a Schwarzschild {\tt rdblur(reflion)} model, which also
nicely accounts for the additional curvature noted in the spectrum at
higher energies by Vaughan \etal, as well as the soft excess we had
initially modeled with phenomenological blackbody components.  
%From this fit, we have constrained the spin of the BH
%to be $a=0.63-0.68$ and $r_{\rm min}=1.00-1.09\,r_{\rm ms}$.  
The best-fit parameter values and error bars are presented in
Table~\ref{tab:tab8.tex}.
Fig.~\ref{fig:ark120_figs} shows the iron line
residual, best fit to this residual, and relative contributions of the
individual model components for the best fit.  The {\tt diskline} and {\tt laor}
fits are shown in Table~\ref{tab:tab10.tex}, while comparisons of the
ionized disk reflection spectrum with and without relativistic
smearing can be found in Table~\ref{tab:tab11.tex}.  
%
%Though this result is intriguing for its intermediate BH spin value,
%one must interpret it with caution.  This source, in particular, seems
%to have a weak emission feature present in the spectrum just below the
%neutral iron line in energy that sticks out as a ``bump'' in the
%spectrum.  It is unclear what this feature may represent, and left
%unmodeled, the spectral fit seems to smooth over it rather than fit it
%properly.  This may affect the parameter values and error bars we
%obtain for the {\tt kdblur(reflion)} model.  More detailed, longer
%observations with higher spectral resolution will be required to solve
%this puzzle and more reliably determine the BH spin in Ark~120.  

\begin{table}
\begin{center}
%\begin{scriptsize}
\begin{tabular} {|l|l|l|} 
\hline \hline
{\bf Model Component} & {\bf Parameter} & {\bf Value} \\
\hline \hline
{\tt phabs}    & $N_{\rm H}\,(\pcmsq)$ & $1.26 \times 10^{21}$ \\
\hline
{\tt zpo}       & $\Gamma_{\rm po}$ & $2.11^{+0.03}_{-0.07}$ \\
               & ${\rm flux}\,(\phpcmsqps)$ & $1.64^{+0.05}_{-0.12} \times 10^{-2}$ \\
\hline
%{\tt bbody 1}  & $kT\,(\keV)$ & $0.19^{+0.01}_{-0.01}$ \\
%               & ${\rm flux}\,(\phpcmsqps)$ & $1.85^{+0.12}_{-0.12} \times 10^{-4}$ \\
%\hline
%{\tt bbody 2}  & $kT\,(\keV)$ & $0.08^{+0.01}_{-0.00}$ \\
%               & ${\rm flux}\,(\phpcmsqps)$ & $6.57^{+0.68}_{-1.04} \times 10^{-4}$ \\
%\hline
{\tt zgauss}   & ${\rm E}\,(\keV)$ & $6.40^{+0.01}_{-0.01}$ \\
               & ${\rm flux}\,(\phpcmsqps)$ & $1.65^{+0.32}_{-0.35} \times 10^{-5}$ \\
               & ${\rm EW}\,(\eV)$ & $36^{+17}_{-8}$ \\
\hline
{\tt zgauss}   & ${\rm E}\,(\keV)$ & $6.75^{+0.04}_{-0.07}$ \\
               & ${\rm flux}\,(\phpcmsqps)$ & $5.73^{+3.08}_{-1.82} \times 10^{-6}$ \\
               & ${\rm EW}\,(\eV)$ & $14^{+8}_{-4}$ \\
\hline
{\tt zgauss}   & ${\rm E}\,(\keV)$ & $7.00^{+0.02}_{-0.03}$ \\
               & ${\rm flux}\,(\phpcmsqps)$ & $1.01^{+0.26}_{-0.29} \times 10^{-5}$ \\
               & ${\rm EW}\,(\eV)$ & $28^{+7}_{-8}$ \\
\hline
{\tt rdblur} & $\alpha$ & $1.31^{+0.61}_{-1.31}$ \\
               & $i\,(\degmark)$ & $41^{+15}_{-6}$ \\
               & $r_{\rm min}\,(r_{\rm g})$ & $6^{+74}_{-6}$ \\
               & $r_{\rm max}\,(r_{\rm g})$ & $400$ \\
\hline
{\tt reflion}  & {\rm Fe/solar} & $0.50$ \\
               & $\xi_{\rm refl}\,(\ergpcmps)$ & $30^{+5}_{-30}$ \\
               & $\Gamma_{\rm refl}$ & $2.11^{+0.03}_{-0.07}$ \\
               & $R_{\rm refl}$ & $1.75^{+0.67}_{-0.95}$ \\
               & ${\rm flux}\,(\phpcmsqps)$ & $2.09^{+0.80}_{-1.37} \times 10^{-5}$ \\
\hline
{\bf $\chi^2/\nu$} & & $1511/1418\,(1.07)$ \\
\hline \hline 
\end{tabular}
%\end{scriptsize}
\end{center}
\caption[Best fit parameters for Ark~120]{Best-fitting model
  parameters for the $2.5-10.0 \keV$ spectrum of Ark~120,
  including components and parameter values from $0.6-1.5 \keV$ for
  completeness.  The energies from $1.5-2.5 \keV$ were not included
  due to the presence of absorption edges from the {\it XMM-Newton}
  mirrors.  Error bars are quoted at $90\%$ confidence.  We required
  the {\tt zgauss} lines to be intrinsically narrow, i.e.,
  $\sigma=0 \keV$.  Redshifts were frozen at the cosmological value for
  the source, in this case $z=0.0327$.  
%A two-zone blackbody soft
%  excess is present in this object, consistent with findings in
%  several other Sy 1 AGN.
\label{tab:tab8.tex}}
\end{table}

\subsubsection{Fairall~9}
\label{sec:fairall9}

Fairall~9 is a radio-quiet Sy 1 galaxy with an elliptical companion,
both at a moderate redshift of $z=0.047$.  This source has not been
observed to undergo very large changes in X-ray flux over the time it
has been observed: the typical $2-10 \keV$ flux of Fairall~9 is
$F_{2-10} \sim 1.5-5.0 \times 10^{-11} \ergpcmsqps$
\citep{Reynolds1997,Gondoin2001}.  Previous X-ray observations with
{\it ASCA} show that the source has a continuum well-described by a
photoabsorbed power-law with a fluorescent Fe K$\alpha$
line and a high-energy tail.  The latter is a common signature of disk reflection, as
has been discussed previously with respect to a number of other
sources in this Section \citep{Reynolds1997,Nandra1997}.  A soft
excess has also been detected below $\sim 2 \keV$ \citep{Pounds1994},
though no strong evidence of warm absorption has been observed in the
soft spectrum \citep{Gondoin2001}.      

The source was observed with {\it XMM-Newton} in July
2000 for an effective duration of $\sim 29 \ks$ \citep{Jansen2001}, resulting in a
filtered EPIC-pn data set with $\sim 2.0 \times 10^5$ photons.
The EPIC-pn data were presented in full by Gondoin \etal the
following year \citep{Gondoin2001}.
We have followed the general procedures for data reduction and
analysis discussed by Gondoin \etal, using the most up-to-date
calibration files and software as
we have for all other sources analyzed in this work. 

Gondoin \etal find that continuum of Fairall~9 is best modeled with a
photoabsorbed power-law ($\Gamma=1.79^{+0.15}_{-0.12}$), in which the
absorbing column density of neutral
hydrogen gas is approximately equal to its Galactic value along the
line of sight to this source ($N_{\rm H}=3.19 \times 10^{20}
\pcmsq$).  A soft excess is noted below $\sim 2 \keV$, and an RGS
analysis yielded a best-fitting blackbody model for this component:
$kT=0.17^{+0.09}_{-0.07} \keV$.  Incorporating reflection from an
ionized disk ({\tt pexriv}) with
solar abundances and a reflection fraction of $R_{\rm refl}=1.0$, the
authors find that the high-energy tail is quite well modeled by a disk
at an inclination angle of $i=26 \degmark$ \citep{Gondoin2001}.  We have
performed a slightly different fit to the EPIC-pn data only, replacing
the {\tt pexriv} component with a
relativistically-blurred {\tt reflion} model, which also accounts
naturally for the soft excess without employing a phenomenological
blackbody component.  The column density of neutral hydrogen
absorption was again frozen at the Galactic value of $N_{\rm H}=3.19 \times
10^{20} \pcmsq$.  We found that $\Gamma=1.88^{+0.20}_{-0.04}$, and
the disk inclination angle fit to $i \sim 54 \degmark$, though its value 
could not be properly constrained.  Unfortunately, due to the relatively small number
of photons in the $0.5-10.0 \keV$ spectrum, parameter constraints were
more problematic for this source than for many of the others in our sample.

Gondoin \etal also note the presence of a $\sim 6.4 \keV$ Fe K$\alpha$ line
in the spectrum with $EW \sim 120 \eV$, which they modeled with a
Gaussian component.  Due to its relatively narrow profile, the line is
suggested to originate from low-ionization material orbiting
relatively far out in the disk.  Additionally, an Fe K absorption edge
is also noted at $7.64 \keV$ with $\tau=0.18$, consistent with
reflection from cold, optically thick material.  We find that the
inclusion of such an edge is not statistically robust in our model once
reflection is included via {\tt reflion} (which incorporates the iron edge), but
we do find the narrow $6.40^{+0.01}_{-0.02} \keV$ line in emission and model it with a
narrow Gaussian feature representing the core of the Fe K$\alpha$
line.  This feature has an $EW=115^{+20}_{-20} \eV$, but even after its inclusion
residual features remain indicating the presence of other unmodeled
narrow lines as well as a potential broader iron line
component as evidenced by a broadened red wing.  Adding in a second
narrow line at $E=7.06^{+0.02}_{-0.02} \keV$ ($EW=20^{+0}_{-0} \eV$),
representing a
contribution from Fe K$\beta$ emission (and so a flux of $0.16 \times$
that of K$\alpha$), resulted in a marginal statistical improvement
in fit ($\Delta\chi^2/\Delta\nu=-8/-2$), though the visual
improvement in fit was readily apparent.  Attempting to model the
remaining residual feature with a single broad line produced
unsatisfactory results, however.  As such, we added a third narrow
Gaussian line at $6.78^{+0.06}_{-0.05} \keV$ ($EW=29^{+18}_{-9} \eV$), representing ionized
iron, which further improved
the fit by $\Delta\chi^2/\Delta\nu=-7/-2$.  While both the
ionized iron line and Fe K$\beta$ line did not result in statistically
significant improvements in fit, according to the {\it F}-test, they
nonetheless produced substantial visual improvements to the fit and
nicely modeled the residual features that could not be accounted for
by a broad Fe K$\alpha$ line component.  A residual emission line-like feature
remains around $\sim 7.1 \keV$ in the observed frame, which likely
represents a line from Ni\,{\sc x-xvi}, but adding in an additional
Gaussian to model it results in no net improvement in fit so it is
left unmodeled.

It should be noted that
Gondoin \etal did not report the detection of a broad Fe K$\alpha$
line in their analysis, and that our attempts to include one show
that, while the inclusion of a broad component at $6.40 \keV$ does
improve our fit by $\Delta\chi^2/\Delta\nu=-13/-2$, a
relativistic disk line model such as {\tt diskline} or {\tt laor}
shows no improvement over a simple Gaussian with $\sigma=0.16 \pm 0.11
\keV$.  Further, the parameter values of the {\tt diskline} and {\tt
  laor} components are largely unconstrained, as one might expect.
Even so, our best statistical fit was achieved with a {\tt
  rdblur(reflion)} model, albeit with the emissivity index and
inclination angle of the disk unconstrained.  The inner radius of disk
emission was constrained to $r_{\rm min} \leq 29\,r_{\rm g}$.  

\begin{figure}
\begin{center}
\includegraphics[width=0.33\textwidth,angle=270]{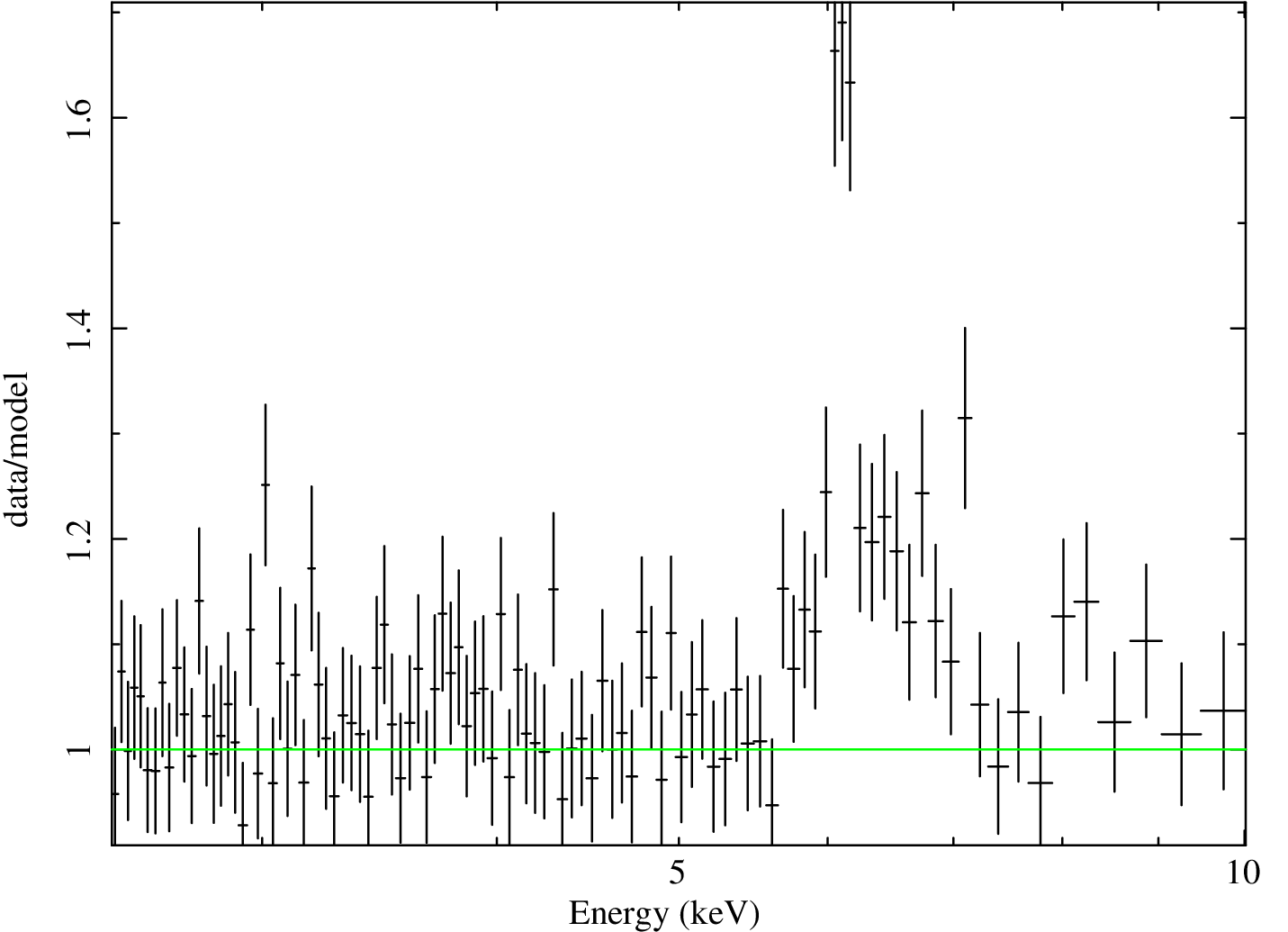}
\includegraphics[width=0.33\textwidth,angle=270]{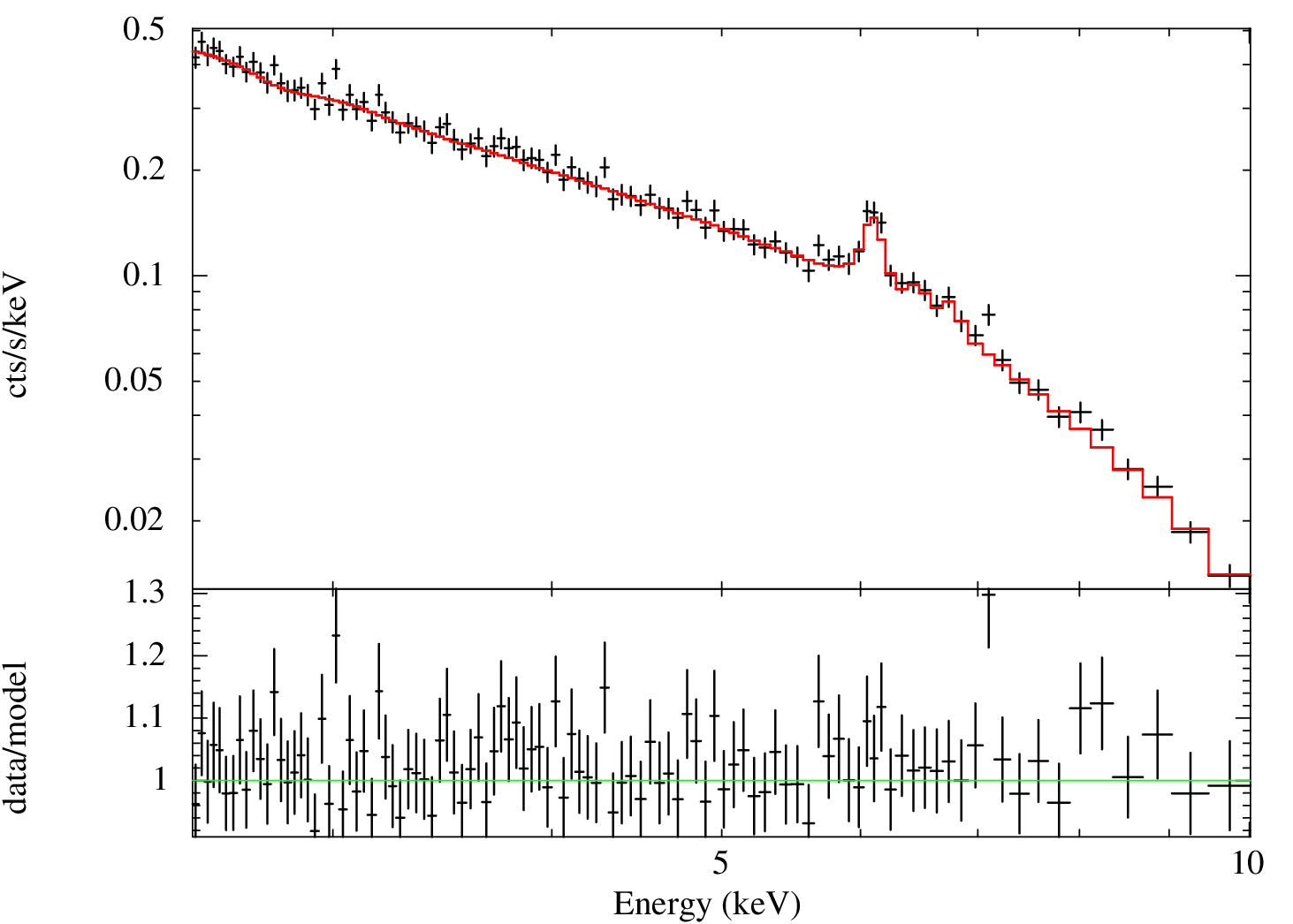}
\end{center}
\begin{center}
\includegraphics[width=0.33\textwidth,angle=270]{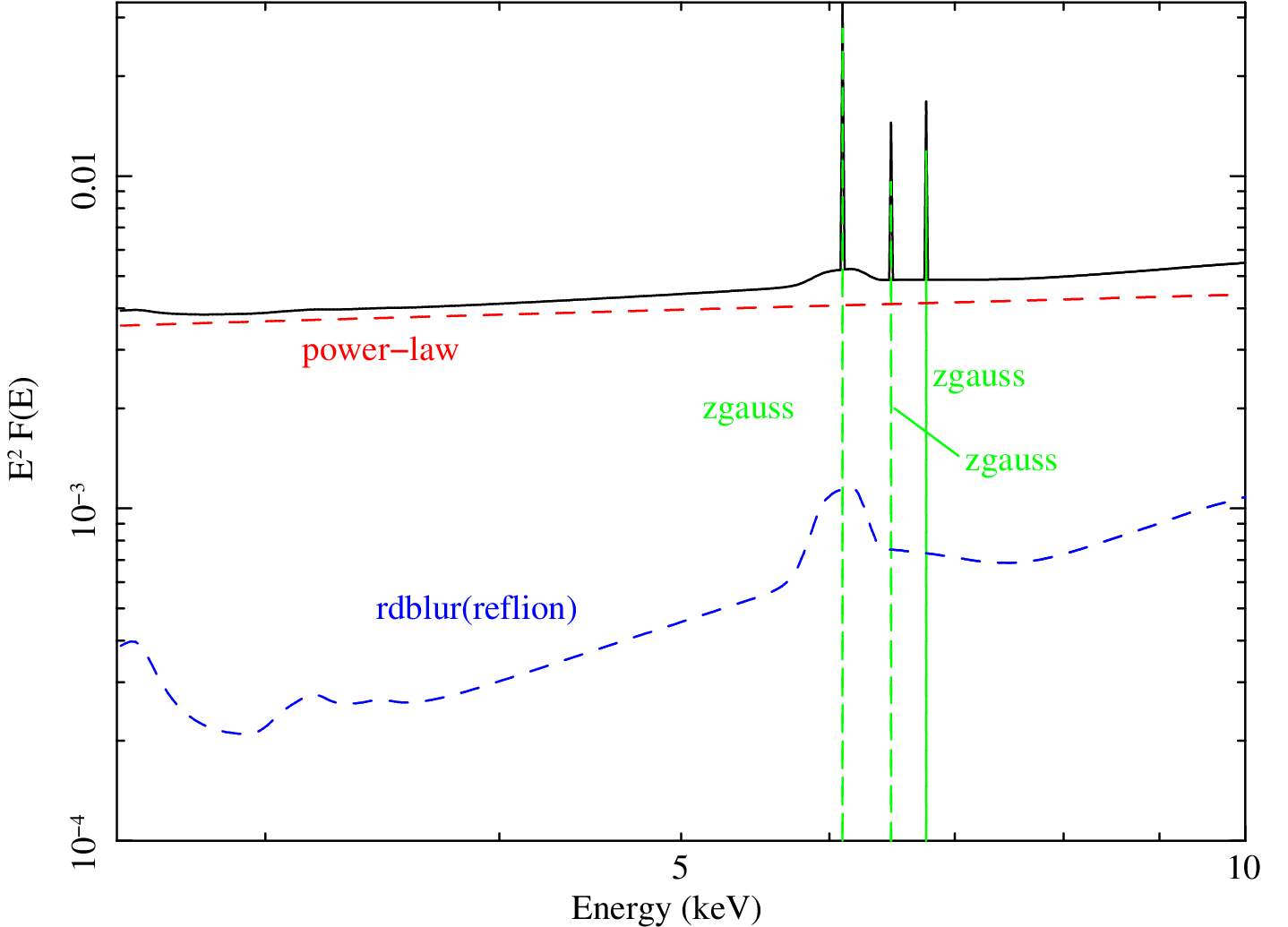}
\end{center}
\caption[Spectral fit for Fairall~9.]{{\it Top left:} The $2.5-10.0
  \keV$ spectrum of Fairall~9 fit with a continuum composed of a power-law model
  modified by Galactic photoabsorption and a blackbody soft excess.
  $\chi^2/\nu=816/699\,(1.17)$.  {\it Top right:} The $2.5-10.0
  \keV$ best-fit model for Fairall~9, including our continuum
  model and an ionized disk reflection spectrum convolved with the
  {\tt rdblur} relativistic smearing kernel.
  $\chi^2/\nu=679/691\,(0.98)$.  {\it Bottom:} An ${\rm
  E}^2\,{\rm F(E)}$ plot depicting the relative flux
  in each of the model components in the spectrum.}
\label{fig:fairall9_figs}
\end{figure}

%with an inner radius
%of emission of $r_{\rm min}\leq 6.27\,r_{\rm ms}$, or $\leq 10.75\,r_{\rm g}$,
%calculated using the fit value of BH spin ($a=0.972$).  The BH spin
%itself, however,
%was unable to be constrained.  
Best-fit values and
error bars for all the parameters in this model are listed in
Table~\ref{tab:tab9.tex}.
Fig.~\ref{fig:fairall9_figs} shows the iron line
residual, best fit to this residual, and relative contributions of the
individual model components, respectively.  The {\tt diskline} and {\tt laor}
fits are shown in Table~\ref{tab:tab10.tex}, while comparisons of the
ionized disk reflection spectrum with and without relativistic
smearing can be found in Table~\ref{tab:tab11.tex}. 

\begin{table}
\begin{center}
%\begin{scriptsize}
\begin{tabular} {|l|l|l|} 
\hline \hline
{\bf Model Component} & {\bf Parameter} & {\bf Value} \\
\hline \hline
{\tt phabs}    & $N_{\rm H}\,(\pcmsq)$ & $3.19 \times 10^{20}$ \\
\hline
{\tt zpo}       & $\Gamma_{\rm po}$ & $1.88^{+0.20}_{-0.04}$ \\
               & ${\rm flux}\,(\phpcmsqps)$ & $3.54^{+0.96}_{-0.41} \times 10^{-3}$ \\
\hline
{\tt zgauss}   & ${\rm E}\,(\keV)$ & $6.40^{+0.01}_{-0.02}$ \\
               & ${\rm flux}\,(\phpcmsqps)$ & $1.54^{+0.27}_{-0.27} \times 10^{-5}$ \\
               & ${\rm EW}\,(\eV)$ & $115^{+20}_{-20}$ \\
\hline
{\tt zgauss}   & ${\rm E}\,(\keV)$ & $6.78^{+0.06}_{-0.05}$ \\
               & ${\rm flux}\,(\phpcmsqps)$ & $3.72^{+2.36}_{-2.38} \times 10^{-6}$ \\
               & ${\rm EW}\,(\eV)$ & $29^{+18}_{-19}$ \\
\hline
{\tt zgauss}   & ${\rm E}\,(\keV)$ & $7.06^{+0.02}_{-0.02}$ \\
               & ${\rm flux}\,(\phpcmsqps)$ & $2.47^{+0.00}_{-0.00} \times 10^{-6}$ \\
               & ${\rm EW}\,(\eV)$ & $20^{+0}_{-0}$ \\
\hline
{\tt rdblur} & $\alpha$ & $10.00^{+0.00}_{-10.00}$ \\
               & $i\,(\degmark)$ & $54^{+36}_{-54}$ \\
               & $r_{\rm min}\,(r_{\rm g})$ & $8.04^{+21}_{-8.04}$ \\
               & $r_{\rm max}\,(r_{\rm g})$ & $400$ \\
\hline
{\tt reflion}  & {\rm Fe/solar} & $0.86^{+0.75}_{-0.48}$ \\
               & $\xi_{\rm refl}\,(\ergpcmps)$ & $50^{+575}_{-50}$ \\
               & $\Gamma_{\rm refl}$ & $1.88^{+0.20}_{-0.04}$ \\
               & $R_{\rm refl}$ & $0.70^{+7.00}_{-0.69}$ \\
               & ${\rm flux}\,(\phpcmsqps)$ & $2.58^{+12.45}_{-2.04} \times 10^{-6}$ \\
\hline
{\bf $\chi^2/\nu$} & & $679/691\,(0.98)$ \\
\hline \hline 
\end{tabular}
%\end{scriptsize}
\end{center}
\caption[Best fit parameters for Fairall~9]{Best-fitting model
  parameters for the $2.5-10.0 \keV$ spectrum of Fairall~9,
  including components and parameter values from $0.6-1.5 \keV$ for
  completeness.  The energies from $1.5-2.5 \keV$ were not included
  due to the presence of absorption edges from the {\it XMM-Newton}
  mirrors.  Error bars are quoted at $90\%$ confidence.  We required
  the $6.4 \keV$ {\tt zgauss} line to be intrinsically narrow, i.e.,
  $\sigma=0 \keV$.  Redshifts were frozen at the cosmological value for
  the source, in this case $z=0.0470$.  A warm absorber is present in
  this object, as in many
  other Sy 1 sources.\label{tab:tab9.tex}}
\end{table}

\section{Results}
\label{sec:compare_all}

We have compiled the relevant parameter constraints from all of our
spectral fits to eight Sy 1 sources: MCG--5-23-16,
Mrk~766, NGC~3783, NGC~4051, Ark~120, Fairall~9, NGC~2992, and 3C~273.
Table~\ref{tab:tab10.tex} presents the results of fitting the broad
iron line with a simple {\tt diskline} or {\tt laor} model as a first
approximation.  Table~\ref{tab:tab11.tex} displays the results of
fitting a more detailed, physical model consisting of many reflection features
from the accretion disk as a whole, both with and without relativistic
smearing via {\tt rdblur} and {\tt kdblur}.  
%Finally, Table~\ref{tab:tab4.tex} shows
%the fitting results for the case where {\tt kdblur} is replaced with
%{\tt kerrconv}.  
Note that, for nearly every source, a model including
reflection from an ionized disk convolved with a relativistic smearing
kernel provides the best statistical fit, as judged by the change in
chi-squared per additional degree of freedom in the model (i.e., the
{\it F}-test), tightness of the parameter
constraints and reduction in the residuals from the data/model ratio.
%In almost every
%case the {\tt kerrconv} model (with arbitrary spin) provides a better
%fit than the {\tt kdblur} model (spin fixed at $a=0.998$).

Previous studies have indicated that broad iron lines may be present
in up to $42\%$ of AGN observed with the {\it XMM-Newton}/EPIC-pn
camera that have $\gtrsim 10^4$ counts \citep{Guainazzi2006}.  To
assess whether evidence exists for a relativistically broadened iron
line in each of our data sets, we first considered the {\tt diskline}
and {\tt laor} fits
in which BH spin is fixed at $a=0.0$ and $a=0.998$, respectively.  While each source
demonstrated a significant improvement in its global fit as compared
to a model fitting only a narrow iron line core with a Gaussian
component, the most important parameter to evaluate in this case is the
inner radius of disk emission.  Roughly speaking, as stated in
\S\ref{sec:Fe_line}, if $r_{\rm min} \lesssim 20\,r_{\rm g}$ we can
say with some confidence that there is substantial emission from the
inner part of the accretion disk where relativistic effects (such as
BH spin) become important in shaping the overall iron line profile.
Out of our eight sources, Fairall~9 and Ark~120 had very poor
constraints on $r_{\rm min}$ based on their {\tt diskline} and
{\tt laor} fits.  Of the six sources where $r_{\rm min}$ could be
constrained, only MCG--5-23-16 did not show a substantial contribution
from $r_{\rm min} \leq
20\,r_{\rm g}$, either in the {\tt diskline} or {\tt laor} fit.  All of
our sources showed little
difference between the {\tt diskline} and {\tt laor} model fitting
results, statistically, as judged by their reduced $\chi^2$ values.
The four sources that did show very modest reductions in reduced
$\chi^2$ with a {\tt laor} model were MCG--5-23-16, Mrk~766, NGC~2992
and 3C~273.  Of these, MCG--5-23-16 showed the most significant improvement
in fit with {\tt laor} as opposed to {\tt diskline}, though it is
doubtful that this difference represents evidence for a spinning BH,
given the constraint on the inner disk radius in this source.

We have also examined the question of how robust the presence of a
broad line is in the spectrum when reflection is
included.  Beginning with a base continuum model including a static ionized
disk reflection spectrum \citep{Ross2005}, we noted the overall
goodness-of-fit and the residual features remaining, especially around
the iron line region.  We then convolved this model with relativistic
effects, first using the {\tt rdblur} smearing kernel,
then substituting in {\tt kdblur} (Table~\ref{tab:tab3.tex}).
Due to the nature of the BH spin parameter
space, the difference between an $a=0.8$ and an $a=0.998$ BH is much
more pronounced than the difference between an $a=0.0$ and an $a=0.8$ BH.
Therefore, while a significant improvement in the global fit with either model
indicates that relativistic smearing is important in the system, 
further significant improvement with {\tt kdblur} instead of {\tt
  rdblur} suggests that not only is relativity important,
but in the best-fit scenario the BH spin likely tends toward the
maximum theoretical value.    
%Whenever possible we allowed the iron abundance and ionization
%parameter to fit freely in the ionized disk model, though in some
%cases these additional degrees of freedom prevented us from obtaining
%a reliable fit.  For these sources it was necessary to assume a
%neutral disk ($\xi=30 \ergpcmps$) and solar iron abundance (Fe/solar=1):
%Ark~120, Fairall~9 and NGC~2992.  We were also forced to freeze the
%iron abundance in the 3C~273 fit.  

The {\tt rdblur} and {\tt kdblur} fits were at least as good or better
than the {\tt diskline} and {\tt laor} model fits in each source.  The
difference is greatest in the cases of MCG--5-23-16 and NGC~4051.
These smearing models also returned similar results in terms of
parameter constraints and overall fit to the iron line profile in each
source, and in nearly all cases also produced an
improvement in the overall goodness-of-fit over the non-smeared case.
More marginal improvements were seen in Fairall~9, NGC~2992 and
3C~273; not coincidentally, these are the three sources for which we have the
lowest photon count (each has $\leq 4 \times 10^5$ photons from $2-10
\keV$).  Note that it was also difficult to constrain several other
parameters in these AGN: the emissivity index and inner radius of disk
emission, particularly, and in the case of Fairall~9 even the
inclination angle of the disk to our line of sight.  Normally
inclination is well-constrained by the position of the blue wing of
the Fe K$\alpha$ line profile, but Fairall~9 has the fewest counts of
any of our sources, and so the blue wing cannot easily be separated
from the underlying continuum.  This underscores the importance of
photon statistics in evaluating the relevance of relativistic effects
in an AGN spectrum. 
Out of the five sources that showed significant
improvement with relativistic smearing, none showed a significant
statistical preference for either {\tt rdblur} or {\tt kdblur}
according to their reduced chi-squared values,
implying that we are not able to robustly distinguish between a non-spinning
and a rapidly-spinning BH in any of our sample members.  MCG--5-23-16,
Ark~120, NGC~2992 and 3C~273 all show a modest preference for {\tt
  kdblur} over {\tt rdblur}, while NGC~3783 shows the opposite.  The
other sources show equally good fits to each model.   

We note that the {\tt reflion} model \citep{Ross2005} does not include
a formal parameter for the
reflection fraction of the disk, so in order to facilitate comparison
between our results and those from studies using other models to
parameterize reflection (e.g., {\tt pexrav}, \citet{MZ1995}), we have
estimated $R_{\rm refl}$ in all of our models shown in
Table~\ref{tab:tab11.tex} and in \S4.
To make these estimates, we have assumed that for $R_{\rm refl}=1$, half
of the radiation from the hard X-ray power-law is received and
reflected by the disk, with the other half traveling directly to the
observer.  Reflection fractions of greater than unity then indicate
that more than half of the power-law radiation illuminates the disk,
implying anisotropic emission from the power-law source or perhaps
strong light-bending of this emission near the BH.  Operationally, we
compute $R_{\rm refl}$ using the following formula:
\begin{equation}
R_{\rm refl} = \frac{N_{\rm refl}}{N_{\rm po}}\, 
\frac{F_{\rm refl}}{F_{\rm po}}\, \left(\frac{\xi}{30}\right)^{-1}.
\end{equation}
Here $N_{\rm refl}$ and $N_{\rm po}$ denote the normalizations of the
{\tt reflion} and power-law components from our best-fit spectral
model, ($\xi$) is the best-fit ionization parameter characterizing the
disk reflection, and $F_{\rm refl}$ and $F_{\rm po}$ are the total
fluxes contained in the {\tt reflion} and power-law components for
unity normalization over the full wavelength range ($0.001-1000
\keV$).  These reflection fractions are often poorly constrained due
to the lack of simultaneous high energy data above $10 \keV$, which
would enable us to constrain the amount of reflection (as well as the
iron abundance and ionization of the disk) more effectively by
measuring the Compton hump, which typically peaks at $20-30 \keV$.
Having this high energy data would therefore allow us better
parameterize many properties of the disk, enabling better constraints
on the nature of the spacetime surrounding the BH (e.g., $r_{\rm min}$
and the BH spin).  For this reason, part II of this work will utilize
{\it Suzaku}/XIS+HXD data to establish robust BH spin constraints in
these AGN.

\clearpage
\begin{sidewaystable}
\begin{center}
\begin{footnotesize}
\begin{tabular} {|l|l|l|l|l|} 
\hline \hline
{\bf AGN} & {\bf $\alpha$} & {\bf $i\,(\degmark)$} & {\bf $r_{\rm
    min}\,(r_{\rm g})$} & {\bf $\chi^2/\nu$} \\
\hline \hline
{\bf MCG--5-23-16} & $1.35-3.04$ & $29-42$ & $\leq 66$ & $1513/1490\,(1.02)$ \\
                   & $1.98-4.30$ & $21-37$ & $18-44$ &
                   $1509/1490\,(1.01)$ \\
\hline
{\bf Mrk~766}      & $2.50-3.84$ & $10-31$ & $\leq 11$ & $1423/1309\,(1.09)$ \\
                   & $---$ & $58-72$ & $1.92-2.78$ &
                   $1421/1309\,(1.09)$ \\
\hline
{\bf NGC~3783}     & $1.94-1.96$ & $21-23$ & $\leq 18$ & $1478/1489\,(0.99)$ \\
                   & $1.38-2.87$ & $21-24$ & $\leq 25$ & $1485/1489\,(1.00)$ \\
\hline  
{\bf NGC~4051}     & $\leq 6.76$ & $12-52$ & $\leq 567$ & $1528/1398\,(1.09)$ \\
                   & $2.45-3.63$ & $15-45$ & $3.96-6.98$ &
                   $1528/1398\,(1.09)$ \\
\hline
{\bf Fairall~9}    & $---$ & $\leq 50$ & $---$ & $685/695\,(0.99)$ \\
                   & $---$ & $\leq 46$ & $---$ &
                   $688/695\,(0.99)$ \\
\hline
{\bf Ark~120}      & $\leq 1.98$ & $31-51$ & $\leq 174$ & $1528/1419\,(1.08)$ \\
                   & $---$ & $50-72$ & $---$ &
                   $1528/1419\,(1.08)$ \\
\hline
{\bf NGC~2992}     & $1.34-4.77$ & $44-62$ & $\leq 23$ & $1400/1311\,(1.07)$ \\
                   & $1.60-2.74$ & $45-63$ & $\leq 19$ & $1399/1311\,(1.07)$ \\
\hline
{\bf 3C~273}        & $\leq 2.97$ & $22-62$ & $9-38$ & $231/265(0.87)$ \\
                    & $---$ & $53-82$ & $\leq 15$ &
                    $228/265\,(0.86)$ \\
\hline \hline 
\end{tabular}
\end{footnotesize}
\end{center}
\caption[Comparison of {\tt diskline} and {\tt laor} model fits for
all sources.]{Comparison of the spectral fitting results for the
  $2.5-10.0 \keV$ spectra using the {\tt diskline} and {\tt laor}
  models.  For each source the
top row represents the {\tt diskline} range in parameter value while the bottom row
represents the {\tt laor} range in parameter value.  Ranges are
given at $90\%$ confidence.  A dashed line rather than a numerical
value indicates that the parameter could not be statistically
constrained in a formal error analysis. 
%For
%all sources, $\alpha_1=\alpha_2$ and $r_{\rm br}=6.0\,r_{\rm g}$.
\label{tab:tab10.tex}}
\end{sidewaystable}

\clearpage
\begin{sidewaystable}
\begin{center}
\begin{footnotesize}
\begin{tabular} {|l|l|l|l|l|l|l|l|} 
\hline \hline
{\bf AGN} & Fe/solar & {\bf $\xi_{\rm refl}\,(\ergpcmps)$} & {\bf
  $R_{\rm refl}$} & {\bf $\alpha$} & {\bf $i\,(\degmark)$} & {\bf $r_{\rm
    min}\,(r_{\rm g})$} & {\bf $\chi^2/\nu$} \\
\hline \hline
{\bf MCG--5-23-16} & $0.53-0.83$ & $\leq 34$ & $0.33-0.50$ & $$ & $$ &
                   $$ & $1555/1492\,(1.04)$ \\
                   & $0.49-0.70$ & $\leq 31$ & $0.33-0.44$ &
                   $1.05-3.13$ & $23-40$ & $\leq 69$ & $1492/1488\,(1.00)$ \\ 
                   & $0.56-0.64$ & $\leq 31$ & $0.39-0.43$ & $1.05-4.40$ &
                   $22-37$ & $\leq 64$ & $1485/1488\,(1.00)$ \\
\hline
{\bf Mrk~766}      & $0.65-4.11$ & $1506-5146$ & $\leq 0.01$ & $$ & $$ & $$ &
                   $1436/1311\,(1.10)$ \\
                   & $---$ & $509-2866$ & $\leq 0.14$ & $1.74-3.25$ &
                   $25-32$ & $\leq 17$ & $1426/1308\,(1.09)$ \\
                   & $---$ & $495-1290$ & $\leq 0.18$ & $1.84-3.64$ &
                   $23-32$ & $\leq 14$ & $1426/1308\,(1.09)$ \\
\hline
{\bf NGC~3783}     & $0.99-1.87$ & $33-72$ & $0.22-0.61$ & $$ & $$ &
                   $$ & $1497/1490\,(1.00)$ \\
                   & $0.71-1.71$ & $\leq 54$ & $0.12-0.71$ & $1.03-3.31$ & $21-26$ &
                   $\leq 46$ & $1480/1487\,(1.00)$ \\
                   & $0.70-1.67$ & $\leq 57$ & $0.12-0.70$ & $\leq 4.50$ & $20-26$ &
                   $\leq 44$ & $1486/1487\,(1.00)$ \\
\hline  
{\bf NGC~4051}     & $\leq 0.18$ & $133-847$ & $0.06-2.25$ & $$ & $$ &
                   $$ & $1485/1395\,(1.06)$ \\
                   & $0.18-0.70$ & $37-120$ & $0.43-24.53$ & $---$ & $26-38$
                   & $\leq 7.84$ & $1468/1392\,(1.05)$ \\
                   & $0.34-0.62$ & $38-125$ & $0.45-13.82$ & $---$ &
                   $23-39$ & $3.99-9.35$ & $1468/1392\,(1.05)$ \\
\hline
{\bf Fairall~9}    & $\leq 1.75$ & $331-2212$ & $\leq 0.10$ & $$ & $$ & $$ &
                   $684/696\,(0.98)$ \\
                   & $0.38-1.61$ & $\leq 625$ & $0.01-7.70$ & $---$ & $---$
                   & $\leq 29$ & $679/692\,(0.98)$ \\
                   & $0.28-2.37$ & $50-2684$ & $\leq 12.94$ & $---$ & $---$ &
                   $---$ & $679/692\,(0.98)$ \\
\hline
{\bf Ark~120}      & $0.93-1.60$ & $117-164$ & $0.13-0.31$ & $$ & $$ & $$ &
                   $1576/1423\,(1.11)$ \\
                   & $0.42-0.68$ & $\leq 35$ & $0.85-2.42$ &
                   $\leq 1.93$ & $35-57$ & $\leq 80$ &
                   $1511/1418\,(1.07)$ \\
                   & $0.42-0.65$ & $\leq 38$ & $0.69-2.07$ & $---$ & $50-55$ &
                   $---$ & $1509/1418\,(1.07)$ \\
\hline
{\bf NGC~2992}     & $0.66-6.06$ & $\leq 226$ & $\leq 0.35$ & $$ & $$ & $$ &
                   $1412/1313\,(1.08)$ \\
                   & $---$ & $\leq 310$ & $0.21-0.39$ & $\leq 3.00$ & $10-27$ &
                   $\leq 304$ & $1398/1310\,(1.07)$ \\
                   & $---$ & $\leq 67$ & $0.12-0.59$ & $---$ & $33-47$ &
                   $2.91-4.93$ & $1396/1310\,(1.07)$ \\
\hline
{\bf 3C~273}       & $---$ & $206-394$ & $0.02-0.30$ & $$ & $$ & $$ &
                   $236/267\,(0.88)$ \\
                   & $---$ & $\leq 1769$ & $0.01-16.75$ & $\leq 2.57$
                   & $30-44$ & $\leq 21$ & $232/264\,(0.88)$ \\
                   & $---$ & $\leq 5689$ & $---$ & $3.97-9.98$ &
                   $35-73$ & $\leq 3.73$ & $229/264\,(0.87)$ \\
\hline \hline 
\end{tabular}
\end{footnotesize}
\end{center}
\caption[Comparison of {\tt reflion}, {\tt rdblur(reflion)} and {\tt kdblur(reflion)} model fits for
all sources.]{Comparison of the spectral fitting results for the
  $2.5-10.0 \keV$ spectra using the ionized reflection model {\tt
    reflion}, {\tt rdblur(reflion)} and {\tt kdblur(reflion)}.  The
  latter two models convolve the reflection
  spectrum with smearing kernels from a non-spinning and
  maximally-spinning BH, respectively.  For each column of data, the
top row represents the {\tt reflion} range in parameter values, the
middle row shows the {\tt rdblur(reflion)} values and the bottom row
represents the {\tt kdblur(reflion)} values.  For the {\tt
  reflion} component, the value of the incident power-law spectral
index is set equal to that of the hard x-ray source.  Reflection fraction is estimated as described above in the text,
based on the relative fluxes and normalizations of the power-law and
{\tt reflion} components of this best-fit model.  Parameter value ranges are given at
$90\%$ confidence.\label{tab:tab11.tex}}
\end{sidewaystable} 

\clearpage

\topmargin=0.0in
\section{Discussion and Conclusions}
\label{chap:conclusions}
\label{sec:concl_whole}

This paper explores the use of broad iron emission lines to constrain
the importance of relativistic effects in a sample of AGN.  Following
our previous work on MCG--6-30-15
(BR06), this study of an additional eight prominent Sy 1 AGN observed
to have prominent iron lines has provided us with some intriguing data
on this subject and has also put us in an excellent position to apply
our spectral fitting techniques to many other sources in the future.

Though this sample is far too small to allow us to draw
any robust conclusions about the nature of the spacetime in the nuclei
of Sy 1
galaxies as a whole, this work has allowed us to design a viable method
for evaluating the presence of a broad iron line in the data and using
a sequence of models to assess the importance of relativistic smearing in shaping the
spectral features emitted from the accretion disk around the BH. 
We have determined that four out of our eight sources show
improvement in their goodness-of-fit parameters when
relativistic effects from the inner disk are taken into account (one
more shows at least marginal improvement).  Of these four, two show
more improvement with a smearing model designed for a
maximally-rotating BH than for a non-spinning BH, which may suggest 
that BH spin is non-zero in these objects.

Access to energies greater than the {\it XMM} limit of $10 \keV$ will help to
better constrain the disk parameters these sources, in turn allowing
for constraints to ultimately be placed on BH spin.  This endeavor
will be the focus of our follow-up {\it Suzaku} work on these AGN.
  
Based on our {\it XMM} data analysis we have
arrived at three important conclusions that we hope will inform
subsequent broad iron line and BH spin surveys:

\begin{enumerate}

\item \citet{Guainazzi2006} found that the fraction of AGN displaying
    broad iron lines in {\it XMM-Newton}/EPIC data increases
    dramatically when the spectra have more than $\sim 10^4$ counts.
    We find that EPIC spectra with $\sim 10^6$ counts are required in
    order to use these iron lines to assess the importance of
    relativistic effects in these sources.

\item Five out of nine sources from our broad Fe K line sample
    (including previous work on MCG--6-30-15) showed a significant improvement in their
    global goodness-of-fit when relativistic smearing was added to a
    static ionized disk reflection spectrum.

\item Though the {\it XMM-Newton}/EPIC-pn instrument allows us to resolve
  the broad iron line in excellent detail, its $10 \keV$ energy cutoff
  limits our ability to accurately model the disk reflection continuum
  due to its inability to measure the peak energy of the Compton
  reflection hump at $\sim 20-30 \keV$.  
\end{enumerate}

Further progress in this direction needs more data with newer, more
sensitive, broader bandwidth instruments in order
to provide constraints on BH spin.  {\it Suzaku}, with its
enhanced spectral resolution in the XIS detector and the HXD/PIN
detector simultaneously covering the energy range from $10-60 \keV$, provides great
advantages for this type of study.  Part II of this work (forthcoming)
will analyze archival {\it Suzaku} observations of these sources,
enabling us to measure the peak of the Compton hump and break the
degeneracies between the parameters of the disk and those governing
the spacetime near the BH.  {\it Suzaku} data will therefore allow us
to constrain BH spin using our {\tt kerrdisk} and {\tt
  kerrconv} models.

\section*{Acknowledgements}

We gratefully acknowledge the guidance and support of the University
of Maryland at College Park's Astronomy Department.  Helpful
contributions and advice for this work were imparted by LB's
dissertation defense committee: R.~Mushotzky, S.~McGaugh, S.~Veilleux,
M.~C.~Miller, and T.~Jacobson.  We are also indebted to Andy Fabian
for his insightful comments that improved this work, and to the
anonymous referee, who provided
several recommendations which have helped us to focus this manuscript.
LB also thanks
the NASA Postdoctoral Program, administered by ORAU, for her current
support at NASA's GSFC.  CSR gratefully acknowledges support from the
National Science Foundation (grant AST 06-07428), as well as the
U.Maryland/NASA-Goddard Center for Research and Exploration on Space
Science and Technology (CRESST).

\bibliographystyle{apj}
\bibliography{adsrefs}

\end{document}